\documentclass[a4paper, 11pt]{article}
\pdfoutput=1
\usepackage{jheppubnew} 
\usepackage{graphicx}
\usepackage{bm}
\usepackage{caption, subcaption}
\usepackage{ifthen}
\usepackage{amsmath}
\usepackage{amsfonts}
\usepackage{amssymb}
\usepackage{amsthm}
\usepackage{mathrsfs}
\usepackage[all]{xy}
\usepackage{xcolor}
\usepackage{comment}
\usepackage[numbers,sort&compress]{natbib}

\newcommand{\Reg}{\mathrm{reg}}
\newcommand{\Ren}{}
\newcommand{\Ct}{\mathrm{ct}}
\newcommand{\Anom}{\mathrm{anomaly}}

\newcommand{\be}{\begin{equation}}
\newcommand{\ee}{\end{equation}}
\renewcommand{\[}{\begin{equation}}
\renewcommand{\]}{\end{equation}}

\newcommand{\D}{\mathrm{d}}

\newcommand{\ep}{\epsilon}

\newcommand{\g}{\gamma}
\newcommand{\<}{\langle}
\renewcommand{\>}{\rangle}

\newcommand{\nn}{\nonumber}
\newcommand{\lla}{\langle \! \langle}
\newcommand{\rra}{\rangle \! \rangle}

\newcommand{\bs}[1]{\boldsymbol{#1}}
\newcommand{\x}{{\bs{x}}}

\newcommand{\p}{\partial}

\DeclareMathOperator{\K}{K}

\DeclareMathOperator{\Lo}{L}
\DeclareMathOperator{\Ro}{R}

\renewcommand{\2}{C}
\renewcommand{\sl}{\2^{(0)}}
\newcommand{\sdt}{D}
\newcommand{\ThreePt}{\empty}
\newcommand{\3}[1]{C_{
		\ifthenelse{\equal{\ThreePt}{\empty}}{#1}{
			\ifthenelse{\equal{#1}{\empty}}{\ThreePt}{\ThreePt,#1}}}}
\newcommand{\redef}[1]{{C'}_{
		\ifthenelse{\equal{\ThreePt}{\empty}}{#1}{
			\ifthenelse{\equal{#1}{\empty}}{\ThreePt}{\ThreePt,#1}}}}
\newcommand{\ren}[1]{C_{
		\ifthenelse{\equal{\ThreePt}{\empty}}{#1}{
			\ifthenelse{\equal{#1}{\empty}}{\ThreePt}{\ThreePt,#1}}}}
\newcommand{\sd}[1]{D_{
		\ifthenelse{\equal{\ThreePt}{\empty}}{#1}{
			\ifthenelse{\equal{#1}{\empty}}{\ThreePt}{\ThreePt,#1}}}}

\newcommand{\ct}{\mathfrak{c}}

\title{Renormalised 3-point functions of stress tensors and conserved currents in CFT}
\author[a,b]{Adam Bzowski,}
\author[c,d]{Paul McFadden}
\author[e]{and Kostas Skenderis.}
\affiliation[a]{Institute for Theoretical Physics, K.U.~Leuven, Belgium.}
\affiliation[b]{Institut de Physique Th\'{e}orique, CEA Saclay, Gif-sur-Yvette, France.}
\affiliation[c]{Theoretical Physics Group, Blackett Laboratory, Imperial College, London, U.K.}
\affiliation[d]{Centre for Particle Theory, Department of Mathematical Sciences, Durham University, U.K.}
\affiliation[e]{STAG Research Centre and Mathematical Sciences, University of Southampton, U.K.}
\emailAdd{adam.bzowski@ipht.fr} 
\emailAdd{paul.l.mcfadden@durham.ac.uk} 
\emailAdd{k.skenderis@soton.ac.uk}

\begin{document}

\abstract{

We present a complete momentum-space prescription for the renormalisation of tensorial correlators in conformal field theories.
Our discussion covers all 3-point functions of stress tensors and conserved currents in arbitrary spacetime dimensions.  
In dimensions three and four, we give explicit results for the renormalised correlators, the anomalous Ward identities they obey, and the conformal anomalies.
For the stress tensor 3-point function in four dimensions, we identify the specific evanescent tensorial structure responsible for the type A Euler anomaly, and show
this anomaly has the form of a double copy of the chiral anomaly.

} 

{
\renewcommand\beforetochook{\pagestyle{myplain}\pagenumbering{roman}\linespread{0.9}}

\maketitle
}

\section{Introduction}
\label{sec:intro}

The classic results for CFT 2- and 3-point functions are formulated in position space \cite{Polyakov:1970xd, Schreier:1971um, Osborn:1993cr,  Erdmenger:1996yc, DiFrancesco:1997nk}.  Nevertheless, for a growing number of modern applications  -- 
from conformal anomalies \cite{Cappelli:2001pz, Giannotti:2008cv, Armillis:2009pq, Coriano:2012wp,   Coriano:2017mux} 
to quantum critical transport 
\cite{Chowdhury:2012km, Huh:2013vga, Huh:2014eea,  Jacobs:2015fiv, Myers:2016wsu, Lucas:2016fju, Lucas:2017dqa}, to holographic cosmology \cite{Maldacena:2011nz, Bzowski:2011ab, Bzowski:2012ih, Schalm:2012pi, Mata:2012bx, Coriano:2012hd, McFadden:2013ria, Kundu:2014gxa, Ghosh:2014kba,  Pajer:2016ieg, Shukla:2016bnu, Isono:2016yyj} -- it is highly desirable to know the counterpart of these results {\it in  momentum space}.  Our aim in this paper, building on  \cite{Bzowski:2013sza}, is to fill this gap for  tensorial   correlators requiring renormalisation. 

Aside from practical applications, 
there are a number of other 
 motivations for developing momentum-space CFT.
One is the need to better integrate CFT methods with those of QFT more broadly. 
Away from the critical point, our 
toolkit is generally limited to perturbative methods such as Feynman diagrams, and 
such calculations are typically performed in momentum space.
Understanding CFT in momentum space would allow us to smoothly interpolate between physics at, and away from, the critical point.

Momentum space is also ideally suited
for the study of 
tensorial correlators, as is our focus here.  
Constructing a set of basis tensors from the metric and independent momenta, any tensorial correlator can be decomposed into a set of 
scalar form factors multiplying 
the elements of this basis.
As the momentum-space trace and transverse Ward identities are algebraic,  
we can moreover 
eliminate all non-transverse-traceless components 
by reducing them to simpler lower-point correlators.
Only transverse-traceless tensors then appear in our basis, and taking into account the 
permutation symmetries associated with exchanging identical operators, we quickly arrive at a minimal 
decomposition for the tensorial structure 
involving the smallest possible 
number of form factors.

For 2- and 3-point correlators, these form factors are simply functions of the corresponding momentum magnitudes.  The complicated tensorial Ward identities deriving from conformal symmetry 
now reduce to a simple set of scalar partial differential equations for the form factors.  
Those corresponding to the special conformal Ward identities factorise and can be solved by elementary separation of variables.  The remaining dilatation Ward identities can be solved by 
performing a Mellin transform to extract the components of appropriate scaling weight.
The resulting form factors 
are then conveniently expressed
in terms of  {\it triple-$K$ integrals}, a 
class of integrals involving three modified Bessel functions.

From a practical perspective, this first-principles approach, based on solving the conformal Ward identities directly in momentum space, is considerably more efficient than attempting to Fourier transform the known expressions for position-space correlators. 
In fact, 
for the correlators we study in this paper, a naive 
attempt to Fourier transform 
fails due to the appearance of 
divergences arising from integrating over configurations 
with coincident operator insertions. 
The position-space expressions for CFT correlators often quoted in textbooks are only valid at non-coincident points: specific contact terms must be added to    
produce 
{\it renormalised} correlation functions which are well-defined distributions, for which the Fourier transform then exists.
Such renormalised position-space expressions (constructed, for example, using differential renormalisation \cite{Freedman:1991tk}) for some of the correlators we study here can be found in \cite{Osborn:1993cr,  Erdmenger:1996yc}.
Even when renormalisation is not required, however, carrying out  the Fourier transform explicitly requires considerable effort, see for example \cite{Bzowski:2013sza}. 

Here we find it easier to work directly in momentum space and proceed from first principles.
For specific spacetime\footnote{In practice, we work in the Wick-rotated theory but will continue to refer to  ``spacetime" dimensions.} and operator dimensions, we find the triple-$K$ integrals representing the scalar form factors in our decomposition contain divergences.  
These divergences can be regulated by infinitesimally shifting the spacetime and operator dimensions. As  these are the only dimensionless parameters appearing in the conformal Ward identities, this scheme  represents the most general universally-applicable regularisation that preserves conformal invariance.
In this scheme the regulated form factors are simply given by triple-$K$ integrals with correspondingly shifted parameters.
To extract their divergences, we use 
 the Mellin mapping theorem, which relates the divergences of triple-$K$ integrals to the poles in a series expansion of their integrand.  
This easy evaluation of divergences is itself another of the advantages of momentum space.
Finally, to cancel the divergences, we add suitable local counterterms.  For the correlators of stress tensors and conserved currents we study here, these counterterms are constructed purely from the corresponding sources.  Their contribution, upon transforming back to position space, corresponds to the missing contact terms in our discussion above.  After removing the regulator, we then recover the finite, renormalised momentum-space correlator.

The counterterms we add introduce a dependence on the renormalisation scale, breaking conformal invariance.  
The renormalised correlators now obey modified (or `anomalous') conformal  Ward identities containing additional inhomogeneous terms  \cite{Deser:1976yx} (see also \cite{Duff:1993wm} and references therein).  As the trace Ward identity is similarly modified, the coefficients appearing in these anomalous Ward identities are exactly those appearing in the trace anomaly.
In the examples we study,
it turns out that both the 2- and the 3-point functions are renormalised by the same counterterm.  The anomaly coefficient we obtain is then related to the overall normalisation of the 2-point function.
Anomalies of this type, corresponding to type B in the classification of \cite{Deser:1993yx}, represent however only half the picture.
In addition, there can be type A anomalies, whose contribution to the trace of the stress tensor vanishes when integrated over all of flat space.
Unlike type B anomalies, type A anomalies are thus {\it scale-invariant}: they break special conformal transformations but not dilatations.
Type A anomalies arise when the regulated correlator features a divergent coefficient multiplying an evanescent tensorial structure that vanishes in the physical spacetime dimension.  The result is then finite, and does not represent a genuine UV divergence. 

Above two dimensions, and assuming parity invariance, the simplest flat-space correlator exhibiting  a type A anomaly is the stress tensor 3-point function in four dimensions.  Understanding the evanescent structure of the  regulated correlator, and its role in generating the Euler contribution to the trace anomaly, is in fact one of the motivations for our present study.
Setting aside the full problem for later, the general flavour of these ideas can already be appreciated from the 2-point function in two dimensions \cite{Deser:1993yx}.  This correlator also features a type A anomaly, and the corresponding tensorial structure is naturally much simpler.
In fact, the transverse and trace Ward identities alone constrain the $d$-dimensional momentum-space regulated correlator  to be
\[
\lla T_{\mu_1\nu_1}(\bs{p})T_{\mu_2\nu_2}(-\bs{p})\rra_{\Reg} = C_{TT}(\ep)\, p^{d}\,\Pi_{\mu_1\nu_1\mu_2\nu_2}(\bs{p}),
\]
where our double-bracket notation 
simply indicates the delta-function of  momentum conservation has been removed.  Here, $C_{TT}(\ep)$ is a ($d$-dependent) constant, and 
\begin{align}\label{projectors1}
\Pi_{\mu \nu \rho\sigma}(\bs{p}) & =  \pi_{\mu(\rho}(\bs{p}) \pi_{\sigma)\nu}(\bs{p}) - \frac{1}{d - 1} \pi_{\mu \nu}(\bs{p}) \pi_{\rho\sigma}(\bs{p}), \qquad
\pi_{\mu\nu}(\bs{p})  = \delta_{\mu\nu} - \frac{p_{\mu} p_{\nu}}{p^2}
\end{align}
are the $d$-dimensional transverse-traceless and transverse projectors respectively.
A short calculation shows this correlator can equivalently be expressed as  
\[
\lla T_{\mu_1\nu_1}(\bs{p})T_{\mu_2\nu_2}(-\bs{p})\rra_{\Reg} = -6 C_{TT}(\ep)\, p^{d-2}\,\Pi_{\mu_1\nu_1}{}^{\alpha_1}{}_{\beta_1}(\bs{p})\Pi_{\mu_2\nu_2}{}^{\alpha_2}{}_{\beta_2}(\bs{p})\,\delta_{[\alpha_1}^{\beta_1}\delta_{\alpha_2}^{\beta_2}p_{\alpha_3]}p^{\alpha_3}.\label{2pt2dttreg}
\]
The right-hand side now involves a 3-form, and hence vanishes in integer dimensions below three where a spacetime index in the antisymmetrisation must necessarily be repeated.  Since  the overall coefficient $C_{TT}(\ep)$ has an $\ep^{-1}$ pole in $d=2+2\ep$ dimensions, the regulated correlator thus has the $0/0$ structure associated with a type A anomaly.\footnote{In higher even dimensions, $C_{TT}(\ep)$ has pole but the 3-form does not vanish; we then have a genuine UV divergence which must be cancelled with a counterterm leading to a type B anomaly, see section \ref{sec:2ptfns}. In higher odd dimensions, $C_{TT}(\ep)$ is finite as $\ep\rightarrow 0$ and there is no anomaly.}  

The remaining steps 
of the renormalisation procedure 
can be found in appendix \ref{app:typeA}. The final  renormalised correlator,
\begin{align}\label{2drenttintro}
\lla T_{\mu_1\nu_1}(\bs{p})T_{\mu_2\nu_2}(-\bs{p})\rra_{\Ren} = C_{TT}\, p^2  \pi_{\mu_1\nu_1}(\bs{p})\pi_{\mu_2\nu_2}(\bs{p}),
\end{align}
has a non-vanishing trace associated with anomalous Ward identity, $\<T\>_s= (c/24\pi) R$, where the subscript $s$ indicates an expectation value in the presence of sources and the central charge is given by  $c= 12 \pi C_{TT}$.
Through two-dimensional identities \cite{Deser:1993yx}, 
we can re-write this trace anomaly 
as the square of the chiral anomaly:
\[
\lla T(\bs{p})T_{\mu\nu}(-\bs{p})\rra = \frac{c}{12\pi}\, (\ep_{\mu\alpha}p^\alpha)(\ep_{\nu\beta}p^\beta).
\]
Strikingly, as we will see, exactly the same is true for the four-dimensional Euler anomaly!  

Ultimately, type A anomalies always originate from such a 0/0 structure \cite{Deser:1993yx}. In position space, the $d=4$ case has been discussed in 
 \cite{Erdmenger:1996yc}.
Here we will see that the mechanism is particularly transparent in momentum space.  Along the way we will also explain that the 0/0 limit can be explicitly evaluated without counterterms (see appendix \ref{app:DS}), as one would expect for a type A anomaly \cite{Deser:1993yx}. 
 
Earlier  discussions of tensorial CFT correlators in momentum space include \cite{Armillis:2009pq, Coriano:2012wp}.
In these works, the momentum-space correlators in four dimensions were obtained through 1-loop Feynman diagram calculations, utilising the observation from position space \cite{Osborn:1993cr} that a mixture of free conformal scalars,  fermions and vectors is sufficient to generate all the tensor structures permitted by conformal symmetry.  
For general spacetime and operator dimensions, however, this approach is not always available due to the absence of a corresponding free field realisation of the CFT.  In such cases, a direct solution of the conformal Ward identities, 
as we develop here, 
appears to be the only possibility.\footnote{Where a holographic description is available, Witten diagram calculations effectively reduce to the triple-$K$ integrals we study here.}

Our plan for this paper is thus to solve for the renormalised 3-point correlators of tensorial operators in a general CFT, focusing on correlators of the stress tensor and conserved currents.  The extension to mixed correlators involving scalars involves new issues and will be discussed in a sequel \cite{Bzowski:2018fql}.
The key ingredients of our approach have been developed over a number of papers, beginning with \cite{Bzowski:2013sza} which introduced our tensorial decomposition and solution of the conformal Ward identities. (Related methods for purely scalar correlators were developed independently in \cite{Coriano:2013jba}.) 
The results of \cite{Bzowski:2013sza} are sufficient to understand all tensorial and scalar correlators in the general case where renormalisation is {\it not} required.\footnote{These results are presented in the published version of \cite{Bzowski:2013sza}; the original arXiv version 1 contains an additional discussion of renormalisation which is  superseded by the present work.}  
For the special cases where the operator and spacetime dimensions are such that divergences arise, we proposed a suitable renormalisation prescription  in \cite{Bzowski:2015pba}, focusing on purely scalar correlators.
Here, we extend this renormalisation prescription to tensorial correlators.  This requires a generalisation of our regularisation procedure, and an understanding of new issues such as the tensorial degeneracies giving rise to type A anomalies.
 All the triple-$K$ integrals we encounter in this paper can be evaluated using the reduction scheme presented in \cite{Bzowski:2015yxv}.

We begin in section \ref{sec:method} with an extended summary of our method for constructing renormalised tensorial 3-point functions.
We review the trace and transverse Ward identities, their role in decomposing the  tensorial structure of correlators into scalar form factors, and the classification of conformal Ward identities into primary and secondary.  These identities can be solved in terms of triple-$K$ integrals yielding the full momentum-space 3-point functions for generic values of the operator and spacetime dimensions. 
As we discuss, for specific values of these parameters, divergences then arise necessitating regularisation and renormalisation.

Our main results for renormalised 3-point functions are presented in section \ref{sec:results}.
After reviewing our conventions for 2-point functions, we proceed to analyse the correlators of three currents, of one stress tensor and two currents, and of three stress tensors.  For one current and two stress tensors, the correlator is trivial ({\it i.e.,} vanishes up to contact terms) and is omitted from our discussion here; see instead \cite{Costa:2011a} and section 9.10 of \cite{Bzowski:2013sza}.
For each correlator we list the relevant  tensorial decomposition, the Ward identities and their solution, plus the divergences and counterterms arising in different spacetime dimensions.  We give explicit results for the regulated form factors, valid in arbitrary spacetime dimensions, and carry out the full renormalisation procedure in dimensions three and four.  This enables a precise identification of the anomalies and the corresponding anomalous Ward identities satisfied by the renormalised correlators.

In section \ref{sec:Eulerdeg}, we discuss the type A Euler anomaly for the stress tensor 3-point function.  This presents the four-dimensional counterpart to our discussion of the two-dimensional type A anomaly above.
In section \ref{sec:findconsts}, we describe how to extract the physical, scheme-independent constants appearing in the renormalised correlators, before concluding in section \ref{sec:discussion}.
Five appendices present supplementary material.
Appendix \ref{app:typeA} completes our analysis of the two-dimensional anomaly, appendix \ref{app:DS} discusses the evaluation of 0/0 structures without the use of counterterms,
while appendix  \ref{sec:conversion} relates alternative definitions of the 3-point function. 
Appendix \ref{sec:evalctcontr} evaluates the 
counterterm contributions to the stress tensor 3-point function, and appendix \ref{app:3ddeg} reviews the   
evanescent tensorial 
operators appearing in three spacetime dimensions.

We put special effort into making this paper self-contained, and the different sections can be read independently 
according to the interests of the reader.
Section \ref{sec:method} reviews the relevant background material, and may be skipped by those already familiar with our method.
The different subsections of section \ref{sec:results} are then independent, so readers interested only in a specific correlator may head directly to the relevant subsection.  
Readers interested only in our results may do likewise,  after briefly reviewing our conventions in section \ref{sec:method}.

\section{Renormalisation of CFTs in momentum space}
\label{sec:method}

\subsection{Notations and conventions for momenta} 

In this paper, we consider CFTs in $d \geq 3$ Euclidean dimensions.  For simplicity, we will restrict ourselves to the parity-even sector.\footnote{Extensions to the parity-odd sector are nevertheless of considerable interest, see {\it e.g.,}  \cite{Closset:2012vg, Closset:2012vp, Bonora:2014qla, Bonora:2015odi, Bonora:2015nqa, Bonora:2017gzz, Almeida:2017lrq, Chowdhury:2017vel}.}
To get started, in this subsection we begin by recollecting some of our main notations and conventions.

Firstly, to denote correlators with the overall delta function of momentum conservation removed, we employ a double bracket notation, {\it e.g.,}
\begin{align}
\<T_{\mu_1\nu_1}(\bs{p}_1)T_{\mu_2\nu_2}(\bs{p}_2)T_{\mu_3\nu_3}(\bs{p}_3)\> &=  (2\pi)^d\delta(\bs{p}_1+\bs{p}_2+\bs{p}_3)\,\lla T_{\mu_1\nu_1}(\bs{p}_1)T_{\mu_2\nu_2}(\bs{p}_2)T_{\mu_3\nu_3}(\bs{p}_3)\rra, \nn\\[1ex]
\<T_{\mu_1\nu_1}(\bs{p}_1)T_{\mu_2\nu_2}(\bs{p}_2)\>&=  (2\pi)^d\delta(\bs{p}_1+\bs{p}_2)\,\lla T_{\mu_1\nu_1}(\bs{p}_1)T_{\mu_2\nu_2}(-\bs{p}_1)\rra.
\end{align}
Due to momentum conservation, only three of the six Lorentz scalars $\bs{p}_i\cdot\bs{p}_j$ for a given 3-point function are independent.  To preserve  symmetry under permutations of operators, we choose these to be 
the momentum magnitudes
\begin{equation}
p_j = | \bs{p}_j | = \sqrt{ \bs{p}_j^2 }, \qquad j = 1, 2, 3.
\end{equation}
To obtain compact expressions, we define the following symmetric polynomials of the momentum magnitudes, 
\begin{align}
a_{123} &= p_1 + p_2 + p_3, \qquad b_{123} = p_1 p_2 + p_1 p_3 + p_2 p_3, \qquad c_{123} = p_1 p_2 p_3, \nn\\
 a_{ij} &= p_i + p_j, \qquad\qquad\,\,\,\,\, b_{ij} = p_i p_j, \label{e:variables}
\end{align}
where $i,j = 1,2,3$, as well as the combination
\begin{align}
J^2 & = (p_1 + p_2 + p_3) (- p_1 + p_2 + p_3) (p_1 - p_2 + p_3) (p_1 + p_2 - p_3) \nn\\[0.5ex]
&= -p_1^4-p_2^4-p_3^4+2p_1^2p_2^2+2p_2^2p_3^2+2p_3^2p_1^2. \label{Jsqdef}
\end{align}
By Heron's formula, $\sqrt{J^2}/4$ represents the area of the triangle formed by the momenta.

When decomposing tensor structure, in order to preserve symmetry under permutations,
 we select the independent momenta using a cyclic rule according to the Lorentz index:
\begin{equation} \label{a:momenta}
\bs{p}_1, \bs{p}_2 \text{ for } \mu_1, \nu_1, \quad \bs{p}_2, \bs{p}_3 \text{ for } \mu_2, \nu_2, \quad  \bs{p}_3, \bs{p}_1 \text{ for }\mu_3, \nu_3.
\end{equation}
Here, the numbering of a given Lorentz index derives from the operator insertion it belongs to.
Use of this convention is assumed whenever we refer to the coefficient of a particular tensorial structure, {\it e.g.,} the coefficient of $p_2^{\mu_1}p_3^{\mu_2}p_1^{\mu_3}$ in $\lla J^{\mu_1 a_1}(\bs{p}_1)J^{\mu_2 a_2}(\bs{p}_2)J^{\mu_3 a_3}(\bs{p}_3)\rra$.  In other words, before the coefficient is read off, we first replace momenta as required (using momentum conservation) so as to ensure that only momenta consistent with \eqref{a:momenta} appear.

\subsection{Transverse and trace Ward identities}

Let us now recap the origin of the transverse and trace Ward identities satisfied by the correlators.
In a renormalised quantum field theory, under a variation of the sources, the variation of the generating functional is
\[\label{variationofW}
\delta W[g_{\mu\nu},A^a_\mu] = \delta \ln Z[g_{\mu\nu},A^a_\mu] = -\int\D^d\x\sqrt{g}
\Big[\frac{1}{2}\<T_{\mu\nu}\>_s\,\delta g^{\mu\nu}+\<J^{\mu a}\>_s\,\delta A^a_{\mu}
\Big],
\]
where the subscript $s$ on the 1-point functions indicates the presence of nontrivial sources.
The gauge field $A^a_\mu$ sources a conserved current $J^{\mu a}$ associated with a (generally non-Abelian) symmetry group $G$, where $a=1,\,\ldots,\,\dim G$ with repeated indices summed.  
To obtain correlation functions, we functionally differentiate this generating functional with respect to the sources, before restoring them to their background values. (Namely, a flat metric with the gauge field switched off, which we denote by a subscript zero.)  We will return to the details of this procedure in section \ref{sec:3ptdefn}.

The transverse Ward identities derive from the invariance of the generating functional under  gauge transformations and diffeomorphisms.
For a gauge field transforming in the adjoint representation,
under a gauge transformation $\alpha^a$ and 
diffeomorphism $\xi^\mu$, 
the variation of the sources is
\begin{align}
\delta g^{\mu\nu} &= -2\nabla^{(\mu}\xi^{\nu)}, \\
\delta A^a_\mu &= \xi^\nu\nabla_\nu A^a_\mu+A_\nu^a\nabla_\mu \xi^\nu - \nabla_{\mu}\alpha^a - g f^{abc}A^b_\mu\alpha^c,
\end{align} 
where $g$ is the gauge coupling, $f^{abc}$ the structure constant of $G$, and the generators $T_R^a$ are normalised such that $\mathrm{tr}(T_R^aT_R^b)=\frac{1}{2}\delta^{ab}$.  Note that $\alpha^a$ is a Lorentz scalar so $\nabla_\mu\alpha^a = \partial_\mu\alpha^a$.
The invariance of the generating functional under each of these transformations then yields the respective Ward identities
\begin{align}\label{p:Jward}
0 &= \nabla_\mu\<J^{\mu a}\>_s+gf^{abc}A^b_\mu\<J^{\mu c}\>_s,\\
0&=\nabla^\mu\<T_{\mu\nu}\>_s-F^a_{\mu\nu}\<J^{\mu a}\>_s,\label{p:Tward}
\end{align}
where the field strength $F^a_{\mu\nu} = 2\nabla_{[\mu}A_{\nu]}^a+gf^{abc}A^b_\mu A^c_\nu$.
Functionally differentiating these identities twice with respect to the various sources, we then obtain the transverse Ward identities for 3-point functions.  In momentum space, these identities are algebraic and fully determine the longitudinal components of correlators, {\it e.g.,}
\begin{align}
& p_{1 \mu_1} \lla J^{\mu_1 a_1}(\bs{p}_1) J^{\mu_2 a_2}(\bs{p}_2) J^{\mu_3 a_3}(\bs{p}_3) \rra  \nn\\
& \qquad = g f^{a_1 b a_3} \lla J^{\mu_3 b}(\bs{p}_2) J^{\mu_2 a_2}(-\bs{p}_2) \rra - g f^{a_1 a_2 b} \lla J^{\mu_2 b}(\bs{p}_3) J^{\mu_3 a_3}(-\bs{p}_3) \rra.
\end{align}

The trace components of correlators involving the stress tensor are determined by the trace Ward identities.  These derive from Weyl transformations, under which\footnote{When scalar operators are present, one can additionally obtain beta functions, see \cite{Bzowski:2018fql}.} 
\begin{align}\label{Wvar1}
\delta g^{\mu\nu} = -2 \sigma g^{\mu\nu},
\qquad \delta A^a_\mu = 0, \qquad \delta W = \int\D^d\x\sqrt{g}\,\mathcal{A}\,\sigma,
\end{align}
where the anomaly $\mathcal{A}$ gives the transformation of the renormalised generating functional under a Weyl variation $\sigma$.
Functionally differentiating twice the generating relation
\[\label{tanom1}
\<T\>_s = \mathcal{A}
\]
then yields the various 3-point trace Ward identities.  We list these identities, and the accompanying transverse Ward identities, in the main results section of the paper.  In particular, we will obtain explicit expressions for all the anomalies that arise.

\subsection{Decomposition of tensor structure}

A key advantage of momentum space is that tensorial correlators can be decomposed in a basis of general tensors constructed from the metric and momenta. 
Each of these tensors appears multiplied by an accompanying scalar form factor, which for 2- and 3-point functions can be written simply as a function of the momentum magnitudes.
Moreover, as all trace and longitudinal components of correlators are completely determined by the trace and transverse Ward identities, this basis need only  include {\it transverse-traceless} tensors.
Thus, choosing these tensors so as to respect the permutation symmetries associated with exchanging identical operators, we obtain an extremely efficient representation of tensorial correlators in terms of a minimal number of scalar form factors.
Even the most complicated 3-point correlator, that of three stress tensors, can in general be represented in terms of just {\it five} scalar form factors.
As a full account of this decomposition may be found in \cite{Bzowski:2013sza}, here we will limit ourselves to a quick review of the main points.

First, it is useful to introduce a notation for the transverse(-traceless) parts of the conserved current and the stress tensor,\footnote{
While technically there is no difference between raised and lowered indices for flat-space correlators, we will retain the index placements inherited from \eqref{variationofW} as a reminder for when we compute counterterm contributions.}
\begin{align}
j^{\mu a}(\bs{p}) = \pi^{\mu}_{\alpha}(\bs{p}) J^{\alpha a}(\bs{p}), \qquad
t_{\mu \nu}(\bs{p}) = \Pi_{\mu \nu}^{\quad\alpha \beta}(\bs{p}) T_{\alpha \beta}(\bs{p}),
\end{align}
where the relevant projectors are defined in \eqref{projectors1}.
Considering, for example, the 3-point function of three currents, the transverse-traceless part then takes the form
\[\label{exampledecomp1}
\lla j^{\mu_1 a_1}(\bs{p}_1)j^{\mu_2 a_2}(\bs{p}_2)j^{\mu_3 a_3}(\bs{p}_3)\rra = \pi^{\mu_1}_{\alpha_1}(\bs{p}_1) \pi^{\mu_2}_{\alpha_2}(\bs{p}_2) \pi^{\mu_3}_{\alpha_3}(\bs{p}_3) X^{\alpha_1\alpha_2\alpha_3 a_1 a_2 a_3},
\]
where $X^{\alpha_1\alpha_2\alpha_3 a_1 a_2 a_3}$ represents the most general third-rank Lorentz tensor that can be constructed from the momenta and the metric, with scalar coefficients depending on the momentum magnitudes and carrying adjoint indices $a_1a_2a_3$.  
In fact, for each Lorentz index $\alpha_j$ with $j=1,2,3$, only a single momentum can appear, since the momentum $\bs{p}_j$ is projected out and the remaining two choices are related by momentum conservation: for example, $\pi^{\mu_2}_{\alpha_2}(\bs{p}_2)p_1^{\alpha_2} =-\pi^{\mu_2}_{\alpha_2}(\bs{p}_2)p_3^{\alpha_2
}$ since  $\bs{p}_1=-\bs{p}_2-\bs{p}_3$.  
This greatly reduces the number of tensor structures that can be written down.

In order to preserve the symmetry under permutations, we choose these independent momenta cyclically according to their Lorentz index following the rule \eqref{a:momenta}, {\it i.e.,} 
\[
p_2^{\alpha_1}, \quad p_3^{\alpha_2},\quad p_1^{\alpha_3}.
\]
We thus write $\pi^{\mu_2}_{\alpha_2}(\bs{p}_2)p_3^{\alpha_2}$ in place of $\pi^{\mu_2}_{\alpha_2}(\bs{p}_2)p_1^{\alpha_2}$, and similarly $\Pi_{\mu_1\nu_1 \alpha_1\beta_1}(\bs{p}_1)p_2^{\alpha_1}p_2^{\beta_1}$ in place of $\Pi_{\mu_1\nu_1\alpha_1\beta_1}(\bs{p}_1)p_2^{\alpha_1}p_3^{\beta_1}$ or $\Pi_{\mu_1\nu_1 \alpha_1\beta_1}(\bs{p}_1)p_3^{\alpha_1}p_3^{\beta_1}$.
Preserving the permutation symmetry in this way leads to a more efficient parametrisation involving fewer form factors, as discussed in section 1 of \cite{Bzowski:2013sza}.
We therefore have the decomposition 
\begin{align}\label{exampledecomp2}
X^{\alpha_1\alpha_2\alpha_3 a_1 a_2 a_3}
 &=
A_1^{a_1 a_2 a_3} p_2^{\alpha_1} p_3^{\alpha_2} p_1^{\alpha_3}  + \: A_2^{a_1 a_2 a_3} \delta^{\alpha_1 \alpha_2} p_1^{\alpha_3} + A_2^{a_1 a_3 a_2}(p_2 \leftrightarrow p_3) \delta^{\alpha_1 \alpha_3} p_3^{\alpha_2} \nn\\
& \qquad\qquad\qquad\qquad  + \: A_2^{a_3 a_2 a_1}(p_1 \leftrightarrow p_3) \delta^{\alpha_2 \alpha_3} p_2^{\alpha_1},
\end{align}
where the two scalar form factors $A_1^{a_1 a_2 a_3}$ and $A_2^{a_1 a_2 a_3}$ are general functions of the momentum magnitudes that can be compactly expressed using the symmetric polynomials \eqref{e:variables}.  By convention, if no arguments are specified, we assume the standard ordering of momenta $A_1^{a_1a_2a_3} = A_1^{a_1a_2a_3}(p_1,p_2,p_3)$, while $p_i\leftrightarrow p_j$ indicates exchanging momenta (but not adjoint indices),  {\it e.g.,} $A_1^{a_1a_2a_3}(p_1\leftrightarrow p_2) = A_1^{a_1a_2a_3}(p_2,p_1,p_3)$.
Various symmetry properties of the form factors under permutations can be deduced from the symmetry of the correlator itself, and are listed in the main results section.

For practical purposes, an important feature of this decomposition is that the form factors can easily be read off from a computation of the full 3-point correlator: one does not need to explicitly separate out the projection operators as per \eqref{exampledecomp1} then manipulate the remainder into the form \eqref{exampledecomp2}.
Instead, the form factors can be extracted simply by looking up the coefficients of certain tensor structures in the full correlator.  In the present case, for example, one finds
\begin{align}
A_1^{a_1 a_2 a_3} & = \text{coefficient of } p_2^{\mu_1} p_3^{\mu_2} p_1^{\mu_3}, \qquad
A_2^{a_1 a_2 a_3}  = \text{coefficient of } \delta^{\mu_1 \mu_2} p_1^{\mu_3}
\end{align}
in the full correlator $\lla J^{\mu_1 a_1}(\bs{p}_1) J^{\mu_2 a_2}(\bs{p}_2) J^{\mu_3 a_3}(\bs{p}_3) \rra$.  Again, we list these relations in the main results section.

We similarly list the `reconstruction formulae' that allow the full correlators to be reconstructed from their transverse-traceless parts, using the trace and transverse Ward identities to fill in the missing pieces.  These formulae typically involve the operator 
\begin{equation} \label{a:T}
\mathscr{T}_{\mu\nu \alpha} (\bs{p}) = \frac{1}{p^2} \left[ 2 p_{(\mu} \delta_{\nu)\alpha} - \frac{p_\alpha}{d-1} \left( \delta_{\mu\nu} + (d-2) \frac{p_\mu p_\nu}{p^2} \right) \right]
\end{equation}
which arises from decomposing a rank-two tensor into transverse-traceless, longitudinal and trace pieces according to 
\[\label{Idecomp}
\delta_{\mu(\alpha}\delta_{\beta)\nu}=\Pi_{\mu\nu\alpha\beta}(\bs{p})+\mathscr{T}_{\mu\nu (\alpha}(\bs{p})\,p_{\beta)}+\frac{1}{d-1}\pi_{\mu\nu}(\bs{p})\delta_{\alpha\beta}.
\]
Finally, we note that in specific spacetime dimensions special tensorial degeneracies exist that allow us to reduce the number of independent form factors.  For example, the number of independent form factors in the stress tensor 3-point function can be reduced from five to two in three dimensions, and from five to four in four dimensions.
We will return to discuss this feature in section \ref{sec:Eulerdeg} and appendix \ref{app:3ddeg}.

\subsection{Solving the conformal Ward identities}

Having decomposed tensorial 3-point functions into scalar form factors, we can now solve for these form factors by imposing the conformal Ward identities.  A comprehensive analysis of the momentum-space conformal Ward identities is presented in \cite{Bzowski:2013sza}, so here we will once again limit ourselves to a summary of the main points.

Inserting the form factor decomposition for a given correlator into the conformal Ward identities, one obtains an individual dilatation Ward identity for each form factor, plus two sets of equations which we refer to as {\it primary} and {\it secondary} conformal Ward identities \cite{Bzowski:2013sza} (or CWI, for short). 
The dilatation Ward identities simply require each form factor to be a homogeneous function of the momentum magnitudes with a certain scaling weight, as discussed in section 4.2 of \cite{Bzowski:2013sza}.
The primary CWI, which closely resemble the CWI for scalar 3-point functions \cite{Bzowski:2015pba}, can be expressed in terms of the second-order differential operators
\begin{align}
\K_{ij} & = \K_i - \K_j, \qquad
\K_j  = \frac{\partial^2}{\partial p_j^2} + \frac{d + 1 - 2 \Delta_j}{p_j} \frac{\partial}{\partial p_j},\label{a:Kij}
\end{align}
where $i,j=1,2,3$ and $\Delta_j$ denotes the conformal dimension of the $j$-th operator in a given 3-point function.  Thus, in $\< T_{\mu_1 \nu_1} J^{\mu_2} J^{\mu_3} \>$ for example, we have $\Delta_1 = d$ and $\Delta_2 = \Delta_3 = d - 1$.
Each primary CWI relates $\K_{ij}$ acting on a given form factor to a linear combination of other form factors (or else to zero).  In total, we obtain two independent primary CWI for each of the form factors present.

The primary CWIs can easily be solved through  elementary separation of variables, in combination with a Mellin transform to satisfy the dilatation Ward identities \cite{Bzowski:2015pba}.
Each form factor is then given by a linear combination of {\it triple-$K$ integrals}, defined as 
\begin{align}
I_{\alpha \{ \beta_1, \beta_2, \beta_3 \}}(p_1, p_2, p_3) & = \int_0^\infty \D x \: x^\alpha \prod_{j=1}^3 p_j^{\beta_j} K_{\beta_j}(p_j x), \label{a:I} 
\end{align}
where $K_{\beta_j}$ is a modified Bessel function of the second kind.  
The parameters $\alpha$ and $\{\beta_j\}=\{\beta_1, \beta_2, \beta_3\}$  vary according to the case at hand, but generally depend on the spacetime dimension $d$ and the conformal dimensions of operators.  
In addition, the solution involves a number of arbitrary constants $C_j$ multiplying the triple-$K$ integrals, which we refer to as primary constants.
To represent the form factors for a given correlator succinctly, we additionally define the `reduced' triple-$K$ integral
\begin{align}
J_{N \{ k_1,k_2,k_3 \}} & = I_{\frac{d}{2} - 1 + N \{ \Delta_1 - \frac{d}{2} + k_1, \Delta_2 - \frac{d}{2} + k_2 , \Delta_3 - \frac{d}{2} + k_3  \}}, \label{a:J}
\end{align}
where $\Delta_j$ is the conformal dimension of the $j$-th operator, as appropriate to the case at hand, and the
$\{k_j\} =\{k_1, k_2, k_3\}$ are integers.

When the triple-$K$ integrals representing a given form factor converge, the solution thus obtained is unique, assuming the absence of collinear singularities \cite{Bzowski:2013sza}.
Where triple-$K$ integrals diverge, we instead need to regulate and renormalise, as we will discuss shortly.  In such cases, the solution is no longer unique reflecting our freedom to change the renormalisation scheme through the addition of finite counterterms \cite{Bzowski:2015pba}.
Where all $\{\beta_j\}$ indices are half integer, triple-$K$ integrals reduce to elementary integrals and may be trivially evaluated.  In many other cases interest, including all those studied here, the triple-$K$ integrals can be evaluated using the reduction scheme presented in \cite{Bzowski:2015yxv}. 

The remaining set of equations derived from the conformal Ward identities, the secondary CWIs, can be expressed in terms of the first-order differential operators\footnote{In the original arXiv version of this paper, we used $\Lo_{s,n}$ and $\Ro_s$ where $s$ is the spin of the first operator.  As this operator is always either $T_{\mu\nu}$ or $J^\mu$, we have simplified by substituting $s=\Delta_1-d+2$ and redefining $\Lo_{s,n}\rightarrow \Lo_{N}$ with $N=s+n$ and $\Ro_s\rightarrow \Ro$.}
\begin{align}
\Lo_{N} & = p_1 (p_1^2 + p_2^2 - p_3^2) \frac{\partial}{\partial p_1} + 2 p_1^2 p_2 \frac{\partial}{\partial p_2} \nn\\
& \qquad + \: \left[ (2d  - \Delta_1-2 \Delta_2 + N) p_1^2 + (2\Delta_1 -d) (p_3^2 - p_2^2) \right], \label{a:L} \\
\Ro & = p_1 \frac{\partial}{\partial p_1} - 2\Delta_1 +d, \label{a:R} \\
\Lo'_{N} & = \Lo_{N} \text{ with } (p_1 \leftrightarrow p_2) \text{ and } (\Delta_1 \leftrightarrow \Delta_2), \label{a:L2} \\
\Ro' & = \Ro \text{ with } (p_1 \rightarrow p_2) \text{ and } (\Delta_1 \rightarrow \Delta_2). \label{a:R2}
\end{align}
Each secondary CWI equates a linear combination of these operators acting on a given form factor to the coefficient of a particular tensor structure appearing in the longitudinal projection of the correlator ({\it i.e.,} its contraction with a momentum).  This coefficient can be read off from the transverse Ward identities.  Plugging in our solution for form factors in terms of triple-$K$ integrals, we simply obtain a set of constraints fixing some of the primary constants $C_j$ appearing in our original solution.
All surviving primary constants are then free parameters characterising the specific CFT at hand.

\subsection{Regularisation}

For specific operator and spacetime dimensions, the triple-$K$ integrals appearing in our solution for form factors diverge.  The construction of a suitable regularisation and renormalisation procedure for dealing with these singularities is our main focus in this paper.
Correlators for which renormalisation is not required were analysed in \cite{Bzowski:2013sza}.

The singularity condition for a general triple-$K$ integral $I_{\alpha\{\beta_j\}}$ is
\[\label{singcond}
\alpha+1 \pm\beta_1\pm\beta_2\pm\beta_3 = -2n,
\]
where $n$ denotes any non-negative integer ({\it i.e.,} $n=0,1,2\ldots$) and any independent choice of $\pm$ sign can be made for each $\beta_j$.
When this condition is satisfied, the triple-$K$ integral diverges as can be seen by expanding about its lower limit.
Whenever this condition is {\it not} satisfied, on the other hand, the triple-$K$ integral either converges absolutely, or else diverges but is uniquely defined by analytic continuation in the parameters $\alpha$, $\{\beta_j\}$ from regions where it does converge \cite{Bzowski:2013sza, Bzowski:2015pba}.

To regulate these divergences, we use the generalised dimensional scheme
\[\label{dimreg}
d \rightarrow \tilde{d} = d+2u\ep, \qquad \Delta_j \rightarrow \tilde{\Delta}_j = \Delta_j + (u+v_j)\ep,
\]
where $u$ and $v_j$ are constants representing a particular choice of  scheme (to which we will return below), and the regulator $\ep$ is a small parameter that will ultimately be set to zero after the divergences have been removed  through renormalisation.
This infinitesimal shift in the spacetime and operator dimensions produces a corresponding shift in the parameters $\alpha$ and $\{\beta_j\}$ appearing in triple-$K$ integrals.  As the singularity condition \eqref{singcond} is no longer satisfied, all triple-$K$ integrals and form factors are now finite.

Crucially, as  $d$ and $\Delta_j$ are the only dimensionless parameters appearing in the conformal Ward identities, the scheme \eqref{dimreg} represents the most general universally-applicable regularisation that preserves conformal invariance.
By preserving conformal invariance, we ensure  the  regulated 3-point functions take a fixed universal form, independent of all details of the theory, which would not be the case had we chosen a regularisation breaking conformal invariance.
The form factors of the regulated theory are therefore given by exactly the same set of triple-$K$ integrals we had before, except that all parameters are now shifted in the manner prescribed by \eqref{dimreg}.

Although these triple-$K$ integrals with shifted parameters are finite, they contain poles in $\ep$ which lead to divergences as $\ep\rightarrow 0$.
In general, these are single poles, although higher-order poles can arise in cases where the singularity condition \eqref{singcond} is multiply satisfied through different choices of the $\pm$ signs.
To proceed, we need to explicitly evaluate the form of these divergences.
This can be accomplished very easily as shown in \cite{Bzowski:2015pba}, since the divergent poles can be directly read off from a series expansion about $x=0$ of the integrand of the regulated triple-$K$ integral.  
Writing this expansion as
\[
x^{\tilde{\alpha}} \prod_{j=1}^3 p_j^{\tilde{\beta}_j} K_{\tilde{\beta}_j}(p_j x) = \sum_{\eta} c_{\eta} x^\eta,
\]
where $\tilde{\alpha}$ and $\tilde{\beta}_j$ represent the shifted parameters of the regulated integral,
the divergences are given by the formula\footnote{When the singularity condition \eqref{singcond} is multiply satisfied, the coefficients $c_{-1+w\ep}$ themselves contain poles in $\ep$, leading to a higher-order overall divergence.}
\[
I^{div}_{\tilde{\alpha},\{\tilde{\beta}_j\}} = \sum_{w}\frac{c_{-1+w\ep}}{w\ep} +O(\ep^0).
\]
In other words, the divergences arise from terms in the expansion of the form $x^{-1+w\ep}$ (for any finite nonzero $w$) that become poles in the limit as $\ep\rightarrow 0$.  In effect, this result follows from the Mellin mapping theorem, which relates  singularities in the Mellin transform of a function to its poles.
The ease with which divergences can be extracted in this fashion represents an additional advantage of momentum-space methods over position-space techniques such as differential regularisation \cite{Freedman:1991tk}.\footnote{See also \cite{PerezVictoria:2001pa, Coriano:2012wp} for an application of differential regularisation to CFT 3-point functions.}

Let us now return to discuss the choice of regularisation scheme specified by the parameters $u$ and $v_j$.
For conserved currents and stress tensors, as we focus on in this paper, gauge and diffeomorphism invariance in fact fix the dimensions of operators.  These canonical dimensions need to be preserved in order for the transverse Ward identities to hold in the regulated theory.
This requirement leads us to the unique regularisation scheme 
\[\label{truescheme}
u=v_1=v_2=v_3, \qquad \Leftrightarrow \qquad
\tilde{\Delta}_J = \tilde{d}-1, \qquad \tilde{\Delta}_T = \tilde{d}.
\]
A subtlety arises, however, which is that for certain correlators this choice of scheme is not always sufficient to regulate all singularities arising in the triple-$K$ integrals for form factors.
In such cases, we can start instead from a more general scheme, such as \eqref{dimreg} with all $v_j=v$ for $j=1,2,3$ but $v\neq u$. 
We then find that triple-$K$ integrals have poles of the form $(u-v)^{-1}$. These poles cancel when we sum up all triple-$K$ integrals that make up the form factors (or they are multiplied by factors that go to zero as $u \to v$) and after these cancellations one can take the limit \eqref{truescheme}.  Perhaps this computation could be re-organised (or the Ward identities solved) in such a way as to avoid introducing the scheme $u \neq v$ in intermediate steps, but we shall not pursue this here.

\subsection{Renormalisation}

\paragraph{Singularity type.}  
For any divergent triple-$K$ integral, there is a close correspondence between the choice of $\pm$ signs appearing in the solution of the singularity condition \eqref{singcond}, which we refer to as the {\it singularity type}, 
and the locality properties of the divergence \cite{Bzowski:2015pba}.  
Singularities of type $(---)$ ({\it i.e.,} those satisfying \eqref{singcond} with three minus signs) are {\it ultralocal}: they are analytic functions of the three squared momenta, corresponding in position space to divergences with support only when all three operator insertions coincide.
Singularities of type $(--+)$ (and its permutations
$(-+-)$ and $(+--)$) are instead {\it semilocal}: they are analytic in only two of the three squared momenta, corresponding in position space to divergences with support only when two of the three operator insertions coincide.
Singularities of type $(-++)$ (including permutations), and also of type $(+++)$, are instead {\it nonlocal}: they are non-analytic in two or more of the squared momenta, and hence in position space are non-vanishing even for separated operator insertions.  
Note these are nonlocal singularities of  individual triple-$K$ integrals, however, and not of the regulated 3-point function itself, which has only ultralocal and/or semilocal singularities.
We will return to the interpretation of these singularities shortly.

For correlators of three scalars, triple-$K$ integrals of all singularity types can arise.  For the correlators we study here, involving only stress tensors and conserved currents, the situation is  more straightforward.   
In fact, in our computations for form factors, we will only encounter triple-$K$ integrals with singularities of type $(---)$ and $(--+)$ (along with permutations), and never singularities of types $(-++)$ or $(+++)$.
Moreover, after imposing the secondary conformal Ward identities, we obtain primary constants such that all  singularities of type $(--+)$ actually cancel out when the triple-$K$ integrals are summed to construct the form factors.
Thus the regulated form factors have only ultralocal divergences, derived from singularities of type $(---)$.

\paragraph{Counterterms.}
The singularity type of a divergence dictates the manner of its removal\footnote{A Wilsonian approach to renormalization of composite operators in CFT can be found in \cite{Lizana:2015hqb,Lizana:2017sjz}.}.
First let us review this procedure in full generality, following \cite{Bzowski:2015pba}, before specialising our discussion to  correlators of stress tensors and conserved currents.
For a general 3-point correlator, we can construct counterterms using either three sources, or else two sources and an operator, along with an appropriate (even) number of covariant derivatives to bring the operator dimension up to the spacetime dimension $d$.  
Counterterms with three sources produce  {\it ultralocal} contributions to the corresponding 3-point function:
they can therefore  be used exclusively to cancel $(---)$ type singularities.
Counterterms of this type give rise to conformal anomalies (which are themselves ultralocal functions of the momenta), through their  dependence on the renormalisation scale.\footnote{ This dependence can be seen explicitly in the regulated theory, where a power of the renormalisation scale is required to bring the counterterm dimension to match the regulated spacetime dimension $\tilde{d}$.}

Counterterms involving two sources and an operator, on the other hand, 
produce {\it semilocal} contributions to 3-point functions.  
As discussed in \cite{Bzowski:2015pba}, their effect is to generate counterterm contributions proportional to the 2-point function of the operator concerned, and thus they can be used exclusively to cancel $(--+)$ type singularities (along with permutations).
Counterterms of this type correspond to a re-definition of the original bare source for the operator: through their dependence on the renormalisation scale, they give rise to a beta function for the new, renormalised source.   

Finally, while it is sometimes possible to construct
terms of classical dimension $d$ featuring two or more operators, terms of this type should be viewed as multi-trace {\it deformations} of the theory.  That is, unlike true counterterms, adding such terms changes the theory.\footnote{Moreover, in the  quantum theory, the renormalisation of UV divergences required to define composite operators potentially introduces anomalous dimensions, shifting the overall dimension away from $d$.}
Our inability to cancel nonlocal divergences through counterterms means that singularities of 
types $(+++)$ and $(-++)$ are in effect spurious: 
they can be removed only by choosing their coefficients (the corresponding primary constants) to vanish as sufficiently high positive powers of the regulator in the limit as $\ep\rightarrow 0$.  Thus, it is the {\it representation} of the 3-point function in terms of triple-$K$ integrals that is singular, and not the 3-point function itself \cite{Bzowski:2015pba}.
After these divergences are removed by selecting coefficients vanishing as sufficiently high powers of $\ep$, 
one recovers nonlocal 3-point functions satisfying all the usual (non-anomalous) conformal Ward identities.
(Note we cannot similarly eliminate singularities of types $(---)$ and/or $(--+)$, however, as this would leave us with ultralocal and/or semilocal 3-point functions, both of which are pure contact terms.)

\label{cancellation}

Returning now to the case at hand, for correlators of stress tensors and conserved currents, the only cubic counterterms we can construct are in fact those involving three sources.
To preserve the transverse Ward identities, all counterterms must be covariant with respect to gauge transformations and diffeomorphisms, and this in turn requires all source dependence to enter in the form of gauge-invariant field strengths and spacetime curvature invariants.  Moreover, in general there are no composite objects transforming as either a metric or a gauge connection, meaning counterterms cannot be constructed by redefining the sources.
On purely dimensional grounds then, we cannot  construct any local covariant counterterms involving two sources and one operator: while counterterms must have dimension $d$, both the field strength and the Riemann tensor have dimension two, and the operators $T_{\mu\nu}$ and $J^{\mu a}$ have dimensions $d$ and $d-1$ respectively.
This absence of counterterms involving two sources and one operator accounts for the observed cancellation of $(--+)$ type singularities in the regulated form factors.

Instead, the $(---)$ type singularities that remain can all be cancelled through covariant counterterms  that are cubic in the sources.
In any given spacetime dimension, it is a straightforward exercise to construct all such possible counterterms and evaluate their contribution to the 3-point function at hand.
After adjusting the counterterm coefficients so as to remove all divergences, the renormalised correlators then follow by sending $\ep\rightarrow 0$ to remove the regulator.  Due to the addition of counterterms, the trace and conformal Ward identities satisfied by the renormalised correlators are potentially anomalous.  The relevant form of these anomalies can easily be determined by inspection in any given case. The conformal anomalies produced in this fashion are the type B anomalies in the terminology of \cite{Deser:1993yx}. In addition, there can be  type A anomalies which, like the chiral anomaly, are produced through a more subtle 0/0 mechanism;  the four-dimensional Euler anomaly we discuss in section \ref{sec:Eulerdeg} is one such example.

\paragraph{Definition of 3-point function.} \label{sec:3ptdefn}

In this paper, we define the 3-point function through functionally differentiating the generating functional three times.  In particular, to preserve symmetry under permutations of the operator insertions, we choose to position all the $\sqrt{g}$ factors {\it outside} all functional derivatives, {\it e.g.,} 
\begin{align}\label{newdef}
&\<T_{\mu_1\nu_1}(\x_1)T_{\mu_2\nu_2}(\x_2)T_{\mu_3\nu_3}(\x_3)\>
\nn\\&\qquad\qquad\qquad
 \equiv \frac{-8}{\sqrt{g(\x_1)g(\x_2)g(\x_3)}}\,\frac{\delta}{\delta g^{\mu_1\nu_1}(\x_1)}\frac{\delta}{\delta g^{\mu_2\nu_2}(\x_2)}\frac{\delta}{\delta g^{\mu_3\nu_3}(\x_3)}W\Big|_0.
\end{align}
Notice that with this positioning, the factors of $\sqrt{g}$ are effectively set to unity.

An alternative definition of the 3-point function, which we employed in \cite{Bzowski:2013sza}, is to consider three insertions of the operator obtained through a {\it single} functional differentiation ({\it i.e.,} three insertions of the operator corresponding to the usual flat-space stress tensor or conserved current). 
With this alternative definition, 
\begin{align}
\label{olddef}
&\Big(\frac{-2}{\sqrt{g(\x_3)}}\frac{\delta}{\delta g^{\mu_3\nu_3}(\x_3)}\Big)\Big(\frac{-2}{\sqrt{g(\x_2)}}\frac{\delta}{\delta g^{\mu_2\nu_2}(\x_2)}\Big)\Big(\frac{-2}{\sqrt{g(\x_1)}}\frac{\delta}{\delta g^{\mu_1\nu_1}(\x_1)}\Big)W\Big|_0\nn\\[1ex]
&\equiv\<T_{\mu_1\nu_1}(\x_1)T_{\mu_2\nu_2}(\x_2)T_{\mu_3\nu_3}(\x_3)\>_{\mathrm{there}} \nn\\[1ex]&\quad
-2\<\frac{\delta T_{\mu_1\nu_1}(\x_1)}{\delta g^{\mu_2\nu_2}(\x_2)}T_{\mu_3\nu_3}(\x_3)\>
-2\<\frac{\delta T_{\mu_1\nu_1}(\x_1)}{\delta g^{\mu_3\nu_3}(\x_3)}T_{\mu_2\nu_2}(\x_2)\>
-2\<\frac{\delta T_{\mu_2\nu_2}(\x_2)}{\delta g^{\mu_3\nu_3}(\x_3)}T_{\mu_1\nu_1}(\x_1)\>,
\end{align}
where the functional derivatives reflect the implicit dependence of the operators on the sources.\footnote{Note that (despite appearances) the right-hand side is symmetric under 
exchanging $(\x_i, \mu_i, \nu_i)$ and $(\x_j, \mu_j, \nu_j)$ for any $i, j=1, 2, 3$. To verify this one uses the fact that $\delta T_{\mu_i \nu_i}(\x_i)/\delta g^{\mu_j \nu_j}(\x_j)$ is related to  $\delta T_{\mu_j \nu_j}(\x_j)/\delta g^{\mu_i \nu_i}(\x_i)$, since both  can be obtained from  $\delta^2S/\delta g^{\mu_i \nu_i}(\x_i)\delta g^{\mu_j \nu_j}(\x_j)$ (see, {\it e.g.}, (4.4) of \cite{McFadden:2011kk}).} 
The relation between these two definitions is then 
\begin{align}
&\<T_{\mu_1\nu_1}(\x_1)T_{\mu_2\nu_2}(\x_2)T_{\mu_3\nu_3}(\x_3)\>_{\mathrm{here}}\nn\\[1ex]
&\quad =\<T_{\mu_1\nu_1}(\x_1)T_{\mu_2\nu_2}(\x_2)T_{\mu_3\nu_3}(\x_3)\>_{\mathrm{there}} 
\nn\\[1ex]&\quad\quad
-2\<\frac{\delta T_{\mu_1\nu_1}(\x_1)}{\delta g^{\mu_2\nu_2}(\x_2)}T_{\mu_3\nu_3}(\x_3)\>
-2\<\frac{\delta T_{\mu_1\nu_1}(\x_1)}{\delta g^{\mu_3\nu_3}(\x_3)}T_{\mu_2\nu_2}(\x_2)\>
-2\<\frac{\delta T_{\mu_2\nu_2}(\x_2)}{\delta g^{\mu_3\nu_3}(\x_3)}T_{\mu_1\nu_1}(\x_1)\>
\nn\\[1ex]&\quad\quad
+\delta_{\mu_3\nu_3}\delta(\x_{13})\<T_{\mu_1\nu_1}(\x_1)T_{\mu_2\nu_2}(\x_2)\>
+\delta_{\mu_3\nu_3}\delta(\x_{23})\<T_{\mu_1\nu_1}(\x_1)T_{\mu_2\nu_2}(\x_2)\>\nn\\[1ex]&\quad\quad
+\delta_{\mu_2\nu_2}\delta(\x_{12})\<T_{\mu_1\nu_1}(\x_1)T_{\mu_3\nu_3}(\x_3)\>.
\end{align}
By expanding the functional derivatives in a local basis of operators, as described in \cite{Bzowski:2013sza}, all correlators involving functional derivatives can be reduced to ordinary 2-point functions.  The difference between these two definitions is thus simply a set of contact terms with support only when two operator insertions coincide.  
These contact terms are at most semilocal {\it i.e.,} individually non-analytic in at most a single squared momentum. 

Where these contact terms are {\it finite}, as was the case for all correlators studied in \cite{Bzowski:2013sza}, which of these two definitions of the 3-point function is adopted is purely a matter of choice.
One can straightforwardly convert from one definition to the other by adding or subtracting the corresponding contact terms, which can be evaluated by decomposing the functional derivatives in a local basis of operators.
The required formulae for this conversion are listed in appendix \ref{sec:conversion}.

In the present paper, however, we will encounter many cases where these contact terms involving functional derivatives are potentially {\it singular}, as the 2-point functions to which they reduce are themselves singular.  
In view of this, we have adopted here the definition \eqref{newdef} based on three functional derivatives.  
This definition is guaranteed to be finite, unlike that in \eqref{olddef}, since it corresponds to taking three functional derivatives of the {\it renormalised} generating functional with respect to the {\it renormalised} sources \cite{Shore:1990wp}.
Moreover, by subsuming all terms involving functional derivatives into the definition of the 3-point function, a number of steps in our computations are simplified; in particular, the secondary CWIs and reconstruction formulae take a simpler form.  
Use of the definition \eqref{newdef} also facilitates comparison with the literature ({\it e.g.}, \cite{Osborn:1993cr}).

\paragraph{Uniqueness.}
Defining the 3-point function as above, the form of all conformal and transverse Ward identities is uniquely fixed \cite{Osborn:1993cr,  Erdmenger:1996yc}.  The solutions we find are likewise unique: in particular,  
no  semi- or ultralocal terms are permitted besides those we list explicitly.  

This is obvious for the cases that do not require renormalisation: as mentioned earlier, in these cases the triple-$K$ integral is the unique solution of the CWIs without collinear singularities \cite{Bzowski:2013sza}.
When the triple-$K$ integral is divergent,  
however, 
there is a possibility of obtaining 
new local solutions
by replacing some
of the Bessel $K$ functions 
with Bessel $I$ 
and extracting the coefficient of the leading divergence.
This coefficient 
by itself solves the primary CWIs 
and is free from collinear singularities. Such cases were discussed for scalar correlators in  appendix A.3 of \cite{Bzowski:2015pba}.
Here, we find that
this substitution leads to non-vanishing results only for $(--+)$ singularities, 
 where 
 a single
  Bessel $K$ corresponding to the $+$ sign is substituted.
While this procedure can generate additional semi- and ultralocal solutions of the {\it primary} CWI, however, we have explicitly checked in all cases that the {\it secondary} CWIs are never satisfied.

\section{Results for renormalised correlators}
\label{sec:results}

Having now reviewed the main points of our approach,
in this section we present our results for the renormalised 2- and 3-point functions of conserved currents and stress tensors.
 In each case, we list the tensorial decomposition of the correlator into form factors (including the associated transverse and trace Ward identities and reconstruction formulae), and the solution for the regulated form factors expressed in terms of triple-$K$ integrals.  We also enumerate the available counterterms and the conformal anomalies they give rise to, along with the anomalous conformal Ward identities satisfied by the renormalised form factors.
Our final results for 3-point functions depend on only a small number of parameters specific to the given CFT at hand.  These are the 2-point function normalisation constants, and any primary constants not fixed by the secondary CWIs.

\subsection{Review of 2-point functions}
\label{sec:2ptfns}

In our analysis of 3-point functions, we will need various results for 2-point functions. 
To this end, and to establish the required notation, let us briefly review the renormalisation of CFT 2-point functions.  Further details may be found in \cite{Bzowski:2013sza, Bzowski:2015pba}.

In spacetimes of general dimension $d>2$, the momentum-space 
2-point functions are\footnote{For $d=2$, see the introduction and appendix \ref{app:typeA}.  
For the currents, we assume a basis where the Killing form of the generators is diagonal.}
\begin{align}
\lla J^{\mu a}(\bs{p}) J^{\nu b}(-\bs{p}) \rra & = \2_{JJ} \delta^{ab} \pi^{\mu\nu}(\bs{p}) p^{d-2}, \label{e:2ptJJ} \\
\lla T_{\mu\nu}(\bs{p}) T_{\rho\sigma}(-\bs{p}) \rra & = \2_{TT} \Pi_{\mu\nu\rho\sigma}(\bs{p}) p^{d}. \label{e:2ptTT}
\end{align}
In even dimensions, however, these expressions are unsuitable.  In position space, they correspond to local functions with support  only when the operator insertions coincide.
By adding local counterterms to the action, as below, these 2-point functions can be set to zero, which in a unitary theory would imply the operators vanish identically.

In reality, new UV divergences arise as we approach even dimensions $d=2N$.  In the dimensionally regulated theory, working in a scheme with $u=v_1=v_2$ so that operators retain their canonical dimensions, we find
\[\label{general2ptcttdef}
C_{JJ}(\ep) = \frac{C_{JJ}}{u\ep} + C_{JJ}^{(0)}+O(\ep), \qquad 
C_{TT}(\ep) = \frac{C_{TT}}{u\ep} + C_{TT}^{(0)}+O(\ep),
\]
so that 
\begin{align}
\lla J^{\mu a}(\bs{p}) J^{\nu b}(-\bs{p}) \rra_{\Reg} & = \delta^{ab} \pi^{\mu\nu}(\bs{p})\, p^{2N-2} \left[ \frac{\2_{JJ}}{u \epsilon} + \2_{JJ} \ln p^2 + \sl_{JJ} + O(\epsilon) \right], \label{e:2ptJJreg} \\
\lla T_{\mu\nu}(\bs{p}) T_{\rho\sigma}(-\bs{p}) \rra_{\Reg} & = \Pi_{\mu\nu\rho\sigma}(\bs{p}) \,p^{2N} \left[ \frac{\2_{TT}}{u \epsilon} + \2_{TT} \ln p^2 + \sl_{TT} + O(\epsilon) \right]. \label{e:2ptTTreg}
\end{align}
The divergences as $\ep\rightarrow 0$ can be removed  through the addition of suitable  counterterms.  To quadratic order in the sources,\footnote{At higher orders, the Laplacians should be replaced by their Weyl-covariant generalisations, see \cite{GJMS, Papadimitriou:2004ap, Aros:2026gms}.} 
the necessary counterterms are 
\begin{align} 
S_{\Ct} & =  \ct_{JJ} \int \D^{2N + 2 u \epsilon} \bs{x} \sqrt{g}  F_{\mu\nu}^a \Box^{N-2} F^{\mu\nu a} \mu^{2 u  \epsilon}, \label{e:SctJJ} \\
S_{\Ct} & = \ct_{TT} \int \D^{2N + 2 u \epsilon} \bs{x} \sqrt{g} W_{\mu\nu\rho\sigma} \Box^{N-2} W^{\mu\nu\rho\sigma} \mu^{2 u  \epsilon}, \label{e:SctTT}
\end{align}
where $W_{\mu\nu\rho\sigma}$ denotes the Weyl tensor and
the renormalisation scale $\mu$ appears on dimensional grounds.
As we are assuming $d\ge 4$, the Laplacians are raised to positive integer powers so these counterterms are local.
 The counterterm coefficients can be expanded as
\begin{align} 
\ct_{JJ}(\ep) & = \frac{(-1)^{N} \2_{JJ}}{4 u \epsilon} + \ct_{JJ}^{(0)} + O(\epsilon), \label{e:ctJJ} \\
\ct_{TT}(\ep) & = (-1)^{N}\frac{(N-1)}{(2N-3)}\frac{ \2_{TT}}{4 u \epsilon} + \ct_{TT}^{(0)} + O(\epsilon), \label{e:ctTT}
\end{align}
where the finite pieces $\ct_{JJ}^{(0)}$ and $\ct_{TT}^{(0)}$ represent a particular choice of renormalisation scheme. After removing the regulator by sending $\ep\rightarrow 0$, we obtain the renormalised correlators 
\begin{align}
\lla J^{\mu a}(\bs{p}) J^{\nu b}(-\bs{p}) \rra & = \delta^{ab} \pi^{\mu\nu}(\bs{p})\, p^{2N-2} \left[ \2_{JJ} \ln \frac{p^2}{\mu^2} + \sdt_{JJ} \right], \label{e:2ptJJren} \\
\lla T_{\mu \nu}(\bs{p}) T_{\rho \sigma}(-\bs{p}) \rra & = \Pi_{\mu\nu\rho\sigma}(\bs{p})\, p^{2N} \left[ \2_{TT} \ln \frac{p^2}{\mu^2} + \sdt_{TT} \right]. \label{e:2ptTTren}
\end{align}
where 
\begin{align}
\sdt_{JJ} & = \sl_{JJ} - 4 (-1)^N \ct_{JJ}^{(0)}, \\[1ex]
\sdt_{TT} & = \sl_{TT} - 4 (-1)^N \frac{(2N-3)}{(N-1)}\ct_{TT}^{(0)}.
\end{align}
From the perspective of the renormalised theory, however, $\sdt_{JJ}$ and $\sdt_{TT}$ are simply scheme-dependent constants whose values can be 
changed arbitrarily by adjusting the renormalisation scale.
Unlike \eqref{e:2ptJJ} and \eqref{e:2ptTT}, the renormalised correlators \eqref{e:2ptJJren} and \eqref{e:2ptTTren} are nonlocal, being non-analytic functions of $p^2$.
Due to the explicit $\mu$-dependence of the counterterms,  they acquire an anomalous scale dependence
\begin{align}
\mu \frac{\partial}{\partial \mu} \lla J^{\mu a}(\bs{p}) J^{\nu b}(-\bs{p}) \rra & = - 2 \delta^{ab} \pi^{\mu\nu}(\bs{p}) p^{2N-2} \2_{JJ}, \\
\mu \frac{\partial}{\partial \mu} \lla T_{\mu \nu}(\bs{p}) T_{\rho \sigma}(-\bs{p}) \rra & = - 2 \Pi_{\mu\nu\rho\sigma}(\bs{p}) p^{2N} \2_{TT}.
\end{align}
As usual, the anomalous terms on the right-hand side here are finite, local, and scheme independent.
They can be derived from the quadratic anomaly action
\begin{align}\label{anomaction}
& \delta_{\sigma} W[g_{\mu\nu},A^a_\mu] = \lim_{\ep\rightarrow 0}\delta_{\sigma} \Big[\ln \< e^{-S_{\Ct}} \>\Big]  \\[1ex]& \qquad = 
\frac{(-1)^{N+1}}{2} \int \D^{2N} \bs{x} \sqrt{g}\sigma \Big[C_{JJ} F_{\mu \nu}^a \Box^{N-2}F^{\mu \nu a} +  \frac{(N-1)}{(2N-3)} C_{TT}W_{\mu\nu\rho\sigma} \Box^{N-2} W^{\mu\nu\rho\sigma} \Big],\nn
\end{align}
which implies the trace anomaly\footnote{In v3, we have corrected a factor  in  
\eqref{e:ctTT} leading to agreement with the results of
 \cite{Aros:2026gms};
 we thank D.~Diaz  for correspondence.}
\[\label{traceanomquad}
\<T\>_s = \frac{(-1)^{N+1}}{2}\Big[C_{JJ} F_{\mu \nu}^a \Box^{N-2}F^{\mu \nu a} +\frac{(N-1)}{(2N-3)}   C_{TT} W_{\mu\nu\rho\sigma} \Box^{N-2} W^{\mu\nu\rho\sigma}\Big].
\]

\subsection{\texorpdfstring{$\<J^{\mu_1}J^{\mu_2}J^{\mu_3}\>$}{<JJJ>}}

\subsubsection{General analysis}

\paragraph{Decomposition.} The transverse Ward identity is
\begin{align}
& p_{1 \mu_1} \lla J^{\mu_1 a_1}(\bs{p}_1) J^{\mu_2 a_2}(\bs{p}_2) J^{\mu_3 a_3}(\bs{p}_3) \rra  \nn\\
& \qquad = g f^{a_1 b a_3} \lla J^{\mu_3 b}(\bs{p}_2) J^{\mu_2 a_2}(-\bs{p}_2) \rra - g f^{a_1 a_2 b} \lla J^{\mu_2 b}(\bs{p}_3) J^{\mu_3 a_3}(-\bs{p}_3) \rra.\label{TWI_JJJ}
\end{align}
The full 3-point function can thus be reconstructed from the transverse-traceless part using
\begin{align}
& \lla J^{\mu_1 a_1}(\bs{p}_1) J^{\mu_2 a_2}(\bs{p}_2) J^{\mu_3 a_3}(\bs{p}_3) \rra\nn\\[1ex]&\qquad = \lla j^{\mu_1 a_1}(\bs{p}_1) j^{\mu_2 a_2}(\bs{p}_2) j^{\mu_3 a_3}(\bs{p}_3) \rra \nn\\[1ex]
& \qquad\qquad + \: \Big(\Big[ \frac{p_1^{\mu_1}}{p_1^2}\Big( g f^{a_1 b a_3} \lla J^{\mu_3 b}(\bs{p}_2) J^{\mu_2 a_2}(-\bs{p}_2) \rra -  g f^{a_1 a_2 b} \lla J^{\mu_2 b}(\bs{p}_3) J^{\mu_3 a_3}(-\bs{p}_3) \rra \Big) \Big] \nn\\
& \qquad\qquad\qquad + \:  [ (\mu_1, a_1, \bs{p}_1) \leftrightarrow (\mu_2, b_2, \bs{p}_2) ] + [ (\mu_1, a_1, \bs{p}_1) \leftrightarrow (\mu_3, a_3, \bs{p}_3) ] \Big)\nn\\[1ex]
& \qquad\qquad + \: \Big(\Big[ \frac{p_1^{\mu_1} p_2^{\mu_2}}{p_1^2 p_2^2} g f^{a_1 a_2 b} p_{2 \alpha} \lla J^{\alpha b}(\bs{p}_3) J^{\mu_3 a_3}(-\bs{p}_3) \rra \Big] \nn\\
& \qquad\qquad\qquad + \: [ (\mu_1, a_1, \bs{p}_1) \leftrightarrow (\mu_3, a_3, \bs{p}_3) ] + [ (\mu_2, a_2, \bs{p}_2) \leftrightarrow (\mu_3, a_3, \bs{p}_3) ]\Big).
\end{align}
This formula is valid in any spacetime dimension $d$.

\paragraph{Form factors.} The tensor decomposition of the transverse-traceless part is
\begin{align}
& \lla j^{\mu_1 a_1}(\bs{p}_1) j^{\mu_2 a_2}(\bs{p}_2) j^{\mu_3 a_3}(\bs{p}_3) \rra \nn\\[1ex]& = \pi^{\mu_1}_{\alpha_1}(\bs{p}_1) \pi^{\mu_2}_{\alpha_2}(\bs{p}_2) \pi^{\mu_3}_{\alpha_3}(\bs{p}_3) \Big[
A_1^{a_1 a_2 a_3} p_2^{\alpha_1} p_3^{\alpha_2} p_1^{\alpha_3}  
\nn\\[1ex]& \quad 
+ A_2^{a_1 a_2 a_3} \delta^{\alpha_1 \alpha_2} p_1^{\alpha_3} 
 + A_2^{a_3 a_1 a_2}(p_3, p_1, p_2) \delta^{\alpha_1 \alpha_3} p_3^{\alpha_2}   + \: A_2^{a_2 a_3 a_1}(p_2, p_3, p_1) \delta^{\alpha_2 \alpha_3} p_2^{\alpha_1} \Big].
\end{align}
The form factors $A_1$ and $A_2$ are functions of the momentum magnitudes. If no arguments are given, then we assume the standard ordering, $A_j = A_j(p_1, p_2, p_3)$.

The $A_1$ factor is completely antisymmetric, \textit{i.e.}, for any permutation $\sigma$ of the set $\{1,2,3\}$ it satisfies
\begin{equation}
A_1^{a_{\sigma(1)} a_{\sigma(2)} a_{\sigma(3)}}(p_{\sigma(1)}, p_{\sigma(2)}, p_{\sigma(3)}) = (-1)^\sigma A_1^{a_1 a_2 a_3}(p_1, p_2, p_3),
\end{equation}
where $(-1)^\sigma$ denotes the sign of the permutation $\sigma$.
Under a permutation of the momenta only, however, the form factor is completely symmetric,
\begin{equation}
A_1^{a_1 a_2 a_3}(p_{\sigma(1)}, p_{\sigma(2)}, p_{\sigma(3)}) = A_1^{a_1 a_2 a_3}(p_1, p_2, p_3).
\end{equation}
The form factor $A_2$ is antisymmetric under $(p_1, a_1) \leftrightarrow (p_2, a_2)$, \textit{i.e.},
\begin{equation}
A_2^{a_2 a_1 a_3}(p_2, p_1, p_3) = -A_2^{a_1 a_2 a_3}(p_1, p_2, p_3).
\end{equation}
Note that the group structure of the form factors requires the existence of tensors of the form $t^{a_1 a_2 a_3}$ with appropriate symmetry properties (fully antisymmetric for the one associated with $A_1$, and antisymmetric in its first two indices for the one associated with $A_2$\footnote{The secondary Ward identities then ensure this tensor is fully antisymmetric, as below.}). One such tensor 
is the structure constant $f^{a_1 a_2 a_3}$. As argued in \cite{Osborn:1993cr}, the correlation function vanishes if the symmetry group is Abelian.

The form factors correspond to the coefficients of the following tensor structures in the full correlator  $\lla J^{\mu_1 a_1}(\bs{p}_1) J^{\mu_2 a_2}(\bs{p}_2) J^{\mu_3 a_3}(\bs{p}_3) \rra$,
\begin{align}
A_1^{a_1 a_2 a_3} & = \text{coefficient of } p_2^{\mu_1} p_3^{\mu_2} p_1^{\mu_3}, \\
A_2^{a_1 a_2 a_3} & = \text{coefficient of } \delta^{\mu_1 \mu_2} p_1^{\mu_3}.
\end{align}
Here, we assume the independent momenta in the correlator have been chosen according to the index rule \eqref{a:momenta}.

\bigskip \noindent \textbf{Primary CWIs.} The primary CWIs are
\begin{equation}
\begin{array}{ll}
\K_{12} A_1^{a_1 a_2 a_3} = 0, & \qquad\qquad \K_{13} A_1^{a_1 a_2 a_3} = 0, \\
\K_{12} A_2^{a_1 a_2 a_3} = 0, & \qquad\qquad \K_{13} A_2^{a_1 a_2 a_3} = 2 A_1^{a_1 a_2 a_3}.
\end{array}
\end{equation}
Their solution in terms of triple-$K$ integrals is
\begin{align}
A_1^{a_1 a_2 a_3} & = \3{1}^{a_1 a_2 a_3} J_{3 \{000\}}, \label{a:JJJ1} \\
A_2^{a_1 a_2 a_3} & = \3{1}^{a_1 a_2 a_3} J_{2 \{001\}} + \3{2}^{a_1 a_2 a_3} J_{1 \{000\}}, \label{a:JJJlast}
\end{align}
where the reduced triple-$K$ integrals are defined in \eqref{a:J} and 
$\3{1}^{a_1 a_2 a_3}$ and $\3{2}^{a_1 a_2 a_3}$ are constants.  We refer to all such constants arising in solutions of the primary CWI as primary constants.

\bigskip \noindent \textbf{Secondary CWIs.} The independent secondary CWIs,
\begin{align}\label{eqn320}
& (*) \ \Lo_{3} A_1^{a_1 a_2 a_3} + 2 \Ro \left[ A_2^{a_1 a_2 a_3} - A_2^{a_3 a_1 a_2}(p_3, p_1, p_2) \right]  \nn\\
& \qquad = 2 (\Delta_1 - 1) \cdot \text{coefficient of } p_3^{\mu_2} p_1^{\mu_3} \text{ in } p_{1 \mu_1} \lla J^{\mu_1 a_1}(\bs{p}_1) J^{\mu_2 a_2}(\bs{p}_2) J^{\mu_3 a_3}(\bs{p}_3) \rra, \\[1ex]
& \Lo_{1} \left[ A_2^{a_2 a_3 a_1}(p_2, p_3, p_1) \right] + 2 p_1^2 \left[ A_2^{a_3 a_1 a_2}(p_3, p_1, p_2) - A_2^{a_1 a_2 a_3} \right]  \nn\\
& \qquad = 2 (\Delta_1 - 1) \cdot \text{coefficient of } \delta^{\mu_2 \mu_3} \text{ in } p_{1 \mu_1} \lla J^{\mu_1 a_1}(\bs{p}_1) J^{\mu_2 a_2}(\bs{p}_2) J^{\mu_3 a_3}(\bs{p}_3) \rra,
\end{align}
where the operators $\Lo_{N}$ and $\Ro$ are defined in \eqref{a:L} and \eqref{a:R}.
In each of the last lines above, we assume that the independent momenta in the correlator have been selected according to the index rule \eqref{a:momenta}.
To evaluate these right-hand sides, we apply the transverse Ward identity \eqref{TWI_JJJ}.  The right-hand side of \eqref{eqn320} vanishes: this equation  determines the symmetry properties of $\3{2}^{a_1 a_2 a_3}$, enforcing $\3{2}^{a_1 a_3 a_2} = -\3{2}^{a_1 a_2 a_3}$ in any dimension. Given its antisymmetry in the first two indices, $\3{2}^{a_1 a_2 a_3}$ is thus fully antisymmetric. Otherwise, this equation is redundant and places no further constraints on the primary constants.

\paragraph{Divergences.} The triple-$K$ integrals appearing in \eqref{a:JJJ1} and \eqref{a:JJJlast} have singularities which can be identified by looking for solutions of the singularity condition \eqref{singcond}.  
We have compiled the results in table \ref{tab:JJJ} below, using $n=0,1,2\ldots$ to represent any non-negative integer.
The 3-point function is clearly finite in all odd spacetime dimensions and has at most a single pole in $\ep$ for even dimensions $d \geq 4$.  (To obtain higher-order divergences would require the singularity condition to be multiply satisfied \cite{Bzowski:2015pba}.) 
One can check that the regularisation scheme $u=v_j$ for $j=1,2,3$  is sufficient to regularise all these singularities.
As this scheme preserves the canonical dimensions of currents ({\it i.e.,} $\tilde{\Delta}_j = \tilde{d}-1$), as required for the transverse Ward identities to hold in the regulated theory, we will adopt it in the following.

\begin{center}
\begin{tabular}{|c|c|c|c|} \hline
Form factor & Integral & $(---)$ & $(--+)$  \\ \hline
$A_1$ & $J_{3 \{000\}}$ & $d=6 + 2n$ & never \\ \hline
$A_2$ & $J_{2 \{001\}}, J_{1\{000\}}$ & $d=4 + 2n$ & never \\ \hline
\end{tabular}
\captionof{table}{Singularities of triple-$K$ integrals in the form factors for $\< J^{\mu_1} J^{\mu_2} J^{\mu_3} \>$.\label{tab:JJJ}}
\end{center}

\paragraph{Counterterms.} All relevant counterterms for the parity-even sector are constructed from the field strength $F_{\mu\nu}^a$.  For 3-point functions, we are interested in terms that are cubic in the source $A_\mu^a$, which arise from counterterms with two or three $F_{\mu\nu}^a$, such as
\begin{equation}
\int \D^4 \bs{x} \sqrt{g} F_{\mu\nu}^a F^{\mu\nu a}, \qquad \int \D^6 \bs{x} \sqrt{g} F_{\mu\nu}^a \Box F^{\mu\nu a}, \qquad f^{abc} \int \D^6 \bs{x}\sqrt{g} F_{\mu}^{\ \nu a} F_{\nu}^{\ \rho b} F_{\rho}^{\ \mu c}.
\end{equation}
Clearly, the available counterterms are dependent on the spacetime dimension. In four dimensions, only the left-hand counterterm shown is available; crucially, this same counterterm also contributes to the renormalisation of the 2-point function (see \eqref{e:SctJJ}), so its coefficient is already fixed as given in section \ref{sec:2ptfns}.  Above four dimensions, more than one counterterm may be present.

\subsubsection{Odd spacetime dimensions} 

In any spacetime dimension $d \neq 2 N$ for integer $N \geq 1$, the secondary CWI leads to
\begin{equation}
\3{2}^{a_1 a_2 a_3} + (d-2) \3{1}^{a_1 a_2 a_3} = \frac{2^{4 - \frac{d}{2}}  g f^{a_1 a_2 a_3} \2_{JJ} \sin \left( \frac{d \pi}{2} \right)}{\pi \Gamma \left( \frac{d}{2} - 1 \right)},
\end{equation}
where $\2_{JJ}$ is the 2-point function normalisation in \eqref{e:2ptJJ}. In particular, for $d = 2N + 1$, this expression simplifies to
\begin{align}
\3{2}^{a_1 a_2 a_3} + (d-2) \3{1}^{a_1 a_2 a_3} & = \frac{(-1)^N 2^{\frac{7}{2} - N} g f^{a_1 a_2 a_3} \2_{JJ}}{\pi \Gamma \left( N - \frac{1}{2} \right)}  =  \frac{(-1)^N 2 g f^{a_1 a_2 a_3} \2_{JJ}}{(d-4)!!}\left( \frac{\pi}{2} \right)^{-\frac{3}{2}},
\end{align}
where we use the convention $(-1)!! = 1$.

\paragraph{Three dimensions.} After re-scaling $C_1^{a_1a_2a_3}$ to remove an overall numerical factor, we find the form factors 
\begin{align}
A_1^{a_1 a_2 a_3} & = \frac{2 \ren{1}^{a_1 a_2 a_3}}{a_{123}^3}, \\
A_2^{a_1 a_2 a_3} & = \ren{1}^{a_1 a_2 a_3} \frac{p_3}{a_{123}^2} - \frac{2 g f^{a_1 a_2 a_3} \2_{JJ}}{a_{123}},
\end{align}
where the symmetric polynomial $a_{123}$ is defined in \eqref{e:variables}.

\subsubsection{Even spacetime dimensions} 

In even dimensions $d \geq 4$, both sides of the secondary CWI diverge as $\ep^{-1}$.  On the left-hand side, the triple-$K$ integrals appearing in the form factors are divergent as discussed above.  On the right-hand side, after making use of the transverse Ward identity, the 2-point functions that arise are similarly divergent.
To handle this, it is useful to expand the primary constants in the regulator $\ep$ as
\[
C_j(\ep) = C_j^{(0)} + \ep C_j^{(1)} + O(\ep^2).
\]
Evaluating the secondary CWIs order by order in $\ep$ then constrains the coefficients $C_j^{(n)}$.

A useful trick here, as discussed in section 6.3 of \cite{Bzowski:2013sza}, is to focus on the zero-momentum limit in which $p_3\rightarrow 0$ and $p_1, p_2\rightarrow p$.  In this limit,
the triple-$K$ integrals reduce to known integrals of only two Bessel functions, yielding\footnote{In this formula it is assumed that $\beta_3>0$ and that all poles of the gamma functions have been removed by dimensional regularisation.} 
\begin{equation}\label{zml1}
I_{\alpha \{\beta_1\beta_2\beta_3\}}(p,p,0) = l_{\alpha \{\beta_1\beta_2\beta_3\}} \, p^{\beta_t - \alpha - 1}, \quad \beta_t = \beta_1 + \beta_2 + \beta_3,
\end{equation}
where
\begin{equation}\label{zml2}
l_{\alpha \{\beta_1\beta_2\beta_3\}} = \frac{2^{\alpha - 3} \Gamma(\beta_3)}{\Gamma(\alpha - \beta_3 + 1)} \prod_{\sigma_1, \sigma_2 \in \{-1, +1\}} \Gamma \left( \frac{\alpha - \beta_3 + 1 + \sigma_1 \beta_1 + \sigma_2 \beta_2}{2} \right).
\end{equation}
To handle derivatives of triple-$K$ integrals with respect to momentum magnitudes, such as arise from acting with the operators $\Lo_{N}$ and $\Ro$ on form factors, we can make use of the recursion relations discussed in section 6.1.2 of \cite{Bzowski:2013sza} (see also section 3.2 of \cite{Bzowski:2015yxv}).  These relations allow us to re-express derivatives of triple-$K$ integrals in terms of pure triple-$K$ integrals, allowing a straightforward evaluation of the zero-momentum limit.

As it turns out, examination of this limiting momentum configuration is sufficient to evaluate the coefficients $C_j^{(n)}$ for general even dimension $d=2N$.
This general solution reads
\begin{align} \label{e:secJJJ1}
& \3{2}^{(0) a_1 a_2 a_3} + 2(N-1) \3{1}^{(0) a_1 a_2 a_3} = \frac{(-1)^N 2^{4 - N}}{(N-2)!} g f^{a_1 a_2 a_3} \2_{JJ}, \\[2ex]
& \3{2}^{(1) a_1 a_2 a_3} + 2(N-1) \3{1}^{(1) a_1 a_2 a_3} + 2 u \3{1}^{(0) a_1 a_2 a_3}  \nn\\
& \qquad = - \frac{(-1)^N 2^{4 - N}}{(N-2)!} u g f^{a_1 a_2 a_3} \left[ \2_{JJ} \left( H_{N-2} + \ln 2 - \gamma_E \right) - \sl_{JJ} \right], \label{e:secJJJ2}
\end{align}
where $H_{n} = \sum_{j=1}^{n} j^{-1}$ is the $n$-th harmonic number and we use the current-conserving regularisation scheme \eqref{truescheme}.
This general solution is valid provided suitable counterterms exist, which as we will see, is always the case.

\paragraph{Four dimensions.}

Substituting $N = 2$ in \eqref{e:secJJJ1} and \eqref{e:secJJJ2}, we find the regulated solution to the secondary CWIs,
\begin{align}
\3{2}^{(0) a_1 a_2 a_3} & = - 2 \3{1}^{(0) a_1 a_2 a_3} + 4 g f^{a_1 a_2 a_3} \2_{JJ}, \\
\3{2}^{(1) a_1 a_2 a_3} & = - 2 (u \3{1}^{(0) a_1 a_2 a_3} + \3{1}^{(1) a_1 a_2 a_3}) - 4 g f^{a_1 a_2 a_3} \left[ \2_{JJ} (\ln 2 - \gamma_E ) - \sl_{JJ} \right].
\end{align}
The form factor $A_1^{a_1 a_2 a_3}$ is then finite, while $A_2^{a_1 a_2 a_3}$ has a single pole,
\begin{align} \label{e:solJJJ}
A_2^{a_1 a_2 a_3} = - \frac{2 g f^{a_1 a_2 a_3} \2_{JJ}}{u \epsilon} + O(\ep^0).
\end{align}
In four dimensions, our only counterterm is \eqref{e:SctJJ} with $N=2$.  Moreover, the  coefficient of this counterterm has already been determined from the renormalisation of the 2-point function, as given in \eqref{e:ctJJ}.  The contribution of this counterterm to the 3-point function must therefore precisely cancel the divergence in the form factor above.
Taking three functional derivatives of $\<e^{-S_{\Ct}}\>$, this counterterm contribution is 
\begin{align}
& \lla J^{\mu_1 a_1}(\bs{p}_1) J^{\mu_2 a_2}(\bs{p}_2) J^{\mu_3 a_3}(\bs{p}_3) \rra_{\Ct} \nn\\
&
\qquad = 4\mathfrak{c}_{JJ} g f^{a_1 a_2 a_3} \mu^{2 u \epsilon}  \left[ \delta^{\mu_1 \mu_2} (p_1^{\mu_3} - p_2^{\mu_3}) + \delta^{\mu_2 \mu_3} (p_2^{\mu_1} - p_3^{\mu_1}) + \delta^{\mu_1 \mu_3} (p_3^{\mu_2} - p_1^{\mu_2}) \right].
\end{align}
In terms of form factors, this yields 
\begin{equation}
A_{1 \text{ ct}}^{a_1 a_2 a_3} = 0, \qquad A_{2 \text{ ct}}^{a_1 a_2 a_3} = 8\mathfrak{c}_{JJ} g f^{a_1 a_2 a_3}.
\end{equation}
Substituting \eqref{e:ctJJ}, we see this cancels the divergence of the second form factor as required. 

The renormalised form factors can now be written
\begin{align}
A_1^{a_1 a_2 a_3} & = \ren{1}^{a_1 a_2 a_3} I_{4\{111\}} \nn\\
& = - \ren{1}^{a_1 a_2 a_3} p_1 p_2 p_3 \frac{\partial^3}{\partial p_1 \partial p_2 \partial p_3} I_{1\{000\}}, \\
A_2^{a_1 a_2 a_3} & = \ren{1}^{a_1 a_2 a_3} p_1 p_2 p_3^2 \frac{\partial^2}{\partial p_1 \partial p_2} I_{1\{000\}} + 4 g f^{a_1 a_2 a_3} \2_{JJ} I_{2\{111\}}^{\text{(fin)}} \nn\\
& \qquad - \: \frac{2}{3} g f^{a_1 a_2 a_3} \2_{JJ} \left[ \ln \frac{p_1^2}{\mu^2} + \ln \frac{p_2^2}{\mu^2} + \ln \frac{p_3^2}{\mu^2} \right] - 2 g f^{a_1 a_2 a_3} \sdt_{JJ},
\end{align}
where for convenience we have re-labelled the ($\ep$-independent) constant $C_1^{(0)a_1a_2a_3}$ as $C_1^{a_1a_2a_3}$, and $\sdt_{JJ}$ is the scheme-dependent constant defined in \eqref{e:2ptJJren}. 
$I_{2\{111\}}^{\text{(fin)}}$ represents the finite part of the integral $I_{2\{111\}}$, and is given by
\begin{align}\label{I2111fin}
& I_{2\{111\}}^{\text{(fin)}} = - \frac{4 p_1^2 p_2^2 p_3^2}{J^2} I_{1\{000\}} - \frac{1}{6 J^2} \left[ p_1^2 (p_2^2 + p_3^2 - p_1^2) \ln \left( \frac{p_1^4}{p_2^2 p_3^2} \right) \right.\nn\\
& \qquad\qquad \left. + \: p_2^2 (p_1^2 + p_3^2 - p_2^2) \ln \left( \frac{p_2^4}{p_1^2 p_3^2} \right) + p_3^2 (p_1^2 + p_2^2 - p_3^2) \ln \left( \frac{p_3^4}{p_1^2 p_2^2} \right) \right].
\end{align}
The integral $I_{1\{000\}}$ can be evaluated   explicitly in terms of dilogarithms \cite{Bzowski:2015yxv},
\begin{align}\label{I1000}
I_{1\{000\}} & = \frac{1}{2\sqrt{-J^2}}\Big[\frac{\pi^2}{6}-2\ln\frac{p_1}{p_3}\ln\frac{p_2}{p_3}+\ln X\ln Y-\mathrm{Li}_2X-\mathrm{Li}_2Y\Big],
\end{align}
where $J^2$ is defined in \eqref{Jsqdef} and 
\begin{align}
X & = \frac{-p_1^2+p_2^2+p_3^2-\sqrt{-J^2}}{2p_3^2}\,, \qquad 
Y = \frac{-p_2^2+p_1^2+p_3^2-\sqrt{-J^2}}{2p_3^2}.
\end{align}
Each renormalised form factor $A_1^{a_1a_2a_3}$ and $A_2^{a_1a_2a_3}$ thus contains a nonlocal piece  involving triple-$K$ integrals.  For $A_1^{a_1a_2a_3}$ this is all we have. 
For $A_2^{a_1a_2a_3}$, we have in addition a scale-violating logarithmic piece containing terms depending explicitly on the renormalisation scale $\mu$, along with a scheme-dependent constant term proportional to $D_{JJ}$.  This constant can be adjusted by changing the renormalisation scale: sending $\mu^2\rightarrow e^{-\lambda}\mu^2$ is equivalent to shifting $D_{JJ}\rightarrow D_{JJ}+\lambda C_{JJ}$.

The anomalous scale-dependence of the 3-point function is
\begin{align}
& \mu \frac{\partial}{\partial \mu} \lla J^{\mu_1 a_1}(\bs{p}_1) J^{\mu_2 a_2}(\bs{p}_2) J^{\mu_3 a_3}(\bs{p}_3) \rra = \lim_{\epsilon \rightarrow 0} \mu \frac{\partial}{\partial \mu} \lla J^{\mu_1 a_1}(\bs{p}_1) J^{\mu_2 a_2}(\bs{p}_2) J^{\mu_3 a_3}(\bs{p}_3) \rra_{\Ct} \nn\\
& \qquad = 2 \2_{JJ} g f^{a_1 a_2 a_3} \left[ \delta^{\mu_1 \mu_2} (p_1^{\mu_3} - p_2^{\mu_3}) + \delta^{\mu_2 \mu_3} (p_2^{\mu_1} - p_3^{\mu_1}) + \delta^{\mu_1 \mu_3} (p_3^{\mu_2} - p_1^{\mu_2}) \right],
\end{align}
or equivalently,
\begin{equation}
\mu \frac{\partial}{\partial \mu} A_1^{a_1 a_2 a_3} = 0, \qquad\qquad \mu \frac{\partial}{\partial \mu} A_2^{a_1 a_2 a_3} =  4 \2_{JJ} g f^{a_1 a_2 a_3}.
\end{equation}
These expressions also follow from the anomaly action \eqref{anomaction}, which in four dimensions extends to cubic order in $A^a_\mu$.  
The same coefficient $C_{JJ}$ thus controls both the 2- and 3-point anomalies.

In the renormalised theory, the primary CWI take the same form as in the regulated theory.  The secondary CWI are anomalous, however, and read
\begin{align}
& \Lo_{3} A_1^{a_1 a_2 a_3} + 2 \Ro \left[ A_2^{a_1 a_2 a_3} - A_2^{a_1 a_2 a_3}(p_2 \leftrightarrow p_3) \right]  \nn\\[1ex]
& \qquad = \Lo_{3} A_1^{a_1 a_2 a_3} + 2 \Ro \left[ A_2^{a_1 a_2 a_3} + A_2^{a_1 a_3 a_2}(p_2 \leftrightarrow p_3) \right] = 0, 
\\[2ex]
& \Lo_{1} \left[ A_2^{a_1 a_2 a_3}(p_1 \leftrightarrow p_3) \right] + 2 p_1^2 \left[ A_2^{a_1 a_2 a_3}(p_2 \leftrightarrow p_3) - A_2^{a_1 a_2 a_3} \right] \nn\\[1ex]
& \qquad = - \Lo_{1} \left[ A_2^{a_3 a_2 a_1}(p_1 \leftrightarrow p_3) \right] - 2 p_1^2 \left[ A_2^{a_1 a_3 a_2}(p_2 \leftrightarrow p_3) + A_2^{a_1 a_2 a_3} \right] \nn\\[1ex]
& \qquad = 4 \cdot \text{coefficient of } \delta^{\mu_2 \mu_3} \text{ in } p_{1 \mu_1} \lla J^{\mu_1 a_1}(\bs{p}_1) J^{\mu_2 a_2}(\bs{p}_2) J^{\mu_3 a_3}(\bs{p}_3) \rra - 4 \2_{JJ} g f^{a_1 a_2 a_3} p_1^2 \nn\\[0.5ex]
& \qquad = 4 g f^{a_1 a_2 a_3} \Big[ p_2^2 \Big( \2_{JJ} \ln \frac{p_2^2}{\mu^2} + \sdt_{JJ} \Big) - p_3^2 \Big( \2_{JJ} \ln \frac{p_3^2}{\mu^2} + \sdt_{JJ} \Big) \Big] - 4 \2_{JJ} g f^{a_1 a_2 a_3} p_1^2.
\end{align}

\subsection{\texorpdfstring{$\<T_{\mu_1\nu_1}J^{\mu_2}J^{\mu_3}\>$}{<TJJ>}}

\subsubsection{General analysis}

\paragraph{Decomposition.} The transverse and trace Ward identities are
\begin{align}\label{TWI_TJJ_1}
& p_1^{\nu_1} \lla T_{\mu_1 \nu_1}(\bs{p}_1) J^{\mu_2 a_2}(\bs{p}_2) J^{\mu_3 a_3}(\bs{p}_3) \rra  \nn\\[0.5ex]
& \qquad = \: 2 \delta^{\mu_3}_{[\mu_1} p_{3\alpha]} \lla J^{\mu_2 a_2}(\bs{p}_2) J^{\alpha a_3}(-\bs{p}_2) \rra + 2 \delta^{\mu_2}_{[\mu_1} p_{2\alpha]} \lla J^{\alpha a_2}(\bs{p}_3) J^{\mu_3 a_3}(-\bs{p}_3) \rra, \\[1ex]\label{TWI_TJJ_2}
& p_{2 \mu_2} \lla T_{\mu_1 \nu_1}(\bs{p}_1) J^{\mu_2 a_2}(\bs{p}_2) J^{\mu_3 a_3}(\bs{p}_3) \rra = 0, \\[1ex]
& \lla T(\bs{p}_1) J^{\mu_2 a_2}(\bs{p}_2) J^{\mu_3 a_3}(\bs{p}_3) \rra = \mathcal{A}^{\mu_2 \mu_3 a_2 a_3}. \label{e:traceW_TJJ}
\end{align}
As we will see, in odd dimensions anomalies are absent and $A^{\mu_2\mu_3a_2a_3}$ can be set to zero.  In even dimensions anomalies are generally present; see \eqref{TJJanom} for the case $d=4$.
Taking the trace of \eqref{TWI_TJJ_2} and swapping $1\leftrightarrow 2$, we see the anomaly is transverse, {\it i.e.,} $p_{2\mu_2}\mathcal{A}^{\mu_2 \mu_3 a_2 a_3}=0$ and similarly $p_{3\mu_3}\mathcal{A}^{\mu_2 \mu_3 a_2 a_3}=0$.

The obtain the full 3-point function from the transverse-traceless part, we use the reconstruction formula
\begin{align}
& \lla T_{\mu_1 \nu_1}(\bs{p}_1) J^{\mu_2 a_2}(\bs{p}_2) J^{\mu_3 a_3}(\bs{p}_3) \rra = \lla t_{\mu_1 \nu_1}(\bs{p}_1) j^{\mu_2 a_2}(\bs{p}_2) j^{\mu_3 a_3}(\bs{p}_3) \rra \nn\\[0.5ex]
& \qquad +  2 \mathscr{T}_{\mu_1 \nu_1}^{\quad\,\,\alpha}(\bs{p}_1)\Big[\delta_{[\alpha}^{\mu_3}p_{3\beta]}  \lla J^{\mu_2 a_2}(\bs{p}_2) J^{\beta a_3}(-\bs{p}_2) \rra
+\delta_{[\alpha}^{\mu_2}p_{2\beta]}  \lla J^{\mu_3 a_3}(\bs{p}_3) J^{\beta a_2}(-\bs{p}_3) \rra\Big]
 \nn\\
& \qquad + \frac{1}{d - 1}\,\pi_{\mu_1 \nu_1}(\bs{p}_1) \mathcal{A}^{\mu_2 \mu_3 a_2 a_3}, \label{a:tjjrecon}
\end{align}
where $\mathscr{T}_{\mu_1 \nu_1\alpha}$ is defined in \eqref{a:T}.

\paragraph{Form factors.} The tensorial decomposition of the transverse-traceless part is
\begin{align}
& \lla t_{\mu_1 \nu_1}(\bs{p}_1) j^{\mu_2 a_2}(\bs{p}_2) j^{\mu_3 a_3}(\bs{p}_3) \rra \nn\\
& \qquad = \Pi_{\mu_1 \nu_1 \alpha_1 \beta_1}(\bs{p}_1) \pi^{\mu_2}_{\alpha_2}(\bs{p}_2) \pi^{\mu_3}_{\alpha_3}(\bs{p}_3) \left[ A_1^{a_2 a_3} p_2^{\alpha_1} p_2^{\beta_1} p_3^{\alpha_2} p_1^{\alpha_3} + A_2^{a_2 a_3} \delta^{\alpha_2 \alpha_3} p_2^{\alpha_1} p_2^{\beta_1} \right.\nn\\
& \qquad \qquad + \: A_3^{a_2 a_3} \delta^{\alpha_1 \alpha_2} p_2^{\beta_1} p_1^{\alpha_3} + A_3^{a_3 a_2}(p_2 \leftrightarrow p_3) \delta^{\alpha_1 \alpha_3} p_2^{\beta_1} p_3^{\alpha_2} \nonumber \\
& \left. \qquad \qquad + \: A_4^{a_2 a_3} \delta^{\alpha_1 \alpha_3} \delta^{\alpha_2 \beta_1} \right],
\end{align}
where all the form factors are functions of the momentum magnitudes $A_j = A_j(p_1, p_2, p_3)$.
In the penultimate term, the momenta but not the adjoint indices are exchanged, \textit{i.e.}, $A_3^{a_3a_2}(p_2 \leftrightarrow p_3) = A_3^{a_3a_2}(p_1, p_3, p_2)$.

The form factors $A_1$, $A_2$ and $A_4$ are symmetric under $(p_2, a_2) \leftrightarrow (p_3, a_3)$, satisfying
\begin{equation}
A_j^{a_3 a_2}(p_1, p_3, p_2) = A_j^{a_2 a_3}(p_1, p_2, p_3), \qquad j \in \{1,2,4\},
\end{equation}
while the form factor $A_3$ does not exhibit any symmetry properties.

Given the 3-point function $\lla T_{\mu_1 \nu_1}(\bs{p}_1) J^{\mu_2 a_2}(\bs{p}_2) J^{\mu_3 a_3}(\bs{p}_3) \rra$ with independent momenta chosen according to \eqref{a:momenta}, the form factors can be read off according to
\begin{align}
A_1^{a_2 a_3} & = \text{coefficient of } p_{2\mu_1} p_{2\nu_1} p_3^{\mu_2} p_1^{\mu_3}, \\
A_2^{a_2 a_3} & = \text{coefficient of } \delta^{\mu_2 \mu_3} p_{2\mu_1} p_{2\nu_1}, \\
A_3^{a_2 a_3} & = 2 \cdot \text{coefficient of } \delta_{\mu_1}^{\mu_2} p_{2\nu_1} p_1^{\mu_3}, \\
A_4^{a_2 a_3} & = 2 \cdot \text{coefficient of } \delta_{\mu_1}^{\mu_2} \delta^{\mu_3}_{\nu_1}.
\end{align}

Finally, as the anomaly is transverse, we can decompose its tensorial structure as
\[\label{TJJanomdecomp}
\mathcal{A}^{\mu_2 \mu_3 a_2 a_3} = \pi^{\mu_2}_{\alpha_2}(\bs{p}_2)\pi^{\mu_3}_{\alpha_3}(\bs{p}_3)\Big[B_1^{a_2a_3} p_3^{\alpha_2}p_1^{\alpha_3}+B_2^{a_2a_3}\delta^{\alpha_2\alpha_3}\Big],
\]
where the form factors $B_j^{a_2a_3}=B_j^{a_2a_3}(p_1,p_2,p_3)$ for $j=1,2$ are functions of the momentum magnitudes.  Explicit expressions for these form factors in $d=4$ are given in \eqref{TJJB1}.

\paragraph{Primary CWIs.} The primary CWIs are
\begin{equation}
\begin{array}{ll}
\K_{12} A_1^{a_2 a_3} = 0, & \qquad\qquad \K_{13} A_1^{a_2 a_3} = 0, \\
\K_{12} A_2^{a_2 a_3} = -2 A_1^{a_2 a_3}, & \qquad\qquad \K_{13} A_2^{a_2 a_3} = - 2 A_1^{a_2 a_3}, \\
\K_{12} A_3^{a_2 a_3} = 0, & \qquad\qquad \K_{13} A_3^{a_2 a_3} = 4 A_1^{a_2 a_3}, \\
\K_{12} A_4^{a_2 a_3} = 2 A_3^{a_2 a_3}, & \qquad\qquad \K_{13} A_4^{a_2 a_3} = 2 A_3^{a_3 a_2}(p_2 \leftrightarrow p_3),
\end{array}
\end{equation}
Their solution in terms of triple-$K$ integrals \eqref{a:J} is
\begin{align}
A_1^{a_2 a_3} & = \3{1}^{a_2 a_3} J_{4 \{000\}}, \label{e:restjj1} \\
A_2^{a_2 a_3} & = \3{1}^{a_2 a_3} J_{3 \{100\}} + \3{2}^{a_2 a_3} J_{2 \{000\}}, \\
A_3^{a_2 a_3} & = 2 \3{1}^{a_2 a_3} J_{3 \{001\}} + \3{3}^{a_2 a_3} J_{2 \{000\}}, \\
A_4^{a_2 a_3} & = 2 \3{1}^{a_2 a_3} J_{2 \{011\}} +  \3{3}^{a_2 a_3} \left( J_{1 \{010\}} + J_{1 \{001\}} \right) + \3{4}^{a_2 a_3} J_{0 \{000\}}, \label{e:restjj4}
\end{align}
where the $\3{j}^{a_2 a_3}$ for $j=1,2,3,4$ represent primary constants. In particular, these constants are all symmetric in the group indices, $\3{j}^{a_3 a_2} = \3{j}^{a_2 a_3}$ for all $j=1,2,3,4$. 
Consequently, they can be represented by any symmetric bilinear form of the algebra, {\it e.g.,} the Killing form $f^{a_2 b}_{\quad\,\, c}f^{a_3 c}_{\quad\,\,b}$, although other choices may also exist.

\paragraph{Secondary CWIs.} The independent secondary CWIs are
\begin{align}
& (*) \ \Lo_{4} A_1^{a_2 a_3} + \Ro \left[ A_3^{a_2 a_3} - A_3^{a_2 a_3}(p_2 \leftrightarrow p_3) \right]  \nn\\
& \qquad = 2 \Delta_1 \cdot \text{coefficient of } p_{2\mu_1} p_3^{\mu_2} p_1^{\mu_3} \text{ in } p_{1}^{\nu_1} \lla T_{\mu_1 \nu_1}(\bs{p}_1) J^{\mu_2 a_2}(\bs{p}_2) J^{\mu_3 a_3}(\bs{p}_3) \rra, \label{e:tjjseccwi1} \\[1ex]
& \Lo'_{3} A_1^{a_2 a_3} + 2 \Ro' \left[ A_3^{a_2 a_3} - A_2^{a_2 a_3} \right]  \nn\\
& \qquad = 2 \Delta_1 \cdot \text{coefficient of } p_{2\mu_1} p_{2\nu_1} p_1^{\mu_3} \text{ in } p_{2\mu_2} \lla T_{\mu_1 \nu_1}(\bs{p}_1) J^{\mu_2 a_2}(\bs{p}_2) J^{\mu_3 a_3}(\bs{p}_3) \rra,
\label{e:tjjseccwi2}
 \\[1ex]
& \Lo_{2} A_2^{a_2 a_3} - p_1^2 \left[ A_3^{a_2 a_3} - A_3^{a_2 a_3}(p_2 \leftrightarrow p_3) \right]  \nn\\
& \qquad = 2 \Delta_1 \cdot \text{coefficient of } \delta^{\mu_2\mu_3} p_{2\mu_1} \text{ in } p_{1 }^{\nu_1} \lla T_{\mu_1 \nu_1}(\bs{p}_1) J^{\mu_2 a_2}(\bs{p}_2) J^{\mu_3 a_3}(\bs{p}_3) \rra, \label{e:tjjseccwi3} \\[1ex]
& \Lo_{4} A_3^{a_2 a_3} - 2 \Ro A_4^{a_2 a_3}  \nn\\
& \qquad = 4 \Delta_1 \cdot \text{coefficient of } \delta_{\mu_1}^{\mu_2} p_1^{\mu_3} \text{ in } p_{1}^{ \nu_1} \lla T_{\mu_1 \nu_1}(\bs{p}_1) J^{\mu_2 a_2}(\bs{p}_2) J^{\mu_3 a_3}(\bs{p}_3) \rra, \label{e:tjjseccwi4}
\end{align}
where $\Lo_{N}$ and $\Ro$ are defined in \eqref{a:L} and \eqref{a:R}. 
All right-hand sides can be determined using the transverse Ward identities \eqref{TWI_TJJ_1}\,-\,\eqref{TWI_TJJ_2}; in particular, those for \eqref{e:tjjseccwi1} and \eqref{e:tjjseccwi2} vanish.
The first identity marked by an asterisk is redundant, \textit{i.e.}, it is trivially satisfied in all cases and does not impose any additional constraints on the primary constants.

\vspace{-1mm}  

\paragraph{Divergences.}  
The triple-$K$ integrals in our solution \eqref{e:restjj1}\,-\,\eqref{e:restjj4} have  singularities that can be identified using the condition \eqref{singcond}.  (A similar procedure is discussed above for $\< J^{\mu_1} J^{\mu_2} J^{\mu_3} \>$.) 
The possible singularities are listed in the following table, where $d$ is the spacetime dimension and $n$ denotes any non-negative integer.

\begin{center}
\begin{tabular}{|c|c|c|c|} \hline
Form factor & Integral & $(---)$ & $(--+)$  \\ \hline
$A_1$ & $J_{4 \{000\}}$ & $d=6 + 2n$ & never \\ \hline
$A_2, A_3$ & $J_{3 \{100\}}, J_{3 \{001\}}, J_{2\{000\}}$ & $d=4 + 2n$ & never \\ \hline
$A_4$ & $J_{2 \{011\}}, J_{1\{010\}}, J_{1\{001\}}, J_{0\{000\}}$ & $d=2 + 2n$ & always \\ \hline
\end{tabular}
\captionof{table}{Singularities of triple-$K$ integrals in the form factors for $\< T_{\mu_1 \nu_1} J^{\mu_2} J^{\mu_3} \>$.\label{tab:TJJ}}
\end{center}

\paragraph{Regularisation.} 
We cannot immediately work in the desired scheme \eqref{truescheme}, as this fails to regulate the $(--+)$ singularities in the triple-$K$ integrals for the form factor $A_4^{a_2a_3}$.
Instead, we begin in the more general scheme $v_j=v\neq u$ for $j=1,2,3$, in which all triple-$K$ integrals are regulated, and then redefine the primary constants as necessary in order to obtain a finite limit as $v\rightarrow u$.

\paragraph{Counterterms.} 

As there are no counterterms capable of removing $(--+)$ singularities,\footnote{See the discussion on page \pageref{cancellation}.} in order for the form factor $A_4^{a_2 a_3}$ to be finite, all the $(--+)$ singularities must necessarily cancel against one another.  
The regulated form factors are therefore finite in odd spacetime dimensions, while in even dimensions they contain at most $\ep^{-1}$ poles, as only the $(---)$ singularities survive.

These surviving $(---)$ singularities can be removed by  counterterms constructed from two gauge-invariant field strengths $F_{\mu\nu}^a$ and at most a single  Riemann tensor, {\it e.g.,}
\begin{equation}
\int \D^4 \bs{x} \sqrt{g} F_{\mu \nu}^a F^{\mu \nu a}, \qquad\qquad \int \D^6 \bs{x} \sqrt{g} R F_{\mu \nu}^a F^{\mu \nu a}.
\end{equation}
Higher powers of the Riemann tensor cannot appear as these would generate a vanishing   contribution to $\<T_{\mu_1\nu_1}J^{\mu_2}J^{\mu_3}\>$ once the metric is restored to flatness.
As both the Riemann tensor and the field strength have scaling dimension two, no counterterms are available in odd dimensions, consistent with the finiteness of the form factors.
In $d = 4 + 2n$ for any non-negative integer $n$, we have counterterms involving two field strengths and $2n$ covariant derivatives, while in $d = 6 + 2n$ we have counterterms involving one Riemann tensor, two field strengths and $2n$ derivatives.  
As the counterterms without a Riemann tensor also contribute to the renormalisation of $\< J^{\mu_2} J^{\mu_3} \>$, their coefficients are already fixed by our discussion in section \ref{sec:2ptfns}. 
In $d=4$, all divergences of the 3-point function  should then be set by the 2-point  normalisation $\2_{JJ}$.
The introduction of counterterms gives rise to conformal anomalies as we will discuss.

\subsubsection{Odd spacetime dimensions}

Our first task is to evaluate the divergences arising in the triple-$K$ integrals for the form factor $A^{a_2a_3}_4$.  From table \ref{tab:TJJ}, we see that only $(--+)$ singularities are present, and hence these triple-$K$ integrals have only single poles in $\ep$.
Regulating in the scheme $v_j=v\neq u$ for $j=1,2,3$,
for odd dimensions $d=2N+1\ge 3$ we find
\begin{align}
J_{0\{000\}} &= I_{\frac{d}{2} - 1 + u \epsilon \{ \frac{d}{2} + v \epsilon, \frac{d}{2} - 1 + v \epsilon, \frac{d}{2} - 1 + v \epsilon \}} \nn\\[1ex]&  = \left( \frac{\pi}{2} \right)^{\frac{3}{2}} \frac{(-1)^N v (d-4)!!}{u - v} p_2^{d-2}  \left[ \frac{1}{v \epsilon} + \ln p_2^2 + \left( 2 H_{d-3} - H_{N-1} - \ln 2 - \gamma_E \right) \right] \nn\\[1ex]&\qquad  + (p_2 \leftrightarrow p_3) + O(\epsilon^0, (u-v)^0),
\end{align}
where $H_n = \sum_{j=1}^n j^{-1}$ denotes the $n$-th harmonic number and $(-1)!! = 1$.  
The divergences in the remaining triple-$K$ integrals can then be obtained using the recursion relations given in section 3.2 of  \cite{Bzowski:2015yxv}, {\it e.g.},
\begin{align}
J_{1\{001\}} & = I_{\frac{d}{2} + u \epsilon \{ \frac{d}{2} + v \epsilon, \frac{d}{2} - 1 + v \epsilon, \frac{d}{2} + v \epsilon \}} = \left(d - 2 + 2 v \epsilon + p_3 \frac{\partial}{\partial p_3} \right) J_{0\{000\}}, \\
J_{2\{011\}} & = I_{\frac{d}{2} + 1 + u \epsilon \{ \frac{d}{2} + v \epsilon, \frac{d}{2} + v \epsilon, \frac{d}{2} + v \epsilon \}} = \left(d - 2 + 2 v \epsilon + p_2 \frac{\partial}{\partial p_2} \right) J_{1\{001\}}.
\end{align}
Assembling the form factor $A_4^{a_2 a_3}$ according to \eqref{e:restjj4}, 
expanding the primary constants as
\[
C_j^{a_2a_3} = C_j^{(0)a_2a_3} + \ep C_j^{(1)a_2a_3}+O(\ep^2), 
\]
we find that the cancellation of the pole in $\epsilon$, as necessitated by the lack of counterterms, requires
\begin{equation} \label{e:tjj_genuvsol0}
\3{4}^{(0) a_2 a_3} = - (d-2) \3{3}^{(0) a_2 a_3}.
\end{equation}
We now have to re-define the remaining primary constants so that the $v \rightarrow u$ limit can be taken, in order to arrive in the scheme \eqref{truescheme}. To this end, it is convenient to set
\begin{equation} \label{e:tjj_genuvsol}
\3{4}^{(1) a_2 a_3} = - (d-2) \3{3}^{(1) a_2 a_3} - 2 v \3{3}^{(0) a_2 a_3} + (u - v) \redef{4}^{(1) a_2 a_3},
\end{equation}
where $\redef{4}^{(1) a_2 a_3}$ is an undetermined, re-defined primary constant.  
With this substitution, the limits $\epsilon \rightarrow 0$ and $v \rightarrow u$ exist and the form factor $A_4^{a_2 a_3}$ becomes finite.

Our next task is to solve the secondary CWI, which are valid in the limit $u=v$.
In fact, it is sufficient to examine only the zero-momentum limit of these equations, making use of  \eqref{zml1} and \eqref{zml2}.
First, we note the form factors $A_1^{a_2 a_3}$, $A_2^{a_2 a_3}$ and $A_3^{a_2 a_3}$ are finite and hence do not depend on any of the regulating parameters.  As the  secondary CWIs \eqref{e:tjjseccwi1}\,-\,\eqref{e:tjjseccwi3} involve only these form factors, and not $A_4^{a_2a_3}$, we will solve these equations first.  Only the leading order $\ep^0$ terms are relevant, yielding
\begin{equation} \label{e:tjj_gensecsol1}
\3{3}^{(0) a_2 a_3} = \3{2}^{(0) a_2 a_3} = - d \3{1}^{(0) a_2 a_3} + \left( \frac{\pi}{2} \right)^{-\frac{3}{2}} \frac{2 (-1)^N \2_{JJ} \delta^{a_2 a_3}}{(d-2)!!}.
\end{equation}
The remaining secondary CWI \eqref{e:tjjseccwi4} then leads to a dependence between the re-defined primary constant $\redef{4}^{(1) a_2 a_3}$ and the other primary constants. Its solution is most easily expressed in terms of the constant $\3{2}^{(0) a_2 a_3}$, and reads
\begin{equation} \label{e:tjj_gensecsol2}
\redef{4}^{(1) a_2 a_3} = \frac{d-4}{2} \3{2}^{(0) a_2 a_3} - \left( \frac{\pi}{2} \right)^{-\frac{3}{2}} \frac{2 (-1)^N \2_{JJ} \delta^{a_2 a_3}}{(d-4)!!}.
\end{equation}

Together, equations \eqref{e:tjj_genuvsol0}\,-\,\eqref{e:tjj_gensecsol2} solve the secondary Ward identities, and guarantee that the correlation function is finite in odd spacetime dimensions.  While these equations involve two undetermined constants, which we can choose as $\3{2}^{(0) a_2 a_3}$ and $\3{3}^{(1) a_2 a_3}$, the final 3-point function depends only on a single constant $\3{2}^{(0) a_2 a_3}$.  
The remaining constant $\3{3}^{(1) a_2 a_3}$ 
appears in \eqref{e:tjj_genuvsol} in such a way that it cancels from the form factor $A_4^{a_2 a_3}$.

\paragraph{Three dimensions.}  Specialising \eqref{e:tjj_genuvsol0}\,-\,\eqref{e:tjj_gensecsol2} to $d = 3$, we find
\begin{align}\label{p3d1}
\3{1}^{(0) a_2 a_3} & = - \frac{1}{3} \3{2}^{(0) a_2 a_3} - \frac{2}{3} \left( \frac{\pi}{2} \right)^{-\frac{3}{2}} \2_{JJ} \delta^{a_2 a_3}, \\
\3{3}^{(0) a_2 a_3} & = \3{2}^{(0) a_2 a_3}, \\
\3{4}^{(0) a_2 a_3} & = -\3{2}^{(0) a_2 a_3}, \\
\3{4}^{(1) a_2 a_3} & = -\3{2}^{(1) a_2 a_3} - 2 v \3{3}^{(0) a_2 a_3} + (u - v) \left( \frac{\pi}{2} \right)^{-\frac{3}{2}} 2 \2_{JJ} \delta^{a_2 a_3}.
\label{p3d2}
\end{align}
For the first three form factors, the relevant  triple-$K$ integrals are finite and can be evaluated setting $\epsilon = 0$.  
For the fourth form factor,
it is convenient to begin in the scheme 
$u = 1$, $v_j = 0$ for $j=1,2,3$, for which all Bessel functions reduce to elementary functions.
After evaluating the relevant triple-$K$ integrals, we can then convert to the scheme $v_j=v\neq u$ used above.  (A general prescription for changing schemes is given in section 4.3.2 of \cite{Bzowski:2015pba}.) The final result is
\begin{align}
A^{a_2 a_3}_1 & = \ren{1}^{a_2 a_3} \frac{2(4 p_1 + a_{23})}{a_{123}^4}, \\
A^{a_2 a_3}_2 & = \ren{1}^{a_2 a_3} \frac{2 p_1^2}{a_{123}^3} - \frac{2(2 p_1 + a_{23})}{a_{123}^2} \2_{JJ} \delta^{a_2 a_3}, \\
A^{a_2 a_3}_3 & = \frac{\ren{1}^{a_2 a_3}}{a_{123}^3} \Big( {-} 2 p_1^2 - p_2^2+p_3^2 - 3 p_1 p_2 + 3 p_1 p_3 \Big) - \frac{2(2 p_1 + a_{23})}{a_{123}^2} \2_{JJ} \delta^{a_2 a_3}, \\
A^{a_2 a_3}_4 & = \ren{1}^{a_2 a_3} \frac{(2 p_1 + a_{23}) (p_1^2 - a_{23}^2 + 4 b_{23})}{2 a_{123}^2} + \:\left( \frac{2 p_1^2}{a_{123}} - a_{23} \right) \2_{JJ} \delta^{a_2 a_3},
\end{align} 
where we have used the symmetric polynomials defined in \eqref{e:variables}.
A straightforward check confirms this solution satisfies all primary and secondary CWIs.

\subsubsection{Even spacetime dimensions}

The procedure for even spacetime dimensions $d=2N\ge 4$ is similar to that for odd dimensions  discussed above.  The two main differences are, firstly, the presence of additional $(- - -)$ singularities, which means the triple-$K$ integrals for the form factor $A_4^{a_2a_3}$ now have $\ep^{-2}$ poles.  The primary constants must therefore be expanded to order $\ep^2$.
Secondly, in order to remove the divergences from the regulated form factors, we will need to renormalise by introducing an appropriate counterterm.

Evaluating the divergent part of $A_4^{a_2 a_3}$ in the general scheme $v_j=v\neq u$ for $j=1,2,3$, we find the singularity as $v\rightarrow u$ can be removed by re-defining the primary constants as
\begin{align}
\3{4}^{(0) a_2 a_3} & = -(d-2) \3{3}^{(0) a_2 a_3}, \\
\3{4}^{(n) a_2 a_3} & = -(d-2) \3{3}^{(n) a_2 a_3} -2 v \3{3}^{(n-1) a_2 a_3} + (u - v) \redef{4}^{(n) a_2 a_3} + O((u-v)^2), 
\end{align}
for $n=1,2$.
With these substitutions the limit $v \rightarrow u$ is finite and the regulated secondary Ward identities can be solved. By taking the zero-momentum limit, one arrives at the solution
\begin{align}
\3{3}^{(0) a_2 a_3} = \3{2}^{(0) a_2 a_3} & = -d \3{1}^{(0) a_2 a_3} + \frac{(-1)^N 2^{3-N} \2_{JJ}}{(N-1)!} \delta^{a_2 a_3}, \\[1ex]
\3{3}^{(1) a_2 a_3} = \3{2}^{(1) a_2 a_3} & = -d \3{1}^{(1) a_2 a_3} -2 v \3{1}^{(0) a_2 a_3} \nn\\
& \qquad + \frac{(-1)^N 2^{3-N} v}{(N-1)!} \delta^{a_2 a_3} \left[ \2_{JJ} (\gamma_E - \ln 2 - H_{N-1} ) + \sl_{JJ} \right], \\[1ex]
\redef{4}^{(1) a_2 a_3} & = - \frac{1}{2} d(d - 4) \3{1}^{(0) a_2 a_3} - \delta^{a_2 a_3} \frac{(-1)^N 2^{3-N} N \2_{JJ}}{(N-1)!}, \\[1ex]
\redef{4}^{(2) a_2 a_3} & = - \frac{d(d - 4)}{2} \3{1}^{(1) a_2 a_3} - 2 (d-2) v \3{1}^{(0) a_2 a_3} \nn\\&\qquad  + \delta^{a_2 a_3} \frac{(-1)^N 2^{3-N} v \2_{JJ}}{(N-1)!}  \left[ -1 + N(H_{N-1} - \gamma_E + \ln 2) \right] \nn\\
& \qquad - \frac{(-1)^N 2^{3-N} N v \sl_{JJ}}{(N-1)!} \delta^{a_2 a_3}.
\end{align}

\paragraph{Four dimensions.} The primary constants for $d=4$ are
\begin{align}
\3{3}^{(0) a_2 a_3} &= \3{2}^{(0) a_2 a_3}  = 2 (\2_{JJ} \delta^{a_2 a_3} - 2 \3{1}^{(0) a_2 a_3}), \\
\3{3}^{(1) a_2 a_3} &= \3{2}^{(1) a_2 a_3}  = 2 v \2_{JJ} ( \gamma_E  - 1 - \ln 2) + 2 v (\sl_{JJ} \delta^{a_2 a_3} {-} \3{1}^{(0) a_2 a_3}) - 4 \3{1}^{(1) a_2 a_3}, \\
\3{4}^{(0) a_2 a_3} & = - 2 \3{3}^{(0) a_2 a_3}, \\
\3{4}^{(1) a_2 a_3} & = - 2 ( v \3{3}^{(0) a_2 a_3} + \3{3}^{(1) a_2 a_3}) - 4 (u - v) \2_{JJ} \delta^{a_2 a_3}, \\
\3{4}^{(2) a_2 a_3} & = - 2 ( v \3{3}^{(1) a_2 a_3} + \3{3}^{(2) a_2 a_3}) \nn\\
& \qquad + \: 2 (u - v) v \left[ \2_{JJ} \delta^{a_2 a_3} \left(1 - 2 \gamma_E +\ln 4)  \right)  - 2 \sl_{JJ} \delta^{a_2 a_3} - 2 \3{1}^{(0) a_2 a_3} \right].
\end{align}
With these primary constants, the limit $v \rightarrow u$ is finite and the resulting form factors solve the secondary CWI. All that remains is to deal with the singularities in $\ep$, which are 
\begin{align}\label{AdivTJJ4d}
A_1^{a_2 a_3} & = O(\ep^0), \nn\\
A_2^{a_2 a_3} & = - \frac{2 \2_{JJ} \delta^{a_2 a_3}}{u \epsilon} +  O(\ep^0), \nn\\
A_3^{a_2 a_3} & = - \frac{2 \2_{JJ} \delta^{a_2 a_3}}{u \epsilon} +  O(\ep^0), \nn\\
A_4^{a_2 a_3} & = \frac{\2_{JJ} \delta^{a_2 a_3}}{u \epsilon} \big( p_1^2 - p_2^2 - p_3^2 \big) +  O(\ep^0).
\end{align}
Notice that these surviving singularities are all $\ep^{-1}$ poles proportional to the 2-point normalisation $\2_{JJ}$.  Moreover, there is no dependence on $\3{1}^{(0) a_2 a_3}$, the sole remaining  undetermined constant. This fits with our earlier observation that the only available counterterm has been already fixed by the renormalisation of the 2-point function. 

Specifically, this counterterm is given by \eqref{e:SctJJ}, setting $N=2$.  The divergent part of the counterterm coefficient is fixed by \eqref{e:ctJJ}.
One can easily verify that the contribution from this counterterm cancels all the divergences in \eqref{AdivTJJ4d}.  We then obtain the renormalised form factors
{\allowdisplaybreaks 
\begin{align}
A_1^{a_2 a_3} & = \ren{1}^{a_2 a_3} I_{5 \{211\}} \nn\\
& = \ren{1}^{a_2 a_3} p_1 p_2 p_3 \left( p_1 \frac{\partial}{\partial p_1} - 1 \right) \frac{\partial^3}{\partial p_1 \partial p_2 \partial p_3} I_{1\{000\}} \\[1ex]
A_2^{a_2 a_3} & = 2 \2_{JJ} \delta^{a_2 a_3} \left( 2  - p_1 \frac{\partial}{\partial p_1} \right) I_{2\{111\}}^{\text{(fin)}} - \ren{1}^{a_2 a_3} p_1^3 p_2 p_3 \frac{\partial^3}{\partial p_1 \partial p_2 \partial p_3} I_{1\{000\}} \nn\\
& \qquad - \: \frac{2}{3} \2_{JJ} \delta^{a_2 a_3} \left[ \ln \frac{p_1^2}{\mu^2} + \ln \frac{p_2^2}{\mu^2} + \ln \frac{p_3^2}{\mu^2} \right] + \frac{2}{3} \2_{JJ} \delta^{a_2 a_3} - 2 \delta^{a_2 a_3} \sdt_{JJ}, \\[1ex]
A_3^{a_2 a_3} & = 2 \2_{JJ} \delta^{a_2 a_3} \left( 2  - p_1 \frac{\partial}{\partial p_1} \right) I_{2\{111\}}^{\text{(fin)}} + 2 \ren{1}^{a_2 a_3} p_1 p_2 p_3^2 \left(1 - p_1 \frac{\partial}{\partial p_1} \right) \frac{\partial^2}{\partial p_1 \partial p_2} I_{1\{000\}} \nn\\
& \qquad - \: \frac{2}{3} \2_{JJ} \delta^{a_2 a_3} \left[ \ln \frac{p_1^2}{\mu^2} + \ln \frac{p_2^2}{\mu^2} + \ln \frac{p_3^2}{\mu^2} \right] - 2 (\ren{1}^{a_2 a_3} + \delta^{a_2 a_3} \sdt_{JJ}) + \frac{2}{3} \2_{JJ} \delta^{a_2 a_3}, \\[10ex]
A_4^{a_2 a_3} & = - 2 \2_{JJ} \delta^{a_2 a_3} p_1^2 I_{2\{111\}}^{\text{(fin)}} - 2 \ren{1}^{a_2 a_3} p_2^2 p_3^2 p_1 \left( 1 - p_1 \frac{\partial}{\partial p_1} \right) \frac{\partial}{\partial p_1} I_{1\{000\}} \nn\\
& \qquad + \: \frac{1}{3} \2_{JJ} \delta^{a_2 a_3} p_1^2 \left[ \ln \frac{p_1^2}{\mu^2} + \ln \frac{p_2^2}{\mu^2} + \ln \frac{p_3^2}{\mu^2} \right] -  \2_{JJ} \delta^{a_2 a_3} \left[ p_2^2 \ln \frac{p_2^2}{\mu^2} + p_3^2 \ln \frac{p_3^2}{\mu^2} \right] \nn\\
& \qquad + (\ren{1}^{a_2 a_3} + \delta^{a_2 a_3} \sdt_{JJ})(p_1^2 - p_2^2 - p_3^2) - \delta^{a_2 a_3} \2_{JJ} p_1^2,
\end{align}}\noindent
where $I_{2\{111\}}^{\text{(fin)}}$ and $I_{1\{000\}}$ are given in \eqref{I2111fin} and \eqref{I1000} respectively,
and $\sdt_{JJ}$ is the scheme-dependent constant appearing in the renormalised 2-point function \eqref{e:2ptJJren}.  The solution also depends on the undetermined constant $\ren{1}^{(0)a_2 a_3}$, which we have relabelled as $\ren{1}^{a_2 a_3}$.
Notice that a change of renormalisation scale $\mu^2\rightarrow e^{-\lambda}\mu^2$ is equivalent to sending $D_{JJ}\rightarrow D_{JJ}+\lambda C_{JJ}$, both for the 2-point function \eqref{e:2ptJJren} and for the 3-point function.

Each renormalised form factor contains the following pieces.  First, there is a nonlocal piece involving contributions derived from triple-$K$ integrals.  (For $A_1^{a_2a_3}$ this is all we have.) Next, there is a scale-violating logarithmic piece consisting of terms that depend explicitly on the renormalisation scale $\mu$.  Finally, there is an ultralocal piece containing terms that are polynomial in the squares of momenta.  While the coefficients of certain ultralocal terms involve $D_{JJ}$ and hence are scheme-dependent, one can nevertheless extract scheme-independent information by combining the coefficients of multiple terms. This is a (rare) example of a contact term that is unambiguous.  It would be interesting to understand the physics associated with such unambiguous contact terms.

\paragraph{Anomalous Ward identities.}
The anomalous dilatation Ward identities are 
\begin{align}
\mu \frac{\partial}{\partial \mu} A_1^{a_2 a_3} & = 0, \\
\mu \frac{\partial}{\partial \mu} A_2^{a_2 a_3} & = 4 \delta^{a_2 a_3} \2_{JJ}, \\
\mu \frac{\partial}{\partial \mu} A_3^{a_2 a_3} & = 4 \delta^{a_2 a_3} \2_{JJ}, \\
\mu \frac{\partial}{\partial \mu} A_4^{a_2 a_3} & = 2 \delta^{a_2 a_3} \2_{JJ} \left(-p_1^2 +p_2^2 + p_3^2 \right),
\end{align}
while of the primary CWI, only those in the fourth line below develop an anomaly,
\begin{align}
& \K_{12} A_1^{a_2 a_3} = 0, && \K_{13} A_1^{a_2 a_3} = 0, \\
& \K_{12} A_2^{a_2 a_3} = -2 A_1^{a_2 a_3}, && \K_{13} A_2^{a_2 a_3} = - 2 A_1^{a_2 a_3}, \\
& \K_{12} A_3^{a_2 a_3} = 0, && \K_{13} A_3^{a_2 a_3} = 4 A_1^{a_2 a_3}, \\
& \K_{12} A_4^{a_2 a_3} = 2 A_3^{a_2 a_3} + 8 \delta^{a_2 a_3} \2_{JJ} , && \K_{13} A_4^{a_2 a_3} = 2 A_3^{a_3 a_2}(p_2 \leftrightarrow p_3) + 8 \delta^{a_2 a_3} \2_{JJ},
\end{align}
The anomalous secondary CWIs are
\begin{align}
& \Lo_{4} A_1^{a_2 a_3} + \Ro \left[ A_3^{a_2 a_3} - A_3^{a_2 a_3}(p_2 \leftrightarrow p_3) \right] = 0, \\[1ex]
& \Lo'_{3} A_1^{a_2 a_3} + 2 \Ro' \left[ A_3^{a_2 a_3} - A_2^{a_2 a_3} \right] = 0, \\[1ex]
& \Lo_{2} A_2^{a_2 a_3} - p_1^2 \left[ A_3^{a_2 a_3} - A_3^{a_2 a_3}(p_2 \leftrightarrow p_3) \right]  \nn\\[0.5ex]
& \qquad =  - 4 \delta^{a_2 a_3} \2_{JJ} p_1^2+ 8 \cdot \text{coefficient of } \delta^{\mu_2 \mu_3} p_{2\mu_1} \text{ in } p_{1}^{ \nu_1} \lla T_{\mu_1 \nu_1}(\bs{p}_1) J^{\mu_2 a_2}(\bs{p}_2) J^{\mu_3 a_3}(\bs{p}_3) \rra \nn\\
& \qquad = 8 \delta^{a_2 a_3} \left[ p_2^2 \left( \2_{JJ} \ln \frac{p_2^2}{\mu^2} + \sdt_{JJ} \right) - p_3^2 \left( \2_{JJ} \ln \frac{p_3^2}{\mu^2} + \sdt_{JJ} \right) - \frac{1}{2} \2_{JJ} p_1^2 \right], \\[1ex]
& \Lo_{4} A_3^{a_2 a_3} - 2 \Ro A_4^{a_2 a_3}  \nn\\[0.5ex]
& \qquad = - 8 \delta^{a_2 a_3} \2_{JJ} p_1^2 +16 \cdot \text{coefficient of } \delta_{\mu_1}^{ \mu_2} p_1^{\mu_3} \text{ in } p_{1}^{ \nu_1} \lla T_{\mu_1 \nu_1}(\bs{p}_1) J^{\mu_2 a_2}(\bs{p}_2) J^{\mu_3 a_3}(\bs{p}_3)) \rra \nn\\
& \qquad = 16 \delta^{a_2 a_3} \left[ - p_3^2 \left( \2_{JJ} \ln \frac{p_3^2}{\mu^2} + \sdt_{JJ} \right) - \frac{1}{2} \2_{JJ} p_1^2 \right].
\end{align}
The trace anomaly is given by the first term in \eqref{traceanomquad}, setting $N=2$.  Expanding to quadratic order in the gauge field, we obtain
the anomaly entering the trace Ward identity \eqref{e:traceW_TJJ} and the reconstruction formula \eqref{a:tjjrecon},
\begin{equation}
\mathcal{A}^{\mu_2 \mu_3 a_2 a_3} =C_{JJ} \delta^{a_2 a_3} \left[ (p_1^2 - p_2^2 - p_3^2 ) \delta^{\mu_2 \mu_3}-2p_3^{\mu_2} p_2^{\mu_3} \right].\label{TJJanom}
\end{equation}
We can equivalently express this  in terms of the form factor decomposition \eqref{TJJanomdecomp} as 
\[\label{TJJB1}
B_1^{a_2a_3}=2C_{JJ}\delta^{a_2a_3}, \qquad
B_2^{a_2a_3} = C_{JJ}\delta^{a_2a_3}(p_1^2-p_2^2-p_3^2).
\]

\subsection{\texorpdfstring{$\<T_{\mu_1\nu_1}T_{\mu_2\nu_2}T_{\mu_3\nu_3}\>$}{<TTT>}}

\subsubsection{General analysis}

\paragraph{Decomposition.} The transverse and trace Ward identities are
\begin{align}
& p_1^{\nu_1} \lla T_{\mu_1 \nu_1}(\bs{p}_1) T_{\mu_2 \nu_2}(\bs{p}_2) T_{\mu_3 \nu_3}(\bs{p}_3) \rra  \nn\\[1ex]
& \qquad =
\: 2 p_{1 (\mu_3} \lla T_{\nu_3) \mu_1}(\bs{p}_2) T_{\mu_2 \nu_2}(-\bs{p}_2) \rra
+ 2 p_{1 (\mu_2} \lla T_{\nu_2) \mu_1}(\bs{p}_3) T_{\mu_3 \nu_3}(-\bs{p}_3) \rra \nn\\[1ex]
& \qquad\quad - \: p_{3 \mu_1} \lla T_{\mu_2 \nu_2}(\bs{p}_2) T_{\mu_3 \nu_3}(-\bs{p}_2) \rra
- p_{2 \mu_1} \lla T_{\mu_2 \nu_2}(\bs{p}_3) T_{\mu_3 \nu_3}(-\bs{p}_3) \rra, \label{e:WardTTT} \\[2ex]
& \lla T(\bs{p}_1) T_{\mu_2 \nu_2}(\bs{p}_2) T_{\mu_3 \nu_3}(\bs{p}_3) \rra \nn\\[1ex]& \qquad=  
2\lla T_{\mu_2\nu_2}(\bs{p}_2)T_{\mu_3\nu_3}(-\bs{p}_2)\rra
+2\lla T_{\mu_2\nu_2}(\bs{p}_3)T_{\mu_3\nu_3}(-\bs{p}_3)\rra
+\mathcal{A}_{\mu_2 \nu_2 \mu_3 \nu_3}. \label{e:TWI_TTT}
\end{align}
In odd spacetime dimensions anomalies are absent and $\mathcal{A}_{\mu_2\nu_2\mu_3\nu_3}$ can be set to zero.  In even dimensions, the anomaly is generally present.  Taking the $\delta^{\mu_2\nu_2}$ trace of \eqref{e:WardTTT}, swapping $1\leftrightarrow 2$, then comparing with the $p_2^{\nu_2}$ contraction of \eqref{e:TWI_TTT}, we see that the anomaly is transverse, {\it i.e.,}  $p_2^{\nu_2}\mathcal{A}_{\mu_2\nu_2\mu_3\nu_3}=0$ and similarly $p_3^{\nu_3}\mathcal{A}_{\mu_2\nu_2\mu_3\nu_3}=0$.

To decompose the tensorial structure of the 3-point function, we define the projector
\[
X_{\mu_1\nu_1}{}^{\alpha_1\beta_1}(\bs{p}_1)=\delta_{\mu_1}^{(\alpha_1}\delta_{\nu_1}^{\beta_1)}-\Pi_{\mu_1\nu_1}{}^{\alpha_1\beta_1}(\bs{p}_1)
=\mathscr{T}_{\mu_1 \nu_1}^{\quad\,\,(\alpha_1}(\bs{p}_1) p_{1}^{ \beta_1)} + \frac{\pi_{\mu_1 \nu_1}(\bs{p}_1)}{d-1} \delta^{\alpha_1 \beta_1}, 
\]
where we used \eqref{a:T} and \eqref{Idecomp}.
With the aid of \eqref{e:WardTTT} and \eqref{e:TWI_TTT}, we can now write
\begin{align}
& X_{\mu_1\nu_1}{}^{\alpha_1\beta_1}(\bs{p}_1) \lla T_{\alpha_1 \beta_1}(\bs{p}_1) T_{\mu_2 \nu_2}(\bs{p}_2) T_{\mu_3 \nu_3}(\bs{p}_3) \rra
\nn\\[1ex]&\quad 
 =\mathcal{L}_{\mu_1 \nu_1 \mu_2 \nu_2 \mu_3 \nu_3}(\bs{p}_1, \bs{p}_2, \bs{p}_3)  
 +
 \mathcal{L}_{\mu_1 \nu_1 \mu_3 \nu_3 \mu_2 \nu_2}(\bs{p}_1, \bs{p}_3, \bs{p}_2)
 +\frac{\pi_{\mu_1\nu_1}(\bs{p}_1)}{d-1}\mathcal{A}_{\mu_2\nu_2\mu_3\mu_3},
\end{align}
where
\begin{align}
 \mathcal{L}_{\mu_1 \nu_1 \mu_2 \nu_2 \mu_3 \nu_3}(\bs{p}_1, \bs{p}_2, \bs{p}_3)  
&=
\left[ 2\mathscr{T}_{\mu_1 \nu_1}^{\quad\,\, \alpha_3}(\bs{p}_1) p_{1(\mu_3} \delta_{\nu_3)}^{ \beta_3}  - p_3^{\sigma} \mathscr{T}_{\mu_1 \nu_1\sigma}(\bs{p}_1) \delta_{\mu_3}^{ \alpha_3} \delta_{\nu_3}^{ \beta_3} \right. 
\nn\\& \qquad
\left. + \: \frac{2 \pi_{\mu_1 \nu_1}(\bs{p}_1)}{d-1} \delta_{\mu_3}^{ \alpha_3} \delta_{\nu_3}^{ \beta_3} \right] \lla T_{\alpha_3 \beta_3}(\bs{p}_2) T_{\mu_2 \nu_2}(- \bs{p}_2) \rra. \label{e:L_TTT}
\end{align}
Next, we proceed to expand out the trivial identity
\begin{align}
&\lla t_{\mu_1 \nu_1}(\bs{p}_1) t_{\mu_2 \nu_2}(\bs{p}_2) t_{\mu_3 \nu_3}(\bs{p}_3) \rra \nn\\[1ex]&
=\Big(\delta_{\mu_1}^{\alpha_1}\delta_{\nu_1}^{\beta_1}-X_{\mu_1\nu_1}{}^{\alpha_1\beta_1}(\bs{p}_1)\Big)
\Big(\delta_{\mu_2}^{\alpha_2}\delta_{\nu_2}^{\beta_2}-X_{\mu_2\nu_2}{}^{\alpha_2\beta_2}(\bs{p}_2)\Big)
\Big(\delta_{\mu_3}^{\alpha_3}\delta_{\nu_3}^{\beta_3}-X_{\mu_3\nu_3}{}^{\alpha_3\beta_3}(\bs{p}_3)\Big)\nn\\[1ex]&\qquad
\qquad\times\lla T_{\alpha_1 \beta_1}(\bs{p}_1) T_{\alpha_2 \beta_2}(\bs{p}_2) T_{\alpha_3 \beta_3}(\bs{p}_3) \rra. 
\end{align}
Since the 2-point function is transverse and traceless, while the anomaly is transverse (but not traceless), we find that 
\begin{align}
&X_{\mu_1\nu_1}{}^{\alpha_1\beta_1}(\bs{p}_1) 
X_{\mu_2\nu_2}{}^{\alpha_2\beta_2}(\bs{p}_2)
\lla T_{\alpha_1 \beta_1}(\bs{p}_1) T_{\alpha_2 \beta_2}(\bs{p}_2) T_{\mu_3 \nu_3}(\bs{p}_3) \rra
\\[1ex]
&=X_{\mu_2\nu_2}{}^{\alpha_2\beta_2}(\bs{p}_2)
 \mathcal{L}_{\mu_1 \nu_1  \mu_3 \nu_3\alpha_2 \beta_2}(\bs{p}_1,  \bs{p}_3, \bs{p}_2)  +\frac{\pi_{\mu_1\nu_1}(\bs{p}_1)\pi_{\mu_2\nu_2}(\bs{p}_2)}{(d-1)^2}\,\delta^{\alpha_2\beta_2}\mathcal{A}_{\alpha_2\beta_2\mu_3\mu_3}
\end{align}
and similarly for permutations.  
Moreover, 
\begin{align}
&X_{\mu_1\nu_1}{}^{\alpha_1\beta_1}(\bs{p}_1) 
X_{\mu_2\nu_2}{}^{\alpha_2\beta_2}(\bs{p}_2)
X_{\mu_3\nu_3}{}^{\alpha_3\beta_3}(\bs{p}_3)
\lla T_{\alpha_1 \beta_1}(\bs{p}_1) T_{\alpha_2 \beta_2}(\bs{p}_2) T_{\mu_3 \nu_3}(\bs{p}_3) \rra
\\[1ex]
&=\frac{1}{(d-1)^3}\pi_{\mu_1\nu_1}(\bs{p}_1)\pi_{\mu_2\nu_2}(\bs{p}_2)\pi_{\mu_3\nu_3}(\bs{p}_3)\delta^{\alpha_2\beta_2}\delta^{\alpha_3\beta_3}\mathcal{A}_{\alpha_2\beta_2\alpha_3\beta_3}.
\end{align}
The full 3-point function can thus be reconstructed from its transverse-traceless part using
\begin{align}
& \lla T_{\mu_1 \nu_1}(\bs{p}_1) T_{\mu_2 \nu_2}(\bs{p}_2) T_{\mu_3 \nu_3}(\bs{p}_3) \rra  \nn\\[1ex]&\quad = \lla t_{\mu_1 \nu_1}(\bs{p}_1) t_{\mu_2 \nu_2}(\bs{p}_2) t_{\mu_3 \nu_3}(\bs{p}_3) \rra   + \: \sum_{\sigma} \mathcal{L}_{\mu_{\sigma(1)} \nu_{\sigma(1)} \mu_{\sigma(2)} \nu_{\sigma(2)} \mu_{\sigma(3)} \nu_{\sigma(3)}}(\bs{p}_{\sigma(1)}, \bs{p}_{\sigma(2)}, \bs{p}_{\sigma(3)}) \nn\\
& \qquad  - X_{\mu_3\nu_3}{}^{\alpha_3\beta_3}(\bs{p}_3) \mathcal{L}_{\mu_1 \nu_1 \mu_2 \nu_2 \alpha_3 \beta_3}(\bs{p}_1, \bs{p}_2, \bs{p}_3) 
-X_{\mu_1\nu_1}{}^{\alpha_1\beta_1}(\bs{p}_1) \mathcal{L}_{\mu_2 \nu_2 \mu_3 \nu_3 \alpha_1 \beta_1}(\bs{p}_2, \bs{p}_3, \bs{p}_1) 
\nn\\[2ex]
& \qquad 
-X_{\mu_2\nu_2}{}^{\alpha_2\beta_2}(\bs{p}_2) \mathcal{L}_{\mu_3 \nu_3 \mu_1 \nu_1 \alpha_2 \beta_2}(\bs{p}_3, \bs{p}_1, \bs{p}_2) 
+\lla T_{\mu_1 \nu_1}(\bs{p}_1) T_{\mu_2 \nu_2}(\bs{p}_2) T_{\mu_3 \nu_3}(\bs{p}_3) \rra_{\Anom}
\label{re_TTT}
\end{align}
where the sum is taken over all six permutations $\sigma$ of the set $\{1,2,3\}$.  The contribution coming from the anomaly is
\begin{align}\label{p:anomdecomp}
& \lla T_{\mu_1 \nu_1}(\bs{p}_1) T_{\mu_2 \nu_2}(\bs{p}_2) T_{\mu_3 \nu_3}(\bs{p}_3) \rra_{\Anom}  \nn\\[1ex]&= \Big[\Big(
\frac{1}{d-1}\,\pi_{\mu_1 \nu_1}(\bs{p}_1) \mathcal{A}_{\mu_2\nu_2\mu_3\nu_3} 
-\frac{1}{(d-1)^2}\,\pi_{\mu_1\nu_1}(\bs{p}_1)\pi_{\mu_2\nu_2}(\bs{p}_2)\delta^{\alpha_2\beta_2}\mathcal{A}_{\alpha_2\beta_2\mu_3\mu_3}\Big)
\nn\\[1ex]&\qquad
+[(\mu_1,\nu_1,\bs{p}_1)\rightarrow(\mu_2,\nu_2,\bs{p}_2)\rightarrow(\mu_3,\nu_3,\bs{p}_3)\rightarrow(\mu_1,\nu_1,\bs{p}_1)]\nn\\[1ex]&\qquad
+[(\mu_1,\nu_1,\bs{p}_1)\rightarrow(\mu_3,\nu_3,\bs{p}_3)\rightarrow(\mu_2,\nu_2,\bs{p}_2)\rightarrow(\mu_1,\nu_1,\bs{p}_1)]\,
\Big]
\nn\\[1ex]
&\quad +\frac{1}{(d-1)^3}\pi_{\mu_1\nu_1}(\bs{p}_1)\pi_{\mu_2\nu_2}(\bs{p}_2)\pi_{\mu_3\nu_3}(\bs{p}_3)\delta^{\alpha_2\beta_2}\delta^{\alpha_3\beta_3}\mathcal{A}_{\alpha_2\beta_2\alpha_3\beta_3}.
\end{align}

\paragraph{Form factors.} The tensorial decomposition of the transverse-traceless part is
\begin{align}
& \lla t_{\mu_1 \nu_1}(\bs{p}_1) t_{\mu_2 \nu_2}(\bs{p}_2) t_{\mu_3 \nu_3}(\bs{p}_3) \rra \nonumber\\[1ex]
& \qquad = \Pi_{\mu_1 \nu_1\alpha_1 \beta_1}(\bs{p}_1) \Pi_{\mu_2 \nu_2\alpha_2 \beta_2}(\bs{p}_2) \Pi_{\mu_3 \nu_3\alpha_3 \beta_3}(\bs{p}_3) \left[
A_1 p_2^{\alpha_1} p_2^{\beta_1} p_3^{\alpha_2} p_3^{\beta_2} p_1^{\alpha_3} p_1^{\beta_3} \right. \nonumber \\[1ex]
& \qquad \qquad + \: A_2 \delta^{\beta_1 \beta_2} p_2^{\alpha_1} p_3^{\alpha_2} p_1^{\alpha_3} p_1^{\beta_3} + A_2(p_1 \leftrightarrow p_3) \delta^{\beta_2 \beta_3} p_2^{\alpha_1} p_2^{\beta_1} p_3^{\alpha_2} p_1^{\alpha_3} \nonumber \\[1ex]
& \qquad \qquad \qquad \qquad + \: A_2(p_2 \leftrightarrow p_3) \delta^{\beta_1 \beta_3} p_2^{\alpha_1} p_3^{\alpha_2} p_3^{\beta_2} p_1^{\alpha_3} \nonumber \\[1ex]
& \qquad \qquad + \: A_3 \delta^{\alpha_1 \alpha_2} \delta^{\beta_1 \beta_2} p_1^{\alpha_3} p_1^{\beta_3} + A_3(p_1 \leftrightarrow p_3) \delta^{\alpha_2 \alpha_3} \delta^{\beta_2 \beta_3} p_2^{\alpha_1} p_2^{\beta_1} \nonumber \\[1ex]
& \qquad \qquad \qquad \qquad + \: A_3(p_2 \leftrightarrow p_3) \delta^{\alpha_1 \alpha_3} \delta^{\beta_1 \beta_3} p_3^{\alpha_2} p_3^{\beta_2} \nonumber \\[1ex]
& \qquad \qquad + \: A_4 \delta^{\alpha_1 \alpha_3} \delta^{\alpha_2 \beta_3} p_2^{\beta_1} p_3^{\beta_2} + A_4(p_1 \leftrightarrow p_3) \delta^{\alpha_1 \alpha_3} \delta^{\alpha_2 \beta_1} p_3^{\beta_2} p_1^{\beta_3} \nonumber \\[1ex]
 & \qquad \qquad \qquad \qquad + \: A_4(p_2 \leftrightarrow p_3) \delta^{\alpha_1 \alpha_2} \delta^{\alpha_3 \beta_2} p_2^{\beta_1} p_1^{\beta_3} \nonumber \\[1ex]
& \left. \qquad \qquad + \: A_5 \delta^{\alpha_1 \beta_2} \delta^{\alpha_2 \beta_3} \delta^{\alpha_3 \beta_1} \right]. \label{e:TTTdecomp}
\end{align}
The form factors $A_j$, $j=1, \ldots, 5$ are functions of the momentum magnitudes. If no arguments are specified then the standard ordering is assumed, $A_j = A_j(p_1, p_2, p_3)$, while by $p_i \leftrightarrow p_j$ we denote the exchange of the two momenta, \textit{e.g.}, $A_1(p_1 \leftrightarrow p_3) = A_2(p_3, p_2, p_1)$.

The form factors $A_1$ and $A_5$ are symmetric under any permutation of momenta, \textit{i.e.}, for any permutation $\sigma$ of the set $\{1,2,3\}$,
\begin{equation}
A_j(p_{\sigma(1)}, p_{\sigma(2)}, p_{\sigma(3)}) = A_j(p_1, p_2, p_3), \qquad j \in \{1,5\}.
\end{equation}
The remaining form factors are symmetric under $p_1 \leftrightarrow p_2$, \textit{i.e.}, they satisfy
\begin{equation}
A_j(p_2, p_1, p_3) = A_j(p_1, p_2, p_3), \qquad j \in \{2,3,4\}.
\end{equation}

Given the full correlator  $\lla T_{\mu_1 \nu_1}(\bs{p}_1) T_{\mu_2 \nu_2}(\bs{p}_2) T_{\mu_3 \nu_3}(\bs{p}_3) \rra$ with independent momenta chosen according to the index rule \eqref{a:momenta}, the form factors can be read off as
\begin{align}
A_1 & = \text{coefficient of } p_{2\mu_1} p_{2\nu_1} p_{3\mu_2} p_{3\nu_2} p_{1\mu_3} p_{1\nu_3}, \label{A1coefftof}\\
A_2 & = 4 \cdot \text{coefficient of } \delta_{\nu_1 \nu_2} p_{2\mu_1} p_{3\mu_2} p_{1\mu_3} p_{1\nu_3}, \\
A_3 & = 2 \cdot \text{coefficient of } \delta_{\mu_1 \mu_2} \delta_{\nu_1 \nu_2} p_{1\mu_3} p_{1\nu_3}, \\
A_4 & = 8 \cdot \text{coefficient of } \delta_{\mu_1 \mu_3} \delta_{\mu_2 \nu_3} p_{2\nu_1} p_{3\nu_2}, \\
A_5 & = 8 \cdot \text{coefficient of } \delta_{\mu_1 \nu_2} \delta_{\mu_2 \nu_3} \delta_{\mu_3 \nu_1}.
\end{align}

In addition to decomposing the transverse-traceless part of the 3-point function, we can introduce an analogous form factor decomposition for the anomaly $\mathcal{A}_{\mu_2\nu_2\mu_3\nu_3}$.
Since the anomaly is transverse, but not in general traceless, we first apply \eqref{Idecomp} twice to separate out the transverse-traceless and trace parts.  The tensor structure of the transverse-traceless parts can then be decomposed in terms of form factors, yielding
\begin{align}\label{anomaly_decomp}
 \mathcal{A}_{\mu_2\nu_2\mu_3\nu_3}  &= \Pi_{\mu_2 \nu_2\alpha_2 \beta_2}(\bs{p}_2) \Pi_{\mu_3 \nu_3\alpha_3 \beta_3}(\bs{p}_3) \left[ B_1 p_3^{\alpha_2} p_3^{\beta_2} p_1^{\alpha_3} p_1^{\beta_3}  + \: B_2 \delta^{\beta_2 \beta_3} p_3^{\alpha_2} p_1^{\alpha_3} + B_3 \delta^{\alpha_2 \alpha_3} \delta^{\beta_2 \beta_3} \right] \nn\\[1ex]
& \quad + \:  B_4\,\pi_{\mu_2 \nu_2}(\bs{p}_2) \Pi_{\mu_3 \nu_3\alpha_3 \beta_3}(\bs{p}_3) p^{\alpha_3}_1 p^{\beta_3}_1 +  B_4(p_2 \leftrightarrow p_3)\Pi_{\mu_2 \nu_2\alpha_2 \beta_2}(\bs{p}_2) \pi_{\mu_3 \nu_3}(\bs{p}_3) p^{\alpha_2}_3 p^{\beta_2}_3 \nn\\[1ex]
& \quad + \: B_5\,\pi_{\mu_2 \nu_2}(\bs{p}_2) \pi_{\mu_3 \nu_3}(\bs{p}_3).
\end{align}
Here, the form factors $B_j=B_j(p_1,p_2,p_3)$, $j=1,\ldots 5$ are functions of the momentum magnitudes, with standard ordering assumed unless otherwise specified.
Explicit expressions for the case $d=4$ are given in \eqref{B1eqn}\,-\,\eqref{B5eqn}.
Using this decomposition, we can re-write the anomalous contribution \eqref{p:anomdecomp} to the reconstruction formula \eqref{re_TTT} as 
\begin{align}\label{re_anomaly}
& \lla T_{\mu_1 \nu_1}(\bs{p}_1) T_{\mu_2 \nu_2}(\bs{p}_2) T_{\mu_3 \nu_3}(\bs{p}_3) \rra_{\Anom}  \nn\\[1ex]&= \Big(
\frac{1}{d-1}\,\pi_{\mu_1 \nu_1}(\bs{p}_1) \mathcal{A}_{\mu_2\nu_2\mu_3\nu_3} 
+[ (\bs{p}_1, \mu_1, \nu_1) \leftrightarrow (\bs{p}_2, \mu_2, \nu_2) ] + [ (\bs{p}_1, \mu_1, \nu_1) \leftrightarrow (\bs{p}_3, \mu_3, \nu_3) ] \Big)\nn\\[1ex]
&\quad - \Big(\: \frac{1}{d-1}\,\pi_{\mu_1 \nu_1}(\bs{p}_1) \pi_{\mu_2 \nu_2}(\bs{p}_2) \left[ \Pi_{\mu_3 \nu_3\alpha_3 \beta_3}(\bs{p}_3)p_1^{\alpha_3}p_1^{\beta_3} B_4(p_1, p_2, p_3) + \pi_{\mu_3 \nu_3}(\bs{p}_3) B_5(p_1, p_2, p_3) \right] \nn\\[1ex]
& \qquad\qquad+[ (\bs{p}_1, \mu_1, \nu_1) \leftrightarrow (\bs{p}_3, \mu_3, \nu_3) ] + [ (\bs{p}_2, \mu_2, \nu_2) \leftrightarrow (\bs{p}_3, \mu_3, \nu_3) ] \Big)\nn\\[0.5ex]
& \quad+ \: \frac{1}{d-1}\, B_5\,\pi_{\mu_1 \nu_1}(\bs{p}_1) \pi_{\mu_2 \nu_2}(\bs{p}_2) \pi_{\mu_3 \nu_3}(\bs{p}_3).
\end{align}

\paragraph{Primary CWIs.} The primary CWIs are
\begin{equation}
\begin{array}{ll}
\K_{12} A_1 = 0, &\qquad \K_{13} A_1 = 0, \\
\K_{12} A_2 = 0, &\qquad \K_{13} A_2 = 8 A_1, \\
\K_{12} A_3 = 0, &\qquad \K_{13} A_3 = 2 A_2, \\
\K_{12} A_4 = 4 \left[ A_2(p_1 \leftrightarrow p_3) - A_2(p_2 \leftrightarrow p_3) \right], &\qquad \K_{13} A_4 = - 4 A_2(p_2 \leftrightarrow p_3), \\
\K_{12} A_5 = 2 \left[ A_4(p_2 \leftrightarrow p_3) - A_4(p_1 \leftrightarrow p_3) \right], &\qquad \K_{13} A_5 = 2 \left[ A_4 - A_4(p_1 \leftrightarrow p_3) \right].
\end{array}
\end{equation}
The solution in terms of triple-$K$ integrals \eqref{a:J} is
\begin{align}
A_1 & = \3{1} J_{6 \{000\}}, \label{a:TTT1} \\
A_2 & = 4 \3{1} J_{5 \{001\}} + \3{2} J_{4 \{000\}}, \label{A2:TTT}\\
A_3 & = 2 \3{1} J_{4 \{002\}} + \3{2} J_{3 \{001\}} + \3{3} J_{2 \{000\}}, \\
A_4 & = 8 \3{1} J_{4 \{110\}} - 2 \3{2} J_{3 \{001\}} + \3{4} J_{2 \{000\}}, \\
A_5 & = 8 \3{1} J_{3 \{111\}} + 2 \3{2} \left( J_{2 \{110\}} + J_{2 \{101\}} + J_{2 \{011\}} \right) + \3{5} J_{0 \{000\}}, \label{a:TTTlast}
\end{align}
where $\3{j}$, $j=1,\ldots,5$ are constants.

\paragraph{Secondary CWIs.} The independent secondary CWIs are
\begin{align}
& (*) \ \Lo_{6} A_1 + \Ro \left[ A_2 - A_2(p_2 \leftrightarrow p_3) \right]  \nn\\
& \qquad = 2 \Delta_1 \cdot \text{coeff. of } p_{2\mu_1} p_{3\mu_2} p_{3\nu_2} p_{1\mu_3} p_{1\nu_3} \text{ in } p_{1}^{ \nu_1} \lla T_{\mu_1 \nu_1}(\bs{p}_1) T_{\mu_2 \nu_2}(\bs{p}_2) T_{\mu_3 \nu_3}(\bs{p}_3) \rra, 
\label{e:tttseccwi1}
 \\[1ex]
& \Lo_{6} A_2 + 2 \Ro \left[ 2 A_3 - A_4(p_1 \leftrightarrow p_3) \right]  \nn\\
& \qquad = 8 \Delta_1 \cdot \text{coefficient of } \delta_{\mu_1 \mu_2} p_{3\nu_2} p_{1\mu_3} p_{1\nu_3} \text{ in } p_{1}^{ \nu_1} \lla T_{\mu_1 \nu_1}(\bs{p}_1) T_{\mu_2 \nu_2}(\bs{p}_2) T_{\mu_3 \nu_3}(\bs{p}_3) \rra, 
\label{e:tttseccwi2}
 \\[-1ex]
& (*) \ \Lo_{4} \left[ A_2(p_1 \leftrightarrow p_3) \right] + \Ro  \left[ A_4(p_2 \leftrightarrow p_3) - A_4 \right] + 2 p_1^2 \left[ A_2(p_2 \leftrightarrow p_3) - A_2 \right]  \nn\\
& \qquad = 8 \Delta_1 \cdot \text{coefficient of } \delta_{\mu_2 \mu_3} p_{2\mu_1} p_{3\nu_2} p_{1\nu_3} \text{ in } p_{1}^{ \nu_1} \lla T_{\mu_1 \nu_1}(\bs{p}_1) T_{\mu_2 \nu_2}(\bs{p}_2) T_{\mu_3 \nu_3}(\bs{p}_3) \rra, 
\label{e:tttseccwi3}
 \\[-1ex]
& \Lo_{4} \left[ A_4(p_2 \leftrightarrow p_3) \right] - 2 \Ro A_5 + 2 p_1^2 \left[ A_4(p_1 \leftrightarrow p_3) - 4 A_3 \right]  \nn\\
& \qquad = 16 \Delta_1 \cdot \text{coefficient of } \delta_{\mu_1 \mu_2} \delta_{\mu_3 \nu_2} p_{1\nu_3} \text{ in } p_{1}^{ \nu_1} \lla T_{\mu_1 \nu_1}(\bs{p}_1) T_{\mu_2 \nu_2}(\bs{p}_2) T_{\mu_3 \nu_3}(\bs{p}_3) \rra,  \\[1ex]
& \Lo_{2} \left[ A_3(p_1 \leftrightarrow p_3) \right] + p_1^2 \left[ A_4 - A_4(p_2 \leftrightarrow p_3) \right]  \nn\\
& \qquad = 4 \Delta_1 \cdot \text{coefficient of } \delta_{\mu_2 \mu_3} \delta_{\nu_2 \nu_3} p_{2\mu_1} \text{ in } p_{1}^{ \nu_1} \lla T_{\mu_1 \nu_1}(\bs{p}_1) T_{\mu_2 \nu_2}(\bs{p}_2) T_{\mu_3 \nu_3}(\bs{p}_3) \rra,
\end{align}
where the operators $\Lo_{N}$ and $\Ro$ are defined in \eqref{a:L} and \eqref{a:R}.   All right-hand sides can be determined using the transverse Ward identity \eqref{e:WardTTT}; in particular, those for \eqref{e:tttseccwi1}\,-\,\eqref{e:tttseccwi3} vanish.
The identities denoted by asterisks are redundant, \textit{i.e.}, they are trivially satisfied in all cases and do not impose any additional conditions on the primary constants. 


\paragraph{Divergences.} The analysis of singularities arising in the solution to the primary CWIs is similar to that presented earlier for $\< J^{\mu_1} J^{\mu_2} J^{\mu_3} \>$. The triple-$K$ integrals in the solution may satisfy the $(---)$ and/or various versions of the $(--+)$ singularity conditions. The following table lists all the possibilities, depending on the spacetime dimension $d$. As usual, $n$ denotes an arbitrary non-negative integer.

\begin{center}
\begin{tabular}{|c|c|c|c|} \hline
Form factor & Integral & $(---)$  & $(--+)$  \\ \hline
$A_1$ & $J_{6 \{000\}}$ & $d=6 + 2n$ & never \\ \hline
$A_2$ & $J_{5 \{001\}}, J_{4\{000\}}$ & $d=4 + 2n$ & never \\ \hline
$A_3, A_4$ &  $J_{4 \{002\}}, J_{4\{110\}}, J_{3\{001\}}, J_{2\{000\}}$ & $d=2 + 2n$ & never \\ \hline
$A_5$ & $J_{3 \{111\}}, J_{2\{101\}}, J_{2\{011\}}$ & $d= 2n$ & never \\
& $J_{2\{110\}}, J_{0\{000\}}$ & $d= 2n$ & always \\ \hline
\end{tabular}
\captionof{table}{Singularities of triple-$K$ integrals in the form factors for $\< T_{\mu_1 \nu_1} T_{\mu_2 \nu_2} T_{\mu_3 \nu_3} \>$.\label{tab:TTT}}
\end{center}

\paragraph{Regularisation.} 
We cannot immediately work in the desired scheme \eqref{truescheme}, as this fails to regulate the $(--+)$ singularities in the final two integrals for the form factor $A_5$ in the table above.  Instead, we begin in the more general scheme  say $v_j = v\neq u$ for $j=1,2,3$, in which all triple-$K$ integrals are regulated, and then redefine primary constants as necessary in order to obtain a finite limit as $v \rightarrow u$.

\paragraph{Counterterms.}
As there are no counterterms capable of removing $(--+)$ singularities\footnote{See the discussion on page \pageref{cancellation}.}, for the form factor $A_5$ to be finite, all the $(--+)$ singularities must necessarily cancel against one another.  
The regulated form factors are therefore finite in odd dimensions, while in even dimensions they contain at most $\ep^{-1}$ poles, as only the $(---)$ singularities survive.

To remove these remaining $(---)$ singularities, we can add counterterms constructed from the metric in a covariant fashion. Since the counterterms should not contribute to the 1-point function, they must contain either two or three Riemann tensors, {\it e.g.,}
\begin{equation}
\int \D^4 \bs{x} \sqrt{g} R_{\mu\nu\rho\sigma} R^{\mu\nu\rho\sigma}, \qquad\qquad \int \D^6 \bs{x} \sqrt{g} R_{\mu\nu}^{\ \ \rho\sigma} R_{\rho\sigma}^{\ \ \kappa \tau} R_{\kappa\tau}^{\ \ \mu\nu}.
\end{equation}
In odd spacetime dimensions, no counterterms are present and hence the 3-point function must be finite and non-anomalous.
In even spacetime dimensions, we can construct counterterms with either two or three Riemann tensors, accompanied by an even number of covariant derivatives. The first option requires $d = 4 + 2n$ for non-negative integer $n$, while the second requires $d = 6 + 2n$.  
The introduction of counterterms such as these gives rise to conformal anomalies, as we will discuss.

\subsubsection{Odd spacetime dimensions}
\label{sec:tttodd}

The form factor $A_5$ has singularities both as $\ep\rightarrow 0$ and as $v\rightarrow u$ deriving from the integrals $J_{2\{110\}}$ and $J_{0\{000\}}$.
The former must cancel out, while the latter can be removed by re-defining the corresponding primary constants.  
To achieve this requires  
\begin{align} \label{e:ttt_genuvsol}
\3{5}^{(0)} & = - 2 d^2 \3{2}^{(0)}, \nn\\
\3{5}^{(1)} & = - 2 d (4 v \3{2}^{(0)} + d \3{2}^{(1)} ) + (u - v) \redef{5}^{(1)},
\end{align}
where $\redef{5}^{(1)}$ represents the undetermined, re-defined primary constant.  
With these substitutions, the limits $v\rightarrow u$ and $\epsilon \rightarrow 0$ are now finite.

Next, we solve the secondary Ward identities, which are valid in the limit $u = v$.   Making use of the zero-momentum limit \eqref{zml1}\,-\,\eqref{zml2} to simplify expressions, we obtain the solution
\begin{align}
\3{3}^{(0)} & = - d \left[ 2(d + 2) \3{1}^{(0)} + \3{2}^{(0)} \right] - \left( \frac{\pi}{2} \right)^{-\frac{3}{2}} \frac{2(-1)^N \2_{TT}}{(d-2)!!}, \\
\3{4}^{(0)} & = 2 \3{3}^{(0)} + (3d+2) \3{2}^{(0)}, \\
\redef{5}^{(1)} & = - d \left[ 2d(d+2) \3{1}^{(0)} + \frac{1}{2}(d+6) \3{2}^{(0)} \right] - \left( \frac{\pi}{2} \right)^{-\frac{3}{2}} \frac{2 d (-1)^N \2_{TT} }{(d-2)!!}.
\end{align}

\paragraph{Three dimensions.} In $d=3$, the form factors read
\begin{align}
A_1 &= \frac{8 \ren{1}}{a_{123}^6} \left[ a_{123}^3 + 3 a_{123} b_{123} + 15 c_{123} \right], \label{e:TTTA1}\\[2ex]
A_2 & = \frac{8 \ren{1}}{a_{123}^5} \left[ 4 p_3^4 + 20 p_3^3 a_{12} + 4 p_3^2 (7 a_{12}^2 + 6 b_{12}) + 15 p_3 a_{12} (a_{12}^2 + b_{12}) + 3 a_{12}^2 (a_{12}^2 + b_{12}) \right] \nn\\
& \qquad + \: \frac{2 \ren{2}}{a_{123}^4} \left[ a_{123}^3 + a_{123} b_{123} + 3 c_{123} \right], \label{e:TTTA2}\\[2ex]
A_3 & = \frac{2 \ren{1} p_3^2}{a_{123}^4} \left[ 7 p_3^3 + 28 p_3^2 a_{12} + 3 p_3 (11 a_{12}^2 + 6 b_{12}) + 12 a_{12} ( a_{12}^2 + b_{12} ) \right] \nn\\
& \qquad + \: \frac{\ren{2} p_3^2}{a_{123}^3} \left[ p_3^2 + 3 p_3 a_{12} + 2 (a_{12}^2 + b_{12}) \right] - \frac{2 \2_{TT}}{a_{123}^2} \left[ a_{123}^3 - a_{123} b_{123} - c_{123} \right], \label{e:TTTA3}\\[2ex]
A_4 & = \frac{4 \ren{1}}{a_{123}^4} \left[ -3 p_3^5 - 12 p_3^4 a_{12} - 9 p_3^3 (a_{12}^2 + 2 b_{12}) + 9 p_3^2 a_{12} (a_{12}^2 - 3 b_{12}) \right.\nn\\
&\qquad\qquad \left. + \: (4 p_3 + a_{12}) (3 a_{12}^4 - 3 a_{12}^2 b_{12} + 4 b_{12}^2) \right] \nn\\
& \qquad + \: \frac{\ren{2}}{a_{123}^3} \left[ -p_3^4 - 3 p_3^3 a_{12} - 6 p_3^2 b_{12} + a_{12} (a_{12}^2 - b_{12}) (3 p_3 + a_{12}) \right] \nn\\
& \qquad - \: \frac{4 \2_{TT}}{a_{123}^2} \left[ a_{123}^3 - a_{123} b_{123} - c_{123} \right], \label{e:TTTA4}\\[2ex]
A_5 & = \frac{2 \ren{1}}{a_{123}^3} \left[ -3 a_{123}^6 + 9 a_{123}^4 b_{123} + 12 a_{123}^2 b_{123}^2 - 33 a_{123}^3 c_{123} + 12 a_{123} b_{123} c_{123} + 8 c_{123}^2 \right] \nn\\
& \qquad + \: \frac{\ren{2}}{2 a_{123}^2} \left[ -a_{123}^5 + 3 a_{123}^3 b_{123} + 4 a_{123} b_{123}^2 - 11 a_{123}^2 c_{123} + 4 b_{123} c_{123} \right] \nn\\
& \qquad + \: 2  \2_{TT} (p_1^3 + p_2^3 + p_3^3), \label{e:TTTA5}
\end{align}
making use of the symmetric polynomials defined in \eqref{e:variables}.
A simple check confirms that all primary and secondary CWIs are satisfied.
The solution involves two undetermined constants, 
$\ren{1}^{(0)}$ and $\ren{2}^{(0)}$, which we have relabelled as $\ren{1}$ and $\ren{2}$.
Note however that $C_2$ can be set to zero through the action of three-dimensional tensorial degeneracies, as discussed in \cite{Bzowski:2013sza} and appendix \ref{app:3ddeg}.

\subsubsection{Even spacetime dimensions}
\label{sec:ttteven}

The procedure for even spacetime dimensions $d=2N\ge 4$ is similar to that for odd dimensions discussed above.  The appearance of $(---)$ singularities leads to the additional complication of $\ep^{-2}$ divergent triple-$K$ integrals, meaning that primary constants must be expanded to order $\ep^2$.  In addition, to remove the divergences in regulated form factors we will need to renormalise by introducing counterterms.

Our first task is to remove the singularity as $v\rightarrow u$ in the form factor $A_5$. In addition, as there are no counterterms capable of removing $(--+)$ singularities, all singularities of this type must cancel among themselves so that only $(---)$ singularities remain.  The form factor should then have only an $\ep^{-1}$ divergence in the regulator, meaning the $\ep^{-2}$ divergences must cancel.
These conditions leads to substitutions \eqref{e:ttt_genuvsol} accompanied by a higher-order substitution. In total,
\begin{align}
\3{5}^{(0)} & = - 2 d^2 \3{2}^{(0)}, \\
\3{5}^{(1)} & = - 2 d (4 v \3{2}^{(0)} + d \3{2}^{(1)} ) + (u - v) \redef{5}^{(1)}, \\
\3{5}^{(2)} & = - 8 v^2 \3{2}^{(0) }- 8 v d \3{2}^{(1)} - 2 d^2 \3{2}^{(2)} + (u - v) \redef{5}^{(2)}.
\end{align}
With these substitutions, the  limit $v \rightarrow u$ is finite and the regulated secondary CWI can now be solved. By taking the zero-momentum limit, we arrive at the solution
\begin{align}
\3{3}^{(0)} & = - d \left[ 2(d + 2) \3{1}^{(0)} + \3{2}^{(0)} \right] - \frac{(-1)^N 2^{3-N} \2_{TT}}{(N-1)!}, \\[1ex]
\3{4}^{(0)} & = 2 \3{3}^{(0)} + (3d + 2) \3{2}^{(0)}, \\[1ex]
\3{3}^{(1)} & = - 2 u \left[ 4 (d + 1) \3{1}^{(0)} + \3{2}^{(0)} \right] - d \left[ 2(d + 2) \3{1}^{(1)} + \3{2}^{(1)} \right] \nn\\
& \qquad + \frac{(-1)^N 2^{3-N} u}{(N-1)!} \left[ \2_{TT} ( H_{N-1} + \ln 2 - \gamma_E) - \sl_{TT} \right], \\[1ex]
\3{4}^{(1)} & = 2 \3{3}^{(1)} + 6 u \3{2}^{(0)} + (3 d + 2) \3{2}^{(1)},  \\[1ex]
\redef{5}^{(1)} & = - d \left[ 2d(d+2) \3{1}^{(0)} + \frac{1}{2}(d+6) \3{2}^{(0)} \right] - \frac{(-1)^N 2^{4-N} N \2_{TT}}{(N-1)!}, \\[1ex]
\redef{5}^{(2)} & = - 2 u \left[ 2 d(3d + 4) \3{1}^{(0)} + (d+3) \3{2}^{(0)} \right] - d \left[ 2d(d+2) \3{1}^{(1)} + \frac{1}{2}(d+6) \3{2}^{(1)} \right] \nn\\
& \qquad + \frac{(-1)^N 2^{4-N} N u}{(N-1)!} \left[  \2_{TT} \left( H_{N-1} - \frac{1}{N} + \ln 2 - \gamma_E \right) -  \sl_{TT} \right].
\end{align}

\paragraph{Four dimensions.} 

Re-labelling $C_1^{(0)}\rightarrow C_1$ and $C_2^{(0)}\rightarrow C_2$ for convenience, with the substitutions above, the remaining singularities are
\begin{align}
A_1 & = O(\ep^0), \label{A1TTTsing} \\
A_2 & = - \frac{4}{u \epsilon}(16 C_1 + C_2) + O(\ep^0), \label{A2TTTsing}
\\
A_3 & = - \frac{2}{u \epsilon} \left[  \2_{TT}(p_1^2 + p_2^2 + p_3^2) + (16C_1+C_2) p_3^2 \right] + O(\ep^0), \\
A_4 & = - \frac{2}{u \epsilon} \left[ 2 \2_{TT}(p_1^2 + p_2^2 + p_3^2) + (16C_1+C_2) (p_1^2 + p_2^2 - p_3^2) \right] + O(\ep^0), \\
A_5 & = \frac{1}{u \epsilon} \left[ 2C_{TT}(p_1^4 + p_2^4 + p_3^4) - (16C_1+C_2) J^2 \right] + O(\ep^0). \label{A5TTTsing}
\end{align}

To remove these singularities, we have at our disposal the counterterms
\begin{align} \label{SctTTT}
S_{\Ct} = \int \D^{4 + 2 u \epsilon} \bs{x} \sqrt{g} \mu^{2 u \epsilon} (\mathfrak{a} E_4 +\mathfrak{b}R^2+  \mathfrak{c} W^2),
\end{align}
where the four-dimensional Euler density $E_4$ and squared Weyl tensor $W^2$ are given by 
\begin{align}
E_4 & \equiv R_{\mu\nu\alpha\beta}R^{\mu\nu\alpha\beta}-4R_{\mu\nu}R^{\mu\nu}+R^2
= 6 R^{\mu_1\mu_2}_{\qquad [\mu_1\mu_2}R^{\mu_3\mu_4}_{\qquad \mu_3\mu_4]}
,\label{Eulerdef}\\
\label{Weyldef}
W^2 &\equiv R_{\mu\nu\alpha\beta}R^{\mu\nu\alpha\beta}-2 R_{\mu\nu}R^{\mu\nu}+\frac{1}{3}R^2.
\end{align}
Together with $R^2$, these form a complete basis of quadratic curvature invariants.
  
The contribution from these counterterms to the Weyl variation of the renormalised generating functional is
\begin{align}
&\delta_{\sigma} W_{\Ct}[g_{\mu\nu}]  = \lim_{\ep\rightarrow 0}\delta_{\sigma} \Big[\ln \< e^{-S_{\Ct}} \> \Big] = 
-\lim_{\ep\rightarrow 0}\delta_{\sigma} S_{\Ct} \nn\\[1ex] &= 
\lim_{\ep\rightarrow 0}\int \D^{4+2u\ep} \bs{x} \sqrt{g}\mu^{2u\ep} \sigma \Big[-2u\ep\,\Big(\mathfrak{a} E_4 +\mathfrak{b} R^2+ \mathfrak{c} W^2\Big) + \Big(4(3+2u\ep)\mathfrak{b}-\frac{4}{3}u\ep \mathfrak{c}\Big) \Box R\Big]. 
\end{align}
In the first line, the variation of the counterterm action can be taken outside the expectation value as it depends only on the sources, and not the dynamical fields.  
The term proportional to $\mathfrak{c}\,\Box R$ in the final line arises as we have chosen in \eqref{Weyldef} to use the four-dimensional square of the Weyl tensor as a counterterm, rather than its $d$-dimensional counterpart.  This choice of scheme has the benefit of simplifying the counterterm contributions.
If we now choose the counterterm coefficients 
\[\label{TTTctexps}
\mathfrak{a} = - \frac{a}{2 u \epsilon} + O(\epsilon^0), \qquad \mathfrak{b} =\frac{1}{12}\Big(b-\frac{2}{3}c\Big)+O(\ep^1),
\qquad\mathfrak{c} = - \frac{c}{2 u \epsilon} + O(\epsilon^0),
\]
we obtain the trace anomaly
\[\label{acdef}
\<T\>_s = a E_4+b\Box R + c W^2.
\]

Since the trace anomaly must be finite, note the $R^2$ counterterm can only contribute at finite order.  In fact, as the $b\Box R$ trace anomaly would generate a nonzero trace for the stress tensor 2-point function, our choice of scheme in section \ref{sec:2ptfns} requires that $b=0$.
Moreover, since we have already used the Weyl-squared counterterm to renormalise the stress tensor 2-point function, the coefficient $c$ is already fixed. 
Comparing \eqref{SctTTT} with \eqref{e:SctTT} and \eqref{e:ctTT} (or alternatively \eqref{acdef} with \eqref{traceanomquad}), we find 
\begin{equation}\label{cresult}
c = - \frac{\2_{TT}}{2}.
\end{equation}

The contribution of the counterterms  \eqref{SctTTT} to the transverse traceless form factors can be evaluated as described in appendix \ref{sec:evalctcontr}.  The $R^2$ counterterm makes no contribution, while the others contribute
\begin{align}\label{ct_TTT_1}
A^{\Ct}_1&= 0,\\
A^{\Ct}_2 &= -16 ( \mathfrak{c}+\mathfrak{a})\mu^{2 u \epsilon},\\
A^{\Ct}_3 &= 8  \mathfrak{c}\mu^{2 u \epsilon}(p_1^2+p_2^2)-8 \mathfrak{a} \mu^{2 u \epsilon} p_3^2,\\
A^{\Ct}_4 &= 8 \mathfrak{c} \mu^{2 u \epsilon}(p_1^2 + p_2^2 + 3 p_3^2) - 8 \mathfrak{a} \mu^{2 u \epsilon}(p_1^2 + p_2^2 - p_3^2),\\
A^{\Ct}_5 &= -4 \mathfrak{c} \mu^{2 u \epsilon} (p_1^2 + p_2^2 + p_3^2)^2 - 4 \mathfrak{a}  \mu^{2 u \epsilon} J^2.\label{ct_TTT_5}
\end{align}
Using \eqref{cresult} and \eqref{TTTctexps}, it is easy to check that the Weyl-squared counterterms cancel the divergences proportional to $C_{TT}$ in \eqref{A1TTTsing}\,-\,\eqref{A5TTTsing}.
The remaining divergences can be cancelled by setting
\begin{equation}
\mathfrak{a} = - \frac{1}{4 u \epsilon} (\2_{TT} + 16 \3{1} + \3{2} ) +O(\ep^0).
\end{equation}
Despite appearances, however, these are not genuine UV divergences as we will discuss in section \ref{sec:Eulerdeg}.
From \eqref{TTTctexps}, we identify the Euler anomaly coefficient 
\begin{equation}\label{aC12}
a = -c +8C_1+\frac{1}{2}C_2.
\end{equation}
Using this relation to eliminate $C_2$, we can now express the renormalised form factors as
{\allowdisplaybreaks
\begin{align}\label{A1TTTren}
A_1 & = \3{1} I_{7\{222\}}, \\[1ex]
A_2 & = 2 \Big[ a + c - 2 \3{1} p_3 \frac{\partial}{\partial p_3} \Big] I^{\text{(fin)}}_{5\{222\}} \nn\\[0.5ex]
& \quad - \frac{8}{3} (a+c) \Big[ \ln \frac{p_1^2}{\mu^2} + \ln \frac{p_2^2}{\mu^2} + \ln \frac{p_3^2}{\mu^2} \Big] \nn\\[0.5ex]
&\quad
- 16 (a + c) - \frac{64}{3} \ren{1} + 4 \sdt_{TT} -8D_1, \\[2ex]
A_3 & = 2 \Big[ 2 c - (a + c + \ren{1}) p_3 \frac{\partial}{\partial p_3} + \ren{1} p_3^2 \frac{\partial^2}{\partial p_3^2} \Big] I_{3\{222\}}^{\text{(fin)}} \nn\\[0.5ex]
& \quad + 2c \Big[ p_2^2 \ln \frac{p_1^2}{\mu^2} +  p_1^2 \ln \frac{p_2^2}{\mu^2} + ( p_1^2 + p_2^2 ) \ln \frac{p_3^2}{\mu^2}\Big] 
-2ap_3^2 \Big[\ln\frac{p_1^2}{\mu^2}+\ln \frac{p_2^2}{\mu^2}\Big]
\nn\\[1ex]
& \quad 
- 2 (a + 3c + 2 \ren{1} + \sdt_{TT}) (p_1^2 + p_2^2) - 4 (c + 4 \ren{1}+D_1) p_3^2, \\[2ex]
A_4 & = 4 \Big[ c-a + (a+c) p_3 \frac{\partial}{\partial p_3}+2 \ren{1} \Big( 8 - 4 \sum_{j=1}^3 p_j \frac{\partial}{\partial p_j} + p_1 p_2 \frac{\partial^2}{\partial p_1 \partial p_2} \Big) \Big] I_{3\{222\}}^{\text{(fin)}} \nn\\[0.5ex]
& \quad + 2 c \Big[(p_2^2+3p_3^2) \ln \frac{p_1^2}{\mu^2} + (p_1^2+3p_3^2) \ln \frac{p_2^2}{\mu^2}  +  (p_1^2 + p_2^2) \ln \frac{p_3^2}{\mu^2}\Big] \nn\\[1ex]
& \quad 
 +2a \Big[(p_3^2-p_2^2) \ln \frac{p_1^2}{\mu^2} 
 + (p_3^2-p_1^2) \ln \frac{p_2^2}{\mu^2}
 -(p_1^2+p_2^2)\ln \frac{p_3^2}{\mu^2} 
 \Big]
 \nn\\[1ex]&\quad
-2 (4 c + \sdt_{TT}+2D_1) (p_1^2 + p_2^2) - 2 (2 a + 6 c - 8 \ren{1} + 3 \sdt_{TT}-2D_1) p_3^2, \\[2ex]
A_5 & = 2(a+c) \Big[ 32 - 8 \sum_{j=1}^3 p_j \frac{\partial}{\partial p_j} + 2 \sum_{i<j} p_i p_j \frac{\partial^2}{\partial p_i \partial p_j} \Big] I_{1\{222\}}^{\text{(fin)}} - 8 \ren{1} p_1^3 p_2^3 p_3^3 \frac{\partial^3}{\partial p_1 \partial p_2 \partial p_3} I_{1\{000\}} \nn\\[0.5ex]
& \quad - 2 c \Big[ \Big(p_1^4 + p_2^4-p_3^4 + p_3^2 (p_1^2 + p_2^2) \Big) \ln \frac{p_3^2}{\mu^2} + (p_1 \leftrightarrow p_3) + (p_2 \leftrightarrow p_3) \Big] \nn\\[0.5ex]
& \quad
- 2 a \Big[ 
\Big(p_1^4 + p_2^4-3p_3^4 + p_3^2 (p_1^2 + p_2^2) \Big) \ln \frac{p_3^2}{\mu^2} + (p_1 \leftrightarrow p_3) + (p_2 \leftrightarrow p_3)
 \Big]
\nn\\[1ex]
& \quad -  (a + c + 8 \ren{1} - \sdt_{TT}+2D_1)J^2 + 2 ( 2a + 2c + \sdt_{TT})(p_1^4 + p_2^4 + p_3^4).
\label{A5TTTren}
\end{align}}
Here
\begin{align}
I_{7\{222\}} & =  - \left( 2 - p_1 \frac{\partial}{\partial p_1} \right) \left( 2 - p_2 \frac{\partial}{\partial p_2} \right) \left( 2 - p_3 \frac{\partial}{\partial p_3} \right) p_1 p_2 p_3 \frac{\partial^3}{\partial p_1 \partial p_2 \partial p_3} I_{1\{000\}}, \\
I_{5\{222\}}^{\text{(fin)}} & = \left( 2 - p_1 \frac{\partial}{\partial p_1} \right) \left( 2 - p_2 \frac{\partial}{\partial p_2} \right) \left( 2 - p_3 \frac{\partial}{\partial p_3} \right) I_{2\{111\}}^{\text{(fin)}}, \\
I_{3\{222\}}^{\text{(fin)}} & = \left( 2 - p_1 \frac{\partial}{\partial p_1} \right) \left( 2 - p_2 \frac{\partial}{\partial p_2} \right) \left( 2 - p_3 \frac{\partial}{\partial p_3} \right) \left( \frac{1}{4} J^2 I_{1\{000\}} \right), \\
I_{1\{222\}}^{\text{(fin)}} & = \left[ p_1^2 p_2^2 p_3^2 - \frac{1}{4} J^2 (p_1^2 + p_2^2 + p_3^2 ) \right] I_{1\{000\}},
\end{align}
where $I_{2\{111\}}^{\text{(fin)}}$ and $I_{1\{000\}}$ are given in \eqref{I2111fin} and \eqref{I1000}.
The scheme-dependent constant $\sd{1}$ is related to the finite part of the counterterm coefficients in the regulated theory. In fact, as we will discuss in section \ref{sec:Eulerdeg}, $D_1$ can be removed through a degeneracy of the form factor basis.
Note that a change of renormalisation scale $\mu^2 \rightarrow e^{-\lambda}\mu^2$ is equivalent to sending $D_{TT} \rightarrow D_{TT} - 2\lambda c$ and $D_1 \rightarrow D_1 + \lambda a$;
the same applies to the 2-point function \eqref{e:2ptTTren}, making use of \eqref{cresult}.

Each renormalised form factor contains the following pieces.
First, there is a nonlocal piece involving contributions derived from triple-$K$ integrals.  (For $A_1$ this is all we have.)  Next, there is a scale-violating logarithmic piece consisting of  terms that depend explicitly on the renormalisation scale $\mu$.  Finally, there is an ultralocal piece containing terms that are polynomial in the squares of momenta.  While the coefficients appearing in this ultralocal piece are scheme-dependent, since each involves $D_{TT}$ and/or $D_1$, 
one can nevertheless 
extract scheme-independent information by combining the coefficients of multiple terms.
This is another (rare) instance in which an ultralocal contact term is unambiguous and  it would be interesting to understand the physics associated with such terms.

\paragraph{Anomalous Ward identities.}
The renormalised form factors satisfy the anomalous dilatation Ward identities
\begin{align}\label{aDWI_TTT1}
\mu \frac{\partial}{\partial \mu} A_1 & = 0, \\
\mu \frac{\partial}{\partial \mu} A_2 & = 16 (c+a), \\
\mu \frac{\partial}{\partial \mu} A_3 & = -8 c(p_1^2 + p_2^2) +8a p_3^2, \\
\mu \frac{\partial}{\partial \mu} A_4 & = -8 c (p_1^2+p_2^2+3 p_3^2) +8a(p_1^2+p_2^2-p_3^2), \\
\mu \frac{\partial}{\partial \mu} A_5 & = 4c (p_1^2+p_2^2+p_3^2)^2+4aJ^2.\label{aDWI_TTT5}
\end{align}
As we will discuss in section \ref{sec:Eulerdeg}, the presence of the Euler coefficient $a$ in these identities is rather surprising and signals the presence of a hidden degeneracy in our form factor basis.

The renormalised form factors satisfy the anomalous primary Ward identities 
\begin{align}
& \K_{12} A_1 = 0, && \K_{13} A_1 = 0, \\[0.5ex]
& \K_{12} A_2 = 0, && \K_{13} A_2 = 8 A_1, \\[0.5ex]
& \K_{12} A_3 = 0, && \K_{13} A_3 = 2 A_2 + 16 (a+c), \\[0.5ex]
& \K_{12} A_4 = 4 \left[ A_2(p_1 \leftrightarrow p_3) - A_2(p_2 \leftrightarrow p_3) \right], 
 && \K_{13} A_4 = - 4 A_2(p_2 \leftrightarrow p_3)
 - 32(a+c), \nn\stepcounter{equation}\\[0.5ex]
&\K_{12} A_5  = 2 \left[ A_4(p_2 \leftrightarrow p_3) - A_4(p_1 \leftrightarrow p_3) \right]   
&& \K_{13} A_5  = 2 \left[ A_4 - A_4(p_1 \leftrightarrow p_3) \right]\nn \\
&\qquad \qquad + 32 a (p_1^2 - p_2^2) && \qquad \qquad + 32 a (p_1^2 - p_3^2) 
\end{align}
The anomalous secondary Ward identities are
\begin{align}
& \Lo_{6} A_1 + \Ro \left[ A_2 - A_2(p_2 \leftrightarrow p_3) \right] = 0, \\[2ex]
& \Lo_{6} A_2 + 2 \Ro \left[ 2 A_3 - A_4(p_1 \leftrightarrow p_3) \right] = - 32(a + c) p_1^2, \\[2ex]
& \Lo_{4} \left[ A_2(p_1 \leftrightarrow p_3) \right] + \Ro  \left[ A_4(p_2 \leftrightarrow p_3) - A_4 \right]  
+ 2 p_1^2 \left[ A_2(p_2 \leftrightarrow p_3) - A_2 \right] 
\nn\\[1ex]&\qquad
= -16 (a + c) p_1^2, \\[2ex]
& \Lo_{4} \left[ A_4(p_2 \leftrightarrow p_3) \right] - 2 \Ro A_5 + 2 p_1^2 \left[ A_4(p_1 \leftrightarrow p_3) - 4 A_3 \right] \nn\\[1ex]
& \qquad = 64 \cdot \text{coefficient of } \delta_{\mu_1 \mu_2} \delta_{\mu_3 \nu_2} p_{1\nu_3} \text{ in } p_{1}^{ \nu_1} \lla T_{\mu_1 \nu_1}(\bs{p}_1) T_{\mu_2 \nu_2}(\bs{p}_2) T_{\mu_3 \nu_3}(\bs{p}_3) \rra \nn\\[1ex]
& \qquad \qquad + 16 p_1^2 \left[ a (-p_1^2 + p_2^2 + p_3^2) + c (p_1^2 + 3 p_2^2 + p_3^2) \right] \nn\\[1ex]
& \qquad = 32 p_2^4 \Big( \2_{TT} \ln \frac{p_2^2}{\mu^2} + \sdt_{TT} \Big) 
+ 16 p_1^2 \left[ a (-p_1^2 + p_2^2 + p_3^2) + c (p_1^2 + 3 p_2^2 + p_3^2) \right], 
\end{align}
\begin{align}
& \Lo_{2} \left[ A_3(p_1 \leftrightarrow p_3) \right] + p_1^2 \left[ A_4 - A_4(p_2 \leftrightarrow p_3) \right]  \nn\\[1ex]
& \qquad = 16 \cdot \text{coefficient of } \delta_{\mu_2 \mu_3} \delta_{\nu_2 \nu_3} p_{2\mu_1} \text{ in } p_{1}^{ \nu_1} \lla T_{\mu_1 \nu_1}(\bs{p}_1) T_{\mu_2 \nu_2}(\bs{p}_2) T_{\mu_3 \nu_3}(\bs{p}_3) \rra 
\nn\\[1ex]& \qquad \qquad 
+ 8 p_1^2 \left[ - a (p_1^2 + p_2^2) + c p_2^2 + (a + c) p_3^2 \right] \nn\\[1ex]
& \qquad = 8 p_2^4 \Big( \2_{TT} \ln \frac{p_2^2}{\mu^2} + \sdt_{TT} \Big) - 8 p_3^4 \Big( \2_{TT} \ln \frac{p_3^2}{\mu^2} + \sdt_{TT} \Big)\nn\\[0.5ex]
& \qquad \qquad + 8 p_1^2 \left[ - a (p_1^2 + p_2^2 -p_3^2) + c (p_2^2 + p_3^2) \right].
\end{align}

Finally, the anomaly $\mathcal{A}_{\mu_2\nu_2\mu_3\nu_3}$ appearing in the trace Ward identity \eqref{e:TWI_TTT} can be obtained by expanding the trace anomaly \eqref{acdef} to quadratic order in the perturbed metric.  The result can be expressed in terms of the transverse decomposition 
\eqref{anomaly_decomp} through the form factors
\begin{align}\label{B1eqn}
B_1 & = 8(a+c), \\[0.5ex]
B_2 & = 8(p_1^2 - p_2^2 - p_3^2)(a+c) - 4 p_1^2 b, \\[0.5ex]
B_3 & = - 2 J^2 (a + c) + 4 p_2^2 p_3^2 c + b p_1^2 (p_2^2 + p_3^2 - 5 p_1^2), \\[0.5ex]
B_4 & = - \frac{8}{3} a p_3^2 + 4 b \Big( p_2^2 + \frac{4}{3} p_1^2 \Big), \\[0.5ex]
B_5 & = \frac{4}{9} a J^2 + \frac{2}{3} b \Big( 2 ( p_1^2 p_2^2 + 2 p_1^2 p_3^2 + 2 p_2^2 p_3^2 ) + 3 (p_2^4 + p_3^4 - p_1^4) \Big)\label{B5eqn}.
\end{align}
These form factors also enter the 
reconstruction formula \eqref{re_TTT} through the anomalous contribution \eqref{re_anomaly}.
We have included here the terms proportional $b$ purely for completeness: in this paper, $b$ vanishes as otherwise the $b\Box R$ trace anomaly leads to a nonzero trace for the stress tensor 2-point function.

\section{Dimension-dependent degeneracies and the Euler anomaly}
\label{sec:Eulerdeg}

The appearance of the Euler coefficient $a$ in the anomalous dilatation Ward identities \eqref{aDWI_TTT1}\,-\,\eqref{aDWI_TTT5} presents us with a puzzle. 
These equations encode the response of the form factors 
to a change in the renormalisation scale, or equivalently a dilatation.
On the other hand, under a dilatation, the variation of the renormalised generating functional  is given by the integral of the trace anomaly. 
As the integral of the Euler density is a topological invariant, however, all subsequent variations with respect to the metric vanish: the anomalous dilatation Ward identities for the stress tensor 3-point function thus {\it cannot} depend on $a$.  
All renormalisation scale-dependent terms in correlators must therefore be proportional to the  Weyl-squared coefficient $c$, and not the Euler coefficient $a$.  (In other words, to the coefficient of the type B and not the type A anomaly \cite{Deser:1993yx}.)
How then can the form factors contain scale-dependent terms proportional to $a$, but not the correlators?

This apparent contradiction is resolved by the existence of hidden dimension-dependent degeneracies in our form factor basis.  As we will explain in this section, these degeneracies correspond to the existence of nontrivial tensorial structures that vanish in specific spacetime dimensions.    
When the full tensorial 3-point function is reassembled from its constituent form factors, we find that all the scale-dependent terms proportional to $a$ in fact reassemble into precisely one of these vanishing tensorial structures.  Thus, while the individual form factors may contain scale-dependent terms proportional to $a$, the full tensorial 3-point function does not.  

Geometrically, the origin of these dimension-dependent degeneracies is simply 
the fact that any $n$-form vanishes in spacetime dimensions $d<n$.  
In our present context, these $n$-forms are constructed from the momenta and metric tensor,
although a close analogy exists with the Lovelock identities \cite{Lovelock} constructed from spacetime curvature forms.
Stripping out all factors of momenta, the underlying identities can also be viewed as Schouten identities \cite{vanProeyen}.
When suitably contracted and projected into a transverse-traceless basis, these vanishing higher forms yield a set of degenerate form factors that produce a vanishing contribution to the full correlator.  
As these degeneracies involve arbitrary functions, they can even be used to set specific form factors to zero.
In three dimensions, as we discuss in appendix \ref{app:3ddeg}, their effect is to reduce the number of independent form factors for the stress tensor 3-point function from five to two.   In four dimensions, as we will show here, we can 
reduce the number of independent form factors from five to four. 
Our main interest however will be to understand 
the connection between degeneracies and 
the Euler anomaly, supplying
the four-dimensional counterpart to the two-dimensional analysis of the anomaly in the introduction and appendix \ref{app:typeA}.

\subsection{Dimension-dependent degeneracies}

Let us begin with an arbitrary (2,2)-form $K_{\alpha_1 \alpha_2}{}^{\beta_1 \beta_2} = K_{[\alpha_1 \alpha_2]}{}^{[\beta_1 \beta_2]}$.
In any (integer) spacetime dimension $d<5$, the totally antisymmetrised product
\[
\delta_{[\alpha_1}^{\beta_1}\delta_{\alpha_2}^{\beta_2}\delta_{\alpha_3}^{\beta_3}K_{\alpha_4\alpha_5]}{}^{\beta_4 \beta_5}
\]
vanishes as an index must necessarily be repeated.
Upon contracting indices, we find
\[\label{Edgar}
\delta_{[\alpha_1}^{\beta_1}\delta_{\alpha_2}^{\beta_2}\delta_{\alpha_3}^{\beta_3}K_{\alpha_4\alpha_5]}{}^{\alpha_4 \alpha_5}=\frac{1}{10}\Big(
3K_{[\alpha_1\alpha_2}{}^{[\beta_1\beta_2}\delta_{\alpha_3]}^{\beta_3]}-6K_{[\alpha_1}{}^{[\beta_1}\delta_{\alpha_2}^{\beta_2}\delta_{\alpha_3]}^{\beta_3]}+K\delta_{[\alpha_1}^{[\beta_1}\delta_{\alpha_2}^{\beta_2}\delta_{\alpha_3]}^{\beta_3]}\Big),
\]
where, in analogy with the Riemann tensor, we have defined 
\[
K_{\alpha}{}^{\beta} = K_{\alpha\sigma}{}^{\beta\sigma}, \qquad K = K_{\sigma}{}^{\sigma}=K_{\rho\sigma}{}^{\rho\sigma}.
\]
(In fact, choosing $K_{\alpha_1\alpha_2}{}^{\beta_1\beta_2}$ to be the Riemann tensor yields a class of generalised Lovelock identities, as discussed in \cite{Edgar:2001vv}.)
For our present purposes, we will take
\[\label{ourK}
K_{\alpha_1 \alpha_2}{}^{\beta_1 \beta_2} =  p_{1[\alpha_1}p_{2 \alpha_2]}p_1^{[\beta_1}p_2^{\beta_2]}, 
\]
whereupon
\[
K_\alpha^\beta = \frac{1}{8}(2p_1^2 p_{2\alpha}p_2^{\beta}+(p_1^2+p_2^2-p_3^2)(p_{1\alpha}p_2^\beta+p_{2\alpha}p_1^\beta)+2p_2^2 p_{1\alpha}p_1^{\beta}),\qquad
K = \frac{1}{8}J^2.
\]
Despite appearances, note that $K_{\alpha_1 \alpha_2}{}^{\beta_1 \beta_2}$ and its contractions are in fact symmetric under all permutations of the momenta. 

We can now decompose \eqref{Edgar} into its component transverse-traceless and trace pieces by contracting with three copies of \eqref{Idecomp}.  As any contraction of \eqref{Edgar} with a momentum vanishes, there are no longitudinal pieces.  We can therefore write
\begin{align}\label{fullEdgar}
&\delta_{(\mu_1}^{\alpha_1}\delta_{\nu_1)\beta_1}
\delta_{(\mu_2}^{\alpha_2}\delta_{\nu_2)\beta_2}
\delta_{(\mu_3}^{\alpha_3}\delta_{\nu_3)\beta_3}\Big(5!\,
\delta_{[\alpha_1}^{\beta_1}\delta_{\alpha_2}^{\beta_2}\delta_{\alpha_3}^{\beta_3}K_{\alpha_4\alpha_5]}{}^{\alpha_4 \alpha_5}\Big) \nn\\ &= 
\Big(\Pi_{\mu_1\nu_1}{}^{\alpha_1}{}_{\beta_1}(\bs{p}_1)+\frac{1}{d-1}\pi_{\mu_1\nu_1}(\bs{p}_1)\delta_{\beta_1}^{\alpha_1}\Big)
\Big(\Pi_{\mu_2\nu_2}{}^{\alpha_2}{}_{\beta_2}(\bs{p}_2)+\frac{1}{d-1}\pi_{\mu_2\nu_2}(\bs{p}_2)\delta_{\beta_2}^{\alpha_2}\Big)\nn\\&\quad \times
\Big(\Pi_{\mu_3\nu_3}{}^{\alpha_3}{}_{\beta_3}(\bs{p}_3)+\frac{1}{d-1}\pi_{\mu_3\nu_3}(\bs{p}_3)\delta_{\beta_3}^{\alpha_3}\Big)5!\,\delta_{[\alpha_1}^{\beta_1}\delta_{\alpha_2}^{\beta_2}\delta_{\alpha_3}^{\beta_3}K_{\alpha_4\alpha_5]}{}^{\alpha_4 \alpha_5}.
\end{align}
The fully transverse-traceless piece, 
\begin{align}
\Pi_{\mu_1\nu_1}{}^{\alpha_1}{}_{\beta_1}(\bs{p}_1)\Pi_{\mu_2\nu_2}{}^{\alpha_2}{}_{\beta_2}(\bs{p}_2)\Pi_{\mu_3\nu_3}{}^{\alpha_3}{}_{\beta_3}(\bs{p}_3)
\,5!\,\delta_{[\alpha_1}^{\beta_1}\delta_{\alpha_2}^{\beta_2}\delta_{\alpha_3}^{\beta_3}K_{\alpha_4\alpha_5]}{}^{\alpha_4 \alpha_5},
\end{align}
can be expressed in our standard form
factor basis \eqref{e:TTTdecomp} as
\begin{align}\label{Eulerdeg}
A_1 = 0, \quad A_2 = 2, \quad A_3 = p_3^2, \quad A_4 = p_1^2+p_2^2-p_3^2, \quad A_5 = \frac{J^2}{2}.
\end{align}
The trace terms can instead be decomposed in the same form as the anomalous contribution to the 3-point function, as given in \eqref{re_anomaly} and \eqref{anomaly_decomp}.
First, from tracing over a single pair of indices, we find
\begin{align}\label{singletr}
&\frac{1}{d-1}\pi_{\mu_1\nu_1}(\bs{p}_1)\delta_{\beta_1}^{\alpha_1}\,\Pi_{\mu_2\nu_2}{}^{\alpha_2}{}_{\beta_2}(\bs{p}_2)\Pi_{\mu_3\nu_3}{}^{\alpha_3}{}_{\beta_3}(\bs{p}_3)\,5!\,\delta_{[\alpha_1}^{\beta_1}\delta_{\alpha_2}^{\beta_2}\delta_{\alpha_3}^{\beta_3}K_{\alpha_4\alpha_5]}{}^{\alpha_4 \alpha_5} \nn\\[1ex]&
=\frac{(d-4)}{(d-1)}\,\pi_{\mu_1\nu_1}(\bs{p}_1)\Pi_{\mu_2\nu_2\alpha_2\beta_2}(\bs{p}_2)\Pi_{\mu_3\nu_3\alpha_3\beta_3}(\bs{p}_3)\,\Big[
  p_3^{\alpha_2} p_3^{\beta_2} p_1^{\alpha_3} p_1^{\beta_3}  \nn\\[1ex]&\qquad\qquad + (p_1^2-p_2^2-p_3^2) \delta^{\beta_2 \beta_3} p_3^{\alpha_2} p_1^{\alpha_3} -\frac{1}{4}J^2 \delta^{\alpha_2 \alpha_3} \delta^{\beta_2 \beta_3} \Big],
\end{align}
along with two similar terms which can be obtained by permutation of index labels and momenta.  Next, tracing over two pairs of indices, we obtain 
\begin{align}\label{dbltr}
&\frac{1}{(d-1)^2}\pi_{\mu_1\nu_1}(\bs{p}_1)\delta_{\beta_1}^{\alpha_1}\,\pi_{\mu_2\nu_2}(\bs{p}_2)\delta_{\beta_2}^{\alpha_2}\,\Pi_{\mu_3\nu_3}{}^{\alpha_3}{}_{\beta_3}(\bs{p}_3)\,5!\,\delta_{[\alpha_1}^{\beta_1}\delta_{\alpha_2}^{\beta_2}\delta_{\alpha_3}^{\beta_3}K_{\alpha_4\alpha_5]}{}^{\alpha_4 \alpha_5} \nn\\[1ex]&
=-\frac{(d-4)(d-3)}{(d-1)^2}\,\pi_{\mu_1\nu_1}(\bs{p}_1)\pi_{\mu_2\nu_2}(\bs{p}_2)\Pi_{\mu_3\nu_3\alpha_3\beta_3}(\bs{p}_3) p_1^{\alpha_3}p_1^{\beta_3}\, p_3^2,
\end{align}
along with two similar terms obtainable by permutation.  Finally, from tracing over all three pairs of indices, we have 
\begin{align}\label{triptr}
&\frac{1}{(d-1)^3}\pi_{\mu_1\nu_1}(\bs{p}_1)\delta_{\beta_1}^{\alpha_1}\,\pi_{\mu_2\nu_2}(\bs{p}_2)\delta_{\beta_2}^{\alpha_2}\,\pi_{\mu_3\nu_3}(\bs{p}_3)\delta_{\beta_3}^{\alpha_3}
\,5!\,\delta_{[\alpha_1}^{\beta_1}\delta_{\alpha_2}^{\beta_2}\delta_{\alpha_3}^{\beta_3}K_{\alpha_4\alpha_5]}{}^{\alpha_4 \alpha_5} \nn\\[1ex]&
=\frac{(d-4)(d-3)(d-2)}{(d-1)^3}\,\pi_{\mu_1\nu_1}(\bs{p}_1)\pi_{\mu_2\nu_2}(\bs{p}_2)\pi_{\mu_3\nu_3}(\bs{p}_3)\frac{1}{4}J^2.
\end{align}
Putting everything together, the contributions from \eqref{singletr}, \eqref{dbltr} and \eqref{triptr} (along with their respective permutations) can be written in precisely the form   
given in \eqref{re_anomaly} and \eqref{anomaly_decomp}, with
\begin{align}\label{Bformfactors}
&B_1 = (d-4), \qquad B_2 = (d-4)(p_1^2-p_2^2-p_3^2),\qquad
B_3 = -(d-4)\frac{J^2}{4}, \nn\\[1ex]& B_4 = -\frac{(d-4)(d-3)}{(d-1)}p_3^2,\qquad
B_5=\frac{(d-4)(d-3)(d-2)}{(d-1)^2}\frac{J^2}{4}.
\end{align}
Thus, in summary, we can write 
\begin{align}\label{summaryeq}
&\delta_{(\mu_1}^{\alpha_1}\delta_{\nu_1)\beta_1}
\delta_{(\mu_2}^{\alpha_2}\delta_{\nu_2)\beta_2}
\delta_{(\mu_3}^{\alpha_3}\delta_{\nu_3)\beta_3}\Big(5!\,
\delta_{[\alpha_1}^{\beta_1}\delta_{\alpha_2}^{\beta_2}\delta_{\alpha_3}^{\beta_3}
p_{1\alpha_4}p_{2\alpha_5]}p_1^{\alpha_4}p_2^{\alpha_5}
\Big) \nn\\[1ex] &= \lla t_{\mu_1\nu_1}(\bs{p}_1)t_{\mu_2\nu_2}(\bs{p}_2)t_{\mu_3\nu_3}(\bs{p}_3)\rra+\lla T_{\mu_1\nu_1}(\bs{p}_1)T_{\mu_2\nu_2}(\bs{p}_2)T_{\mu_3\nu_3}(\bs{p}_3)\rra_{\Anom},
\end{align}
where on the right-hand side the first term is transverse traceless and 
constructed from the form factors \eqref{Eulerdeg}, and the second term takes the form of an anomaly contribution with form factors  \eqref{Bformfactors}.\footnote{Note these `correlators' are simply a short-hand for the corresponding form factor decompositions, rather than representing the actual 3-point correlator.}

In exactly four dimensions, clearly all the expressions listed in \eqref{Bformfactors} vanish due to the explicit factors of $(d-4)$.  As the left-hand side of \eqref{summaryeq} also vanishes in $d=4$, the form factors in \eqref{Eulerdeg} are therefore {\it degenerate}: they yield a vanishing contribution to the 3-point function. 
For any given tensorial component, this can be verified explicitly by writing out all squared momenta in terms of their individual components.

\subsection{Euler anomaly}
\label{sec:euleranom}

Let us now discuss the implications of these results for the stress tensor 3-point function.

Firstly, any multiple of \eqref{Eulerdeg} that appears in the {\it renormalised} form factors can immediately be removed. Studying the form factors in \eqref{A1TTTren}\,-\,\eqref{A5TTTren}, we see that the scheme-dependent terms proportional to $D_1$ are of precisely this form. 
The same is also true for the terms proportional to $a$ on the right-hand side of the anomalous dilatation Ward identities \eqref{aDWI_TTT1}\,-\,\eqref{aDWI_TTT5}, or equivalently all terms proportional to $a \ln \mu^2$ in the form factors \eqref{A1TTTren}\,-\,\eqref{A5TTTren}.  As claimed, the scale-dependence of the renormalised correlator thus depends only on the Weyl coefficient $c$ and not on the Euler coefficient $a$.

More generally, we can remove any multiple of the degenerate form factors in \eqref{Eulerdeg} by an arbitrary function of the momenta that is symmetric under all permutations.  
 If desired, we can use this freedom to set one of the renormalised form factors to zero, reducing the number of independent form factors from five to four. 
After doing so, 
however, the anomalous Ward identities for the form factors will take a more complicated form, and are no longer given by the homogeneous conformal Ward identities plus inhomogeneous terms.

Let us now consider the role of the degeneracy from the standpoint of the regulated theory, where $d=4+2u\ep$. 
Here, the form factors contain divergent $\ep^{-1}$ poles.  If we multiply   \eqref{summaryeq} by an overall factor of $-4a/u\ep$, 
we find the transverse-traceless form factors
\[\label{ffdiv}
A_1 = 0, \quad A_2=-\frac{8 a}{u\ep}, \quad A_3=-\frac{4a }{u\ep}p_3^2, \quad A_4=-\frac{4a}{u\ep}(p_1^2+p_2^2-p_3^2),\quad A_5 =-\frac{2 a }{u\ep}J^2.
\]
Comparing with the actual regulated form factors for the stress tensor 3-point, which we found earlier in \eqref{A1TTTsing}\,-\,\eqref{A5TTTsing},  after using the identification \eqref{aC12} we see that all the divergences proportional to $a$ are in fact of exactly this form.  

Putting this observation to one side, let us now examine $-4a/u\ep$ times the second term on the right hand side of \eqref{summaryeq}.  The zeros coming from the factors of $(d-4)$ in \eqref{Bformfactors} are cancelled by this pole, yielding the finite form factors
\[\label{Eulerff}
B_1 = -8a, \quad B_2 = -8a(p_1^2-p_2^2-p_3^2),\quad
B_3 = 2aJ^2,\quad B_4 = \frac{8a}{3}p_3^2,\quad
B_5=-\frac{4a}{9}J^2,
\]
along with vanishing corrections of order $\ep$.
Comparing with the terms proportional to $a$ in the actual anomaly contribution we found earlier in \eqref{B1eqn}\,-\,\eqref{B5eqn}, we find an exact match up to an overall sign.  The form factors \eqref{Eulerff} thus represent minus the Euler anomaly contribution of the 3-point function.

Rearranging \eqref{summaryeq}, we see the divergent form factors in \eqref{ffdiv} are equal to the finite Euler anomaly contribution ({\it i.e.,} minus the form factors in \eqref{Eulerff}) plus $-4a/u\ep$ times the left-hand side of \eqref{summaryeq}.
This last quantity is of the form $0/0$, namely, an $\ep^{-1}$ pole multiplying an evanescent tensorial structure that vanishes in $d=4$.  The limit $\ep\rightarrow 0$ is then finite and hence the form factors in \eqref{ffdiv} do not represent a genuine UV divergence.\footnote{This can be shown explicitly as discussed in appendix \ref{app:DS}.}
 Indeed, this is precisely the behaviour we expect for a scale-invariant type A anomaly \cite{Deser:1993yx}. A genuine UV divergence, such as that represented by the terms proportional to $c$, would need to be removed by a divergent counterterm contribution; this counterterm contribution  would however introduce a non-vanishing dependence on the renormalisation scale, describing instead a scale-dependent type B anomaly.

Evaluating the totally antisymmetrised product in dimensional regularisation is in practice rather awkward \cite{Collins}, and the result is moreover scheme-dependent.
Rather than trying to evaluate $-4a/u\ep$ times the left-hand side of \eqref{summaryeq} explicitly, we can simply {\it remove} this term through the addition of an Euler counterterm.  (An alternative prescription is discussed in appendix \ref{app:DS}.)
As we will now show, this counterterm generates a correspondingly finite and scale-invariant contribution of precisely the required form.  Indeed, this can already be anticipated from our discussion in 
section \ref{sec:ttteven}, where we showed that the form factor divergences proportional to $a$ are cancelled by the Euler counterterm in (\ref{SctTTT}). 
\label{Eulertricks}
Evaluating the metric variation of this Euler counterterm about an arbitrary $d$-dimensional background, we find 
\begin{align}\label{Eulervar}
\delta {\int}\D^d \x \sqrt{g} \mu^{d-4} E_4 &= \int\D^d\x\sqrt{g}\mu^{d-4}\,\Big(\frac{1}{2}E_4 g^{\alpha_1\beta_1}\delta g_{\alpha_1\beta_1}+12R^{\alpha_1\alpha_2}_{[\alpha_1\alpha_2}\delta R^{\alpha_3\alpha_4}_{\alpha_3\alpha_4]}\Big)\nn\\&=
\int\D^d\x\sqrt{g}\mu^{d-4}\Big(\frac{1}{2}E_4 \delta_{\alpha_1}^{\beta_1}-12 R^{\beta_1\alpha_2}_{[\alpha_1\alpha_2}R^{\alpha_3\alpha_4}_{\alpha_3\alpha_4]}\Big)g^{\alpha_1\sigma}\delta g_{\sigma \beta_1} \nn\\&
=15\int\D^d\x \sqrt{g}\mu^{d-4} \Big(\delta_{[\alpha_1}^{\beta_1}R_{\alpha_2\alpha_3}^{\alpha_2\alpha_3}R_{\alpha_4\alpha_5]}^{\alpha_4\alpha_5}\Big) g^{\alpha_1\sigma}\delta g_{\sigma\beta_1}.
\end{align}
Here, to go from the first line to the second, we used
\[
\delta R_{\alpha\beta\gamma}{}^\delta = -2\nabla_{[\alpha}\delta \Gamma^\delta_{\beta]\gamma}, \qquad \delta \Gamma^\gamma_{\alpha\beta}=\frac{1}{2}g^{\gamma\delta}\Big(\nabla_{\alpha}\delta g_{\beta\delta}+\nabla_\beta\delta g_{\alpha\delta}-\nabla_\delta \delta g_{\alpha\beta}\Big)
\]
to show that
\[
\delta R_{\alpha\beta}{}^{\gamma\delta}= R_{\alpha\beta}{}^{\rho[\gamma}g^{\delta]\sigma}\delta g_{\rho\sigma}-2\nabla_{[\alpha}\nabla^{[\gamma}(g^{\delta]\sigma}\delta g_{\beta]\sigma}),
\]
and hence
\begin{align}
\int\D^d\x\sqrt{g}\mu^{d-4}R^{\alpha_1\alpha_2}_{[\alpha_1\alpha_2}\delta R^{\alpha_3\alpha_4}_{\alpha_3\alpha_4]}&=\int\D^d\x\sqrt{g}\mu^{d-4}\Big(R^{\alpha_1\alpha_2}_{[\alpha_1\alpha_2}R_{\alpha_3\alpha_4]}^{\rho\alpha_3}g^{\alpha_4\sigma}\delta g_{\rho\sigma}\nn\\&\qquad\qquad\qquad\qquad -2R^{\alpha_1\alpha_2}_{[\alpha_1\alpha_2}\nabla_{\alpha_3}\nabla^{\alpha_3}(g^{\alpha_4\sigma}\delta g_{\alpha_4]\sigma})\Big),
\end{align}
where the last term vanishes by the Bianchi identity upon integrating by parts and extracting the final index from the explicit antisymmetrisation.

Setting $g_{\mu\nu}=\delta_{\mu\nu}+h_{\mu\nu}$ and expanding to cubic order in the  perturbation $h_{\mu\nu}$, we find
\begin{align}\label{Eulercubic}
S_{\Ct}=\mathfrak{a}\int\D^d\x\sqrt{g}\mu^{d-4}E_4 = 20\mathfrak{a}\int\D^d\x\,\mu^{d-4} h^{\alpha_1}_{[\alpha_1}\p_{\alpha_2}\p^{\alpha_2}h_{\alpha_3}^{\alpha_3}\p_{\alpha_4}\p^{\alpha_4}h_{\alpha_5]}^{\alpha_5} + O(h^4),
\end{align}
where all indices are now raised with the background metric ({\it i.e.,} $h^{\alpha}_{\beta}=\delta^{\alpha\sigma}h_{\sigma\beta}$).
From this expression, we then obtain the counterterm contribution to the 3-point function
\begin{align}\label{Eulerctcontr5form}
&\lla T_{\mu_1\nu_1}(\bs{p}_1)T_{\mu_2\nu_2}(\bs{p}_2)T_{\mu_3\nu_3}(\bs{p}_3)\rra_{\Ct}\nn\\&\qquad  = -960\,\mathfrak{a}\,\mu^{d-4}\delta_{(\mu_1}^{\alpha_1}\delta_{\nu_1)\beta_1}
\delta_{(\mu_2}^{\alpha_2}\delta_{\nu_2)\beta_2}
\delta_{(\mu_3}^{\alpha_3}\delta_{\nu_3)\beta_3}\Big(
\delta_{[\alpha_1}^{\beta_1}\delta_{\alpha_2}^{\beta_2}\delta_{\alpha_3}^{\beta_3}p_{1\alpha_4}p_{2\alpha_5]}p_1^{\alpha_4}p_2^{\alpha_5}\Big).
\end{align}
Since $\mathfrak{a}=-a/2u\ep+O(\ep^0)$ (see \eqref{TTTctexps}), this counterterm contribution is thus exactly equal to $-4a/u\ep$ times the left-hand side of \eqref{summaryeq}.
As in our discussion above, it is finite meaning all dependence on the renormalisation scale vanishes in the limit $\ep\rightarrow 0$.\footnote{Explicitly, $\mu(\p/\p\mu)$ of \eqref{Eulerctcontr5form} is of order $\ep$ times a finite $0/0$ piece.  In contrast, a type B counterterm contribution proportional to $\ep^{-1}\mu^{d-4}$ would yield a finite piece when we act with $\mu(\p/\p\mu)$.}

In summary, then, while the regulated form factors \eqref{A1TTTsing}\,-\,\eqref{A5TTTsing} of the stress tensor 3-point function contain apparent divergences of the form \eqref{ffdiv}, these divergent form factors reassemble into a tensorial structure that vanishes in $d=4$.
The resulting contribution to the regulated 3-point function thus has the $0/0$ form expected of type A anomalies, providing an explicit four-dimensional counterpart to the two-dimensional discussion of \cite{Deser:1993yx}.
By adding an Euler counterterm, we can eliminate this $0/0$ piece yielding a finite and unambiguous trace anomaly contribution of the expected Euler type.

Finally, tracing over the first pair of indices in \eqref{Eulerctcontr5form}, note that we can write the Euler contribution to the trace anomaly  \eqref{e:TWI_TTT} as 
\begin{align}\label{epsilonanomaly}
\mathcal{A}^{\mathrm{Euler}}_{\mu_2\nu_2\mu_3\nu_3}
=40 \,a \,\delta_{(\mu_2}^{\alpha_2}\delta_{\nu_2)}^{\beta_2}
\delta_{(\mu_3}^{\alpha_3}\delta_{\nu_3)}^{\beta_3}
(\ep_{\alpha_2\alpha_3\alpha_4\alpha_5}p_1^{\alpha_4}p_2^{\alpha_5})(\ep_{\beta_2\beta_3\beta_4\beta_5}p_1^{\beta_4}p_2^{\beta_5}).
\end{align}
Strikingly, as mentioned in the introduction, this is the ``square'' of the chiral anomaly.
Relaxing parity conservation, 
the transverse Ward identity \eqref{TWI_JJJ} for the current 3-point function acquires an anomalous term
\begin{align}
&p_{1\mu_1}\lla J^{\mu_1a_1}(\bs{p}_1) J^{\mu_2a_2}(\bs{p}_2) J^{\mu_3a_3}(\bs{p}_3) \rra 
\nn\\[1ex]&\quad=
  g f^{a_1 b a_3} \lla J^{\mu_3 b}(\bs{p}_2) J^{\mu_2 a_2}(-\bs{p}_2) \rra 
 - g f^{a_1 a_2 b} \lla J^{\mu_2 b}(\bs{p}_3) J^{\mu_3 a_3}(-\bs{p}_3) \rra +d^{a_1a_2a_3}\mathcal{A}^{\mathrm{chiral}}_{\mu_2 \mu_3},
\end{align}
where $d^{a_1 a_2  a_3}$ is a group theoretic tensor.  This chiral anomaly takes the form 
\[
\mathcal{A}^{\mathrm{chiral}}_{\mu_2 \mu_3}
=\mathcal{N} (\ep_{\mu_2\mu_3\alpha_2\alpha_3}p_1^{\alpha_2}p_2^{\alpha_3}),
\]
where $\mathcal{N}$ depends on the matter content, and hence we have
\[
\mathcal{A}^{\mathrm{Euler}}_{\mu_2\nu_2\mu_3\nu_3}
\propto\delta_{(\mu_2}^{\alpha_2}\delta_{\nu_2)}^{\beta_2}
\delta_{(\mu_3}^{\alpha_3}\delta_{\nu_3)}^{\beta_3}
\mathcal{A}^{\mathrm{chiral}}_{\alpha_2 \alpha_3}\mathcal{A}^{\mathrm{chiral}}_{\beta_2 \beta_3}.
\]
This relation resembles the double-copy relation between Yang-Mills and gravity scattering amplitudes, see \cite{Bern:2017tuc} and references therein.
It is likely that analogous relations for type A anomalies exist in all higher even dimensions.

\section{Determining the scheme-independent 
constants}  
\label{sec:findconsts}

From our analysis above, we now know the renormalised 3-point functions up to only a few constants.  Of these constants, a number are scheme-dependent and can be arbitrarily adjusted by adding finite counterterms or by changing the renormalisation scale.  
The remainder are scheme-independent, and serve to characterise the particular CFT at hand.
These physical, theory-specific constants are: 
(i) the constant $C_1$, which 
as the only new parameter unrelated to anomalies entering the 3-point function, 
effectively determines the OPE coefficient; (ii) the normalisation of the relevant 2-point function  (either $C_{JJ}$ or $C_{TT}$), which also controls the type B anomaly of the 3-point function; and (iii), for the four-dimensional correlator of three stress tensors, the Euler coefficient $a$ parametrising the type A trace anomaly.

Our remaining task is thus to extract these physical constants for some given CFT of interest.  As $C_{JJ}$ and $C_{TT}$ are already known from the 2-point function, we will focus on the new constants $C_1$ and $a$ arising in the 3-point function.  For perturbative theories, these constants can be identified as follows.  First, the scalar Feynman integrals corresponding to the required form factors are found by examining the coefficients of the appropriate tensor structures.  We can then either evaluate these integrals directly, in special kinematic configurations for which they simplify, or else map them to triple-$K$ integrals.  Here, the relations between  Feynman- and/or Schwinger-parametrised integrals and triple-$K$ integrals given in appendix C of \cite{Bzowski:2013sza} are particularly useful.  

In the rest of this section, we illustrate both these approaches, using free Dirac fermions in four dimensions as a generic example.
The corresponding flat-space operators are thus
\begin{equation}
T_{\mu\nu} = - i \bar{\psi} \gamma_{(\mu} \overleftrightarrow{\p}_{\nu)} \psi + i g_{\mu\nu} \bar{\psi} \gamma^{\alpha} \overleftrightarrow{\p}_{\alpha} \psi,
\qquad J^{\mu} = i g \bar{\psi} \gamma^\mu \psi.
\end{equation}
Note however that on a general curved background, $T_{\mu\nu}$ depends on the metric and hence $\delta T_{\mu\nu}/\delta g^{\rho\sigma}$ is nonzero.  
Our stress tensor 3-point function then involves a 2-point contribution from this functional derivative operator, according to our definitions \eqref{nowdefTTT} and \eqref{relatedefTTT}. 
By decomposing this functional derivative operator in a local basis, as in section 4.3 of \cite{Bzowski:2011ab}, one finds however that only the form factor $A_5$ is affected.  
As this form factor is not used in our discussion below, the contribution from  $\delta T_{\mu\nu}/\delta g^{\rho\sigma}$ can be safely ignored.

\subsection{The constant $C_1$}

The constant $C_1$ is most easily isolated from the form factor $A_1$.  
In the 3-point function, this form factor multiplies the tensorial structure of the highest dilatation weight, namely that built entirely from momenta.   In four dimensions, for all the correlators we study, the dilatation weight of $A_1$ itself is then minus two.
As any triple-$K$ integral of negative dilatation weight converges, $A_1$ is therefore finite   
and we can evaluate $C_1$ 
directly in the bare theory.\footnote{More generally, the dilatation weight of $A_1$ is $d-6$ meaning $A_1$ diverges for even dimensions greater than four; $C_1$ can then be extracted from the coefficient of this divergence in the regulated theory.}

Let us consider, for example, the case of $\< T_{\mu_1\nu_1} J^{\mu_2}J^{\mu_3}\>$ for free fermions in $d=4$.  The form factor $A_1$ can be obtained from the 1-loop Feynman-parametrised integral
\begin{align}\label{parametricint}
A_1 &=\text{coefficient of } p_{2\mu_1} p_{2\nu_1} p_3^{\mu_2} p_1^{\mu_3} \text{ in } \lla T_{\mu_1 \nu_1}(\bs{p}_1) J^{\mu_2}(\bs{p}_2) J^{\mu_3}(\bs{p}_1) \rra \nn\\& = - \frac{2g^2}{\pi^2} \int_{[0,1]^3} \D X \: \frac{x_1^2 x_2 x_3}{\Delta},
\end{align}
where
\begin{align}
& \D X = \D x_1 \D x_2 \D x_3 \: \delta(x_1 + x_2 + x_3 - 1), \\
& \Delta = p_1^2 x_2 x_3 + p_2^2 x_1 x_3 + p_3^2 x_1 x_2.
\end{align}
The calculation leading to this result is straightforward and proceeds as discussed in appendix B.3 of \cite{Bzowski:2011ab} and section 7.5 of \cite{Bzowski:2013sza}.\footnote{The result may also be obtained directly from the calculations of \cite{Armillis:2009pq}.  
Starting from the tensors listed in table 2 of \cite{Armillis:2009pq}, substituting $(\mu,\nu,\alpha,\beta,\bs{k},\bs{p},\bs{q})\rightarrow (\mu_1,\nu_1,\mu_2,\mu_3,-\bs{p}_1,\bs{p}_2,\bs{p}_3)$ and putting all momenta into their standard form \eqref{a:momenta}, we find 
$
A_1 = -4(F_3+F_5-F_7)-2p_2^2 F_9-2p_3^2 F_{10} = -C_7+2C_8-C_9
$
after using (180)-(182) and (187)-(188). From (178) and table 3 in \cite{Armillis:2009pq}, one then recovers the parametric integral \eqref{parametricint}.}
Conformal invariance dictates that this parametric integral  {\it must} however be equivalent to a triple-$K$ integral. Using the mapping given in (C.23) of \cite{Bzowski:2013sza}, we can convert the triple-$K$ integral for $A_1$ (see \eqref{e:restjj1}) to a Feynman-parametrised integral, 
\begin{equation} 
A_1 = C_1 J_{4 \{000\}} = C_1 I_{5 \{211\}} = 96 C_1\int_{[0,1]^3} \D X \: \frac{x_1^2 x_2 x_3}{\Delta}.
\end{equation}
Comparing with \eqref{parametricint}, we can immediately identify
\begin{equation}\label{QEDC1}
C_1 = - \frac{g^2}{48 \pi^2}.
\end{equation}

Alternatively, we can look for special kinematic configurations in which the relevant integrals simplify.  The squeezed (or collinear) limit where one momentum magnitude vanishes is especially useful in this regard.
In this limit, triple-$K$ integrals reduce to double-$K$ integrals which can be evaluated via the simple analytic formula\footnote{The integral converges for $\mathrm{Re}\,\alpha>2|\mathrm{Re}\,\beta|-1$ but may be extended beyond this range by analytic continuation as discussed in \cite{Bzowski:2013sza}.} \cite{Prudnikov} 
\begin{align}
\int_0^\infty \D x \,x^{\alpha}K_{\beta}^2(p x) = \frac{2^{\alpha-2}}{\Gamma(\alpha+1)p^{\alpha+1}}\,\Gamma\Big(\frac{\alpha+1+2\beta}{2}\Big)\Gamma\Big(\frac{\alpha+1-2\beta}{2}\Big)\Gamma^2\Big(\frac{\alpha+1}{2}\Big).
\end{align}
For the case at hand, this yields
\begin{align}
A_1(0,p,p) =C_1 I_{5\{211\}}(0,p,p)= 2C_1p^2 \int_0^\infty\D x\,x^{3}K_1^2(px) =\frac{4C_1}{3p^2}.
\end{align}
The parametric integral \eqref{parametricint} also simplifies in this limit, and can be likewise be evaluated:
\[
A_1(0,p,p) = - \frac{2g^2}{\pi^2 p^2} \int_{[0,1]^3} \D X \: \frac{x_1 x_2 x_3}{(x_2+x_3)} = - \frac{g^2}{36 \pi^2 p^2},
\]
Once again, we then recover \eqref{QEDC1}.  

Using the same methods, we can similarly compute   
$C_1$ for all other correlators.  For example, for the stress tensor 3-point function of free fermions in $d=4$, the form factor $A_1$ is given by the 1-loop integral
\begin{align}\label{Feynintresult}
A_1 &=  \text{coefficient of } p_{2\mu_1} p_{2\nu_1} p_{3\mu_2} p_{3\nu_2} p_{1\mu_3} p_{1\nu_3} \text{ in } \lla T_{\mu_1\nu_1}(\bs{p}_1)T_{\mu_2\nu_2}(\bs{p}_2)T_{\mu_3\nu_3}(\bs{p}_3)\rra  \nn\\[1ex]& =
-\frac{2}{\pi^2}\int_{[0,1]^3}\D X\:\frac{(x_1x_2x_3)^2}{\Delta}.
\end{align}
This result can be obtained by adapting the analogous three-dimensional calculation in \cite{Bzowski:2011ab}.\footnote{The required integral is given by (4.13) in \cite{Bzowski:2011ab}, where the loop integral is now four-dimensional rather than three-dimensional.  The right-hand side of (4.14) needs to be multiplied by an additional factor of two as the gamma matrices are now four-dimensional, rather than two-dimensional as is the case in $d=3$.}
Comparing with our present result \eqref{a:TTT1}, using (C.23) of \cite{Bzowski:2011ab} we find
\[
A_1 = C_1 J_{6\{000\}} = C_1 I_{7\{222\}} = 11520 C_1 \int_{[0,1]^3}\D X\:\frac{(x_1x_2x_3)^2}{\Delta},
\]
and hence
\[\label{C1fermions}
C_1=-\frac{1}{5760\pi^2}.
\]
Alternatively, we can arrive at this result by evaluating the triple-$K$ and Feynman parametric integrals in the squeezed limit: 
\[
C_1 = \frac{5}{64}p^2 A_1(0,p,p) = -\frac{5}{32\pi^2} \int_{[0,1]^3}\D X\:\frac{(x_1x_2x_3)^2}{x_1(x_2+x_3)}=-\frac{1}{5760\pi^2}.
\]

\subsection{The Euler coefficient $a$}

While $C_1$ follows from the form factor $A_1$, which is finite in four dimensions, to compute the Euler coefficient $a$ we need to examine the {\it divergences} in one (or more) of the regulated form factors $A_2$ to $A_5$.  
In the appropriate scheme with $u=v_j$ for  $j=1,2,3$, combining \eqref{aC12} and \eqref{A1TTTsing}\,-\,\eqref{A5TTTsing}, we see these divergences take the form
\begin{align}\label{formfactordivergences1}
A_1 &= O(\ep^0),\\[0.5ex]
A_2 &= -\frac{8}{u\ep}(a+c)+O(\ep^0),\\[0.5ex]
A_3 &= \frac{4}{u\ep}\Big[c(p_1^2+p_2^2)-a p_3^2\Big]+O(\ep^0),\\[0.5ex]
A_4 &=  \frac{4}{u\ep}\Big[c(p_1^2+p_2^2+3p_3^2)-a(p_1^2+p_2^2-p_3^2)\Big]+O(\ep^0),\\[0.5ex]
A_5 &= -\frac{2}{u\ep}\Big[c(p_1^2+p_2^2+p_3^2)^2+aJ^2\Big]+O(\ep^0).
\label{formfactordivergences5}
\end{align}
The form factor $A_2$ is particularly convenient in that its divergence is independent of the momenta.
For free fermions in $d=4$, one finds the divergent part of $A_2$ in the scheme\footnote{From appendix C of \cite{Bzowski:2013sza}, the scheme $u=v_j=1$ corresponds to evaluating 1-loop tensorial Feynman integrals in $d=4+2\ep$ while keeping the powers of momenta $\delta_j$ appearing in the denominators fixed.}
 $u=v_j=1$ is given by the parametric integral
\[
A_2= -\frac{1}{2\pi^2\ep}\int_{[0,1]^3}\D X\, x_1x_2x_3(3-2x_3)+O(\ep^0) =-\frac{7}{720\pi^2\ep}+O(\ep^0),
\]
and hence 
\[
a+c = \frac{7}{5760\pi^2}.
\]
We can now either evaluate the divergence of  another form factor, or else simply use the 2-point function to compute $c$ via the relation \eqref{cresult}.
Taking this latter route, we calculate for example
\begin{align}\label{2pttraceddiv}
\lla T_{\mu\nu}(\bs{p})T^{\mu\nu}(-\bs{p})\rra_{\Reg} =\frac{5}{\ep}C_{TT}p^4+O(\ep^0)= -\frac{p^4}{32\pi^2 \ep}+O(\ep^0), 
\end{align}
where the repeated indices are summed over and we used \eqref{e:2ptTTreg}.  We then find
\[\label{acfreefermions}
c = -\frac{C_{TT}}{2} = \frac{1}{320\pi^2}, \qquad a = -\frac{11}{5760\pi^2},
\]
in agreement with standard results obtained by other methods \cite{Osborn:1993cr}, \cite{Birrell_and_Davies}.\footnote{See page 179,  
noting their stress tensor is defined with an overall minus sign relative to ours.} 

With these values of $a$ and $c$, the divergences in the remaining form factors $A_3$, $A_4$ and $A_5$ (including now the contribution from  $\delta T_{\mu\nu}/\delta g^{\rho\sigma}$) take the expected form \eqref{formfactordivergences1}\,-\,\eqref{formfactordivergences5}. 
Indeed, for additional security, we have checked that we were able to reproduce the full renormalised form factors \eqref{A1TTTren}\,-\,\eqref{A5TTTren}.  Recovering the correct dependence on the scheme-dependent constant $D_{TT}$ is a particularly nontrivial check, requiring a consistent regularisation for both the 2- and 3-point functions.

To summarise, for perturbative theories all scheme-independent constants entering the renormalised 3-point functions can be obtained through elementary means.  The overall normalisation $C_1$ can be found from the finite form factor $A_1$; either by relating parametrised Feynman integrals to triple-$K$ integrals,
or else by directly evaluating the form factor in the squeezed limit.
For all correlators apart from that of three stress tensors, the only remaining scheme-independent constants are then the 2-point function normalisations.  For the stress tensor 3-point function, one has in addition the Euler coefficient $a$.  As above, this can be evaluated from the divergences of the regulated form factors.
Alternatively, if the renormalised correlator is already to hand, we can read off the Euler coefficient from the fully-traced 3-point function: 
\[\label{afromtrace}
\lla T(\bs{p}_1)T(\bs{p}_2)T(\bs{p}_3)\rra = 4aJ^2.
\]
This formula follows from the trace Ward identity \eqref{e:TWI_TTT} and the decomposition of the anomaly \eqref{anomaly_decomp} in terms of the form factors \eqref{B1eqn}\,-\,\eqref{B5eqn}.  (We assume here we are working in a scheme where the stress tensor 2-point function is traceless, setting $b=0$ in \eqref{B5eqn}.) 
Geometrically, the right-hand side is equal to $64\,a$ times the  squared area of the triangle formed from the three momenta.

\section{Discussion}
\label{sec:discussion}

We have solved for the 
renormalised, tensorial 3-point functions 
of a general CFT.  Our solution highlights the advantages of working in momentum space.  
First, the tensorial structure of correlators can be decomposed into a minimal set of scalar form factors, which multiply independent basis tensors constructed from the metric and independent momenta.
Second, 
all non-transverse-traceless components of this basis can be eliminated through the trace and transverse Ward identities, whose form in momentum space is algebraic. 
The resulting scalar form factors now obey simple conformal Ward identities whose form is {\it near-identical} to those obtained  
from purely scalar correlators.  Of these momentum-space Ward identities, those corresponding to special conformal transformations factorise and 
can be solved 
through elementary separation of variables.  The remaining dilatation Ward identities are
solved through a Mellin transform leading to triple-$K$ integral solutions.

The divergences of these triple-$K$ integrals are readily understood and can be regulated 
 through infinitesimal shifts of the operator and spacetime dimensions.  This generalised dimensional regularisation preserves conformal invariance,
maintaining our control over the form of the 3-point functions.
For correlators of stress tensors and conserved currents, all divergences can then be satisfactorily removed by the addition of counterterms cubic in the sources. (More generally, for mixed 3-point correlators involving scalar operators, other type of counterterms are required as discussed in \cite{Bzowski:2018fql}.)
The resulting renormalised correlators now obey anomalous conformal Ward identities, whose inhomogeneous terms encode the breaking of conformal invariance by the 
counterterms.
In position space, the contribution of these counterterms represents the missing contact terms whose role is to remove the singularities associated with coincident operator insertions.

Besides scheme-dependent terms, 
only a small number of physical, scheme-independent constants appear in the renormalised 3-point correlators.  These are an overall normalisation associated with the form factor of lowest dilatation weight, and the coefficients appearing the trace anomaly. 
The latter can be split into the coefficients of type B anomalies (which control the 2-point normalisations), and for the correlator of three stress tensors, the Euler coefficient encoding the type A anomaly.  The values of all scheme-independent constants can easily be evaluated for any CFT of interest, either by evaluating the form factors in special kinematic configurations, or else by returning them to their canonical representation as triple-$K$ integrals.

As we saw explicitly in $d=4$, the scale-independence of the type A anomaly is associated with a UV divergent coefficient multiplying an evanescent operator that vanishes in the physical spacetime dimension.
This general structure was predicted long ago in 
\cite{Deser:1993yx}, and our present results supply all the remaining details.
The geometric origin of these evanescent operators, as forms of higher rank than the spacetime dimension, is particularly clear in momentum space.
Interestingly, the Euler contribution to the anomaly of the stress tensor 3-point function takes the form of the square of the chiral anomaly. The same also holds in two dimensions and it is likely to be true in all even dimension. It would be interesting to understand if there is a deeper meaning to this relation, or if it simply follows from kinematics.  Even in the latter case, it may still have non-trivial implications, for example by allowing results established for chiral anomalies ({\it e.g.},  descent relations) to be applied to (the Euler part of) conformal anomalies.

Looking ahead, 
we hope our momentum-space methods will open new approaches; not only to contemporary problems, such as bootstrapping tensorial correlators \cite{Dymarsky:2017yzx}, but to  classic problems whose past investigation has been hindered by the complications of tensorial structure.
In particular, our improved understanding of the stress tensor 3-point function could provide   renewed impetus for the following investigations:

\paragraph{Nonlocal effective actions.} As 2- and 3-point functions are universal, there should exist a 
nonlocal geometric effective action, whose quadratic and cubic parts are universal, reproducing the full stress tensor 2- and 3-point functions.
What is the form of this effective action in four dimensions?  
While the Euler anomaly contribution follows from the Riegert action \cite{Riegert:1984kt, Fradkin:1983tg} (see also \cite{Deser:1993yx,  Erdmenger:1996yc, Erdmenger:1997gy,  Deser:1999zv, Mazur:2001aa, Coriano:2017mux}), this represents just one part of the full 3-point function.  
The situation is thus quite different to that in two dimensions, where, since all two-dimensional metrics are conformally flat, the Polyakov action obtained by integrating the anomaly captures {\it all} the information in stress tensor correlators.
For a four-dimensional CFT, the most general 2- and 3-point functions of the stress tensor can instead be obtained from a combination of free fields.
Using \cite{Barvinsky:1995it, Mirzabekian:1995qf, Barvinsky:1994cg, Barvinsky:1993en}, these computations should then be sufficient to obtain  the general covariant effective action at cubic order. 
It would be interesting to find this action and verify it reproduces the results we report here.

\paragraph{Anomaly matching.}
The matching of anomalies between broken and unbroken phases, proposed by Schwimmer and Theisen in \cite{Schwimmer:2010za}, is a  fundamental ingredient in the proof of the $a$-theorem \cite{Komargodski:2011vj, Komargodski:2011xv}. 
Given their physical importance, it is desirable to put all the arguments of \cite{Schwimmer:2010za}  on as explicit a footing as possible.
Now that we have the full stress tensor 3-point function to hand, it should be possible to clarify a number of remaining ambiguities. What are the precise tensor structures corresponding to the $A$ and $B$ amplitudes of \cite{Schwimmer:2010za}?
Can we verify that the analytic structure of the $B$ amplitude, at the special kinematic point where all Lorentz invariants vanish, degenerates from a branch cut to an apparent pole?  How does this enable us to isolate the Euler anomaly, and how 
sharp an analogy can we make with the chiral anomaly \cite{Coleman:1982yg, Frishman:1980dq}?

\paragraph{A spectral proof of the ${\bs a}$-theorem?}  
In two dimensions, an elegant proof of Zamolodchikov's $c$-theorem \cite{Zamolodchikov:1986gt} can be obtained from the positivity of the K{\"a}ll{\'e}n-Lehmann spectral representation \cite{Cappelli:1990yc}.  
Is there an analogous {\it spectral} proof for the four-dimensional $a$-theorem?\footnote{Unpublished  work concerning such a proof, by Z. Komargodski, A. Schwimmer and S. Theisen, has been presented at a number of conferences in recent years (A. Schwimmer,  private communication).}  
Through standard dispersion relations, we can construct spectral representations for the form factors of the stress tensor 3-point function \cite{Cappelli:2001pz}.
These spectral functions, corresponding to the imaginary parts of form factors, then obey homogeneous ({\it i.e.,} non-anomalous) conformal Ward identities, as all counterterm contributions are analytic in the squared momenta and thus are projected out.
A successful 
spectral approach to the $a$-theorem could provide complementary insight to the 
anomaly matching and dilaton effective action 
arguments of 
\cite{Komargodski:2011vj, Komargodski:2011xv}.
The simplicity of our tensorial decomposition -- involving only four form factors  after removal of the degeneracy -- 
could yield new insight.

\paragraph{Defining an $\bs{a}$-function.}
As we have seen, at fixed points of RG flow we can extract the Euler anomaly coefficient $a$ from various projections of the renormalised 3-point function: from the fully traced correlator as in \eqref{afromtrace}, for example, 
but also from the transverse traceless part. 
Projections such as these are natural candidates for an  $a$-{\it function}: they are well-defined observables, away from the fixed point, that reduce to the Euler coefficient  at the fixed point itself.  It would be interesting to investigate whether any 
of these candidates 
also exhibit the required monotonicity under RG flow.


\section*{Acknowledgements}

We thank Claudio Coriano, Yuri Gusev, Emil Mottola, Adam Schwimmer and Stefan Theisen for discussions.
The work of 
AB is supported in part by the National Science Foundation
of Belgium (FWO) grant G.001.12 Odysseus,  
the European Research
Council grant ERC-2013-CoG 616732 HoloQosmos, 
the COST Action MP1210 ``The String Theory Universe'' and the ANR grant Black-dS-String ANR-16-CE31-0004-01.
The work of PM is supported by the STFC through an Ernest Rutherford Fellowship and the Consolidated Grant, ``M-theory, Cosmology and Quantum Field Theory'', ST/L00044X/1.  
KS is supported in part by the STFC Consolidated Grant, ``Exploring the Limits of the Standard Model and Beyond'', ST/L000296/1, and 
``New Frontiers in Particle Physics and Cosmology", ST/P000711/1.
This project has received funding from the European Union's Horizon 2020 research and innovation programme under the Marie Sk\l{}odowska-Curie grant agreement No 690575.
AB and PM thank the University of Southampton, and PM thanks Marjorie Schillo and the University of Leuven for their kind 
hospitality.

\appendix

\section{Appendices}
\subsection{The type A anomaly in two dimensions}
\label{app:typeA}

Following our discussion in the introduction, in this appendix we resume our analysis of the stress  tensor 2-point function in two dimensions.
This is the simplest example featuring a type A anomaly, and serves as 
a warm-up for  
the  four-dimensional discussion in section \ref{sec:Eulerdeg}.

We begin with the only available counterterm,
\[
S_{\Ct} = \mathfrak{c}\int\D^d\x\sqrt{g}\mu^{d-2}R,
\]
where the renormalisation scale $\mu$ enters on dimensional grounds. 
Setting $g_{\mu\nu}=\delta_{\mu\nu}+h_{\mu\nu}$ and expanding to quadratic order, we find
\[
S_{\Ct} = -3\mathfrak{c}\int\D^d\x\,\mu^{d-2} h_{[\alpha_1}^{\alpha_1}\partial_{\alpha_2}\partial^{\alpha_2}h_{\alpha_3]}^{\alpha_3}+O(h^3)
\]
where all indices are raised with the flat background metric.  The form of this result reflects the fact that the Ricci scalar is the two-dimensional Euler density; the 
manipulations used to derive it can be found in section \ref{sec:Eulerdeg}  (see page \pageref{Eulertricks}).
The 2-point contribution from this counterterm then takes the form
\[
\lla T_{\mu_1\nu_1}(\bs{p})T_{\mu_2\nu_2}(-\bs{p})\rra_{\Ct}= -6 \mathfrak{c}\,\mu^{d-2}\,\delta_{(\mu}^{\alpha_1}\delta_{\nu_1)\beta_1}\delta_{(\mu_2}^{\alpha_2}\delta_{\nu_2)\beta_2}\,\delta_{[\alpha_1}^{\beta_1}\delta_{\alpha_2}^{\beta_2}p_{\alpha_3]}p^{\alpha_3}.
\]
As we see, exactly the same 3-form appears here as in the regulated 2-point function \eqref{2pt2dttreg}.  Thus, in $d=2+2\ep$, even if the counterterm coefficient $\mathfrak{c}$ has a $\ep^{-1}$ divergence as $\ep\rightarrow 0$, the vanishing of the 3-form in two dimensions means that this counterterm contribution is finite.  In addition, the dependence on the renormalisation scale drops out.

To connect the form of this counterterm contribution with that of the regulated 2-point function, we now need to separate out its transverse-traceless and trace pieces.  This can be accomplished using two copies of the projection operator \eqref{Idecomp}.  As the counterterm contribution is purely transverse ({\it i.e.,} vanishes when any free index is contracted with the momentum), we can write this projection using only the transverse-traceless and transverse projectors introduced in \eqref{projectors1}.  We thus have
 \begin{align}
&\lla T_{\mu_1\nu_1}(\bs{p})T_{\mu_2\nu_2}(-\bs{p})\rra_{\Ct}\nn\\[1ex]
&= -6\mathfrak{c}\mu^{d-2}\Big(\Pi_{\mu_1\nu_1}{}^{\alpha_1}{}_{\beta_1}+\frac{1}{d-1}\pi_{\mu_1\nu_1}\delta^{\alpha_1}_{\beta_1}\Big)\Big(\Pi_{\mu_2\nu_2}{}^{\alpha_2}{}_{\beta_2}+\frac{1}{d-1}\pi_{\mu_2\nu_2}\delta^{\alpha_2}_{\beta_2}\Big)\delta_{[\alpha_1}^{\beta_1}\delta_{\alpha_2}^{\beta_2}p_{\alpha_3]}p^{\alpha_3}\nn\\[1ex]
&=-6\mathfrak{c}\mu^{d-2}\Big[
\Pi_{\mu_1\nu_1}{}^{\alpha_1}{}_{\beta_1}(\bs{p})\Pi_{\mu_2\nu_2}{}^{\alpha_2}{}_{\beta_2}(\bs{p})\,\delta_{[\alpha_1}^{\beta_1}\delta_{\alpha_2}^{\beta_2}p_{\alpha_3]}p^{\alpha_3}
+\frac{(d-2)}{6(d-1)} p^2 \pi_{\mu_1\nu_1}(\bs{p})\pi_{\mu_2\nu_2}(\bs{p})\Big].
\label{2pt2dctcontr}
\end{align}
The first transverse-traceless term now matches the form of the regulated 2-point function.   The second term has a nonzero trace, and crucially an extra factor of $(d-2)$ which arises from the tracing over the 3-form:
\[\label{ddimtr3form}
\delta_{\beta_1}^{\alpha_1}\delta_{[\alpha_1}^{\beta_1}\delta_{\alpha_2}^{\beta_2}p_{\alpha_3]}p^{\alpha_3} = \frac{1}{6}(d-2)\,p^2\pi_{\alpha_2}^{\beta_2}(\bs{p}).
\]
This extra factor of $(d-2)$ ensures the second term in \eqref{2pt2dctcontr} is finite as $\ep\rightarrow 0$, even though the corresponding tensor structure is nonvanishing. 
If we choose the counterterm coefficient as minus  that of the regulated 2-point function, 
\[\label{2dcttdef}
C_{TT}(\ep) = \frac{C_{TT}}{2\ep}+O(\ep^0),\qquad
\mathfrak{c}(\ep) = - \frac{C_{TT}}{2\ep}+O(\ep^0),
\]
the transverse-traceless first term of the counterterm contribution \eqref{2pt2dctcontr} now cancels the regulated 2-point function \eqref{2pt2dttreg}.  In the limit $\ep\rightarrow 0$, we obtain 
\begin{align}\label{appTTren}
\lla T_{\mu_1\nu_1}(\bs{p})T_{\mu_2\nu_2}(-\bs{p})\rra_{\Ren} = C_{TT}\, p^2  \pi_{\mu_1\nu_1}(\bs{p})\pi_{\mu_2\nu_2}(\bs{p}).
\end{align}
This renormalised correlator is independent of the renormalisation scale, as we expect for a type A anomaly.
Let us emphasise once again that both the counterterm contribution and the regulated 2-point function are each {\it individually} finite in the limit as $\ep\rightarrow 0$.  Our introduction of the counterterm is simply a convenient device for  evaluating this limit: rather than attempting to evaluate the antisymmetrised terms in $d=2+2\ep$, we simply cancel them against one another.  
An alternative prescription is given below.

\subsection{Evaluating 0/0 limits without counterterms}\label{app:DS}

Type A anomalies arise from 0/0 structures in which an evanescent operator, which vanishes in the physical spacetime dimension, appears with a linearly divergent coefficient.
As we have seen, this occurs both for the stress tensor 2-point function in $d=2$, and for the 3-point function in $d=4$.  In the above, we opted simply to cancel these 0/0 structures in the regulated form factors with a matching 0/0 contribution from the corresponding Euler counterterm. 
In the following, we discuss an alternative approach allowing the $\ep\rightarrow 0$ limit 
to be defined without the introduction of counterterms.  This serves to emphasise the UV finiteness and scheme-independence of type A anomalies.

Following \cite{Deser:1993yx}, the idea is to define the regulated correlators such that all  {\it external} tensorial structures live in the physical spacetime dimension, while their accompanying scalar  coefficients live in the regulated spacetime dimension $d$.
In practice, this means all  external ({\it i.e.,} uncontracted) Lorentz indices  are restricted to take physical values, as are all indices associated with the momenta.  (In the absence of loop integrals, all momenta in our discussion are external hence have vanishing non-physical components.)
The remaining contractions of internal indices are then summed over the full $d$ dimensions.

The relevant 0/0 structures to evaluate are then the regulated 2-point function \eqref{2pt2dttreg}, which substituting for $C_{TT}(\ep)$ using \eqref{general2ptcttdef} reads
\begin{align}\label{2dstpt}
&\lla T_{\mu_1\nu_1}(\bs{p})T_{\mu_2\nu_2}(-\bs{p})\rra_{\Reg} = -\frac{3 C_{TT}}{2u\ep}\, p^{d-2}\,\Pi_{\mu_1\nu_1}{}^{\alpha_1}{}_{\beta_1}(\bs{p})\Pi_{\mu_2\nu_2}{}^{\alpha_2}{}_{\beta_2}(\bs{p})\,\delta_{[\alpha_1}^{\beta_1}\delta_{\alpha_2}^{\beta_2}p_{\alpha_3]}p^{\alpha_3}, 
\end{align}
and the Euler part of the regulated 3-point function, namely all pole terms proportional to $a$.  From our discussion in section \ref{sec:euleranom}, these can be written as
\begin{align}\label{4dstpt}
&\lla T_{\mu_1\nu_1}(\bs{p}_1)T_{\mu_2\nu_2}(\bs{p}_2)T_{\mu_3\nu_3}(\bs{p}_3)\rra_{\mathrm{Euler}} \nn\\[1ex]&=-\frac{480 a}{u\ep}\,
\Pi_{\mu_1\nu_1}{}^{\alpha_1}{}_{\beta_1}(\bs{p}_1)
\Pi_{\mu_2\nu_2}{}^{\alpha_2}{}_{\beta_2}(\bs{p}_2)\Pi_{\mu_3\nu_3}{}^{\alpha_3}{}_{\beta_3}(\bs{p}_3)\,\delta_{[\alpha_1}^{\beta_1}\delta_{\alpha_2}^{\beta_2}\delta_{\alpha_3}^{\beta_3}p_{1\alpha_4}p_{2\alpha_5]} p_1^{\alpha_4}p_2^{\alpha_5}. 
\end{align}
To evaluate these 0/0 structures, we use \eqref{Idecomp} to expand 
\[\label{TTprojexp}
\Pi_{\mu_1\nu_1}{}^{\alpha_1}{}_{\beta_1}(\bs{p}_1) = \delta_{(\mu_1}^{\alpha_1}\delta_{\nu_1)\beta_1}-\frac{1}{d-1}\pi_{\mu_1\nu_1}(\bs{p}_1)\delta^{\alpha_1}_{\beta_1}+\ldots
\] 
and similarly for the other projectors.
The omitted term  has a longitudinal component  and hence makes no contribution when contracted with either the 3-form in \eqref{2dstpt} or the 5-form in \eqref{4dstpt}.  Notice also that the indices $\alpha_1$ and $\beta_1$ in the first term on the right-hand side of \eqref{TTprojexp} are restricted to physical values, since $\mu_1$ and $\nu_1$ are external.  In the second term, however, the indices $\alpha_1$ and $\beta_1$  run over the full $d$ dimensions.  
Thus, when we contract with the forms in \eqref{2dstpt} and \eqref{4dstpt}, the contributions derived from selecting this first term for all the projectors vanishes, since all the indices in the forms are then restricted to physical values.  (Recall the indices on momenta are already restricted to physical values.)

The remaining terms all contain at least one $d$-dimensional trace arising from selecting the second term in \eqref{TTprojexp} for one (or more) of the projectors.
Evaluating this $d$-dimensional trace of the corresponding form then yields the necessary factor of $\ep$ to cancel the explicit $\ep^{-1}$ poles in \eqref{2dstpt} and \eqref{4dstpt}.  For the 2-point function, tracing over the 3-form gives a factor of $(d-2)$ as shown in \eqref{ddimtr3form}, while in the 3-point function tracing over the 5-form similarly gives a factor of $(d-4)$.
After completing this evaluation and sending $\ep\rightarrow 0$, we then recover precisely the renormalised 2-point function 
\eqref{appTTren}, and the 3-point Euler anomaly contribution; namely,  \eqref{p:anomdecomp} with \eqref{epsilonanomaly}, or equivalently \eqref{re_anomaly} with the terms proportional to $a$ in \eqref{B1eqn}\,-\,\eqref{B5eqn}.

\subsection{Relating 3-point function definitions}
\label{sec:conversion}

Here we collect for easy reference the formulae needed to convert the 3-point functions appearing in our previous work \cite{Bzowski:2013sza} to those of the present paper.  As discussed on page \pageref{sec:3ptdefn}, in this paper we define the 3-point functions by
\begin{align}
 &\<J^{\mu_1 a_1}(\x_1)J^{\mu_2 a_2}(\x_2)J^{\mu_3 a_3}(\x_3)\>
\nn\\&\qquad\qquad\qquad
 \equiv \frac{-1}{\sqrt{g(\x_1)g(\x_2)g(\x_3)}}\,\frac{\delta}{\delta A^{\mu_1 a_1}(\x_1)}\frac{\delta}{\delta A^{\mu_2 a_2}(\x_2)}\frac{\delta}{\delta A^{\mu_3 a_3}(\x_3)}W\Big|_0,
 \\[4ex]
 &\<T_{\mu_1\nu_1}(\x_1)J^{\mu_2 a_2}(\x_2)J^{\mu_3 a_3}(\x_3)\>
\nn\\&\qquad\qquad\qquad
 \equiv \frac{-2}{\sqrt{g(\x_1)g(\x_2)g(\x_3)}}\,\frac{\delta}{\delta g^{\mu_1\nu_1}(\x_1)}\frac{\delta}{\delta A^{\mu_2 a_2}(\x_2)}\frac{\delta}{\delta A^{\mu_3 a_3}(\x_3)}W\Big|_0,
 \\[4ex]
&\<T_{\mu_1\nu_1}(\x_1)T_{\mu_2\nu_2}(\x_2)T_{\mu_3\nu_3}(\x_3)\>
\nn\\&\qquad\qquad\qquad
 \equiv \frac{-8}{\sqrt{g(\x_1)g(\x_2)g(\x_3)}}\,\frac{\delta}{\delta g^{\mu_1\nu_1}(\x_1)}\frac{\delta}{\delta g^{\mu_2\nu_2}(\x_2)}\frac{\delta}{\delta g^{\mu_3\nu_3}(\x_3)}W\Big|_0.
\label{nowdefTTT}
\end{align}
Those used in \cite{Bzowski:2013sza} are instead defined through
\begin{align}
&\Big(\frac{-1}{\sqrt{g(\x_3)}}\frac{\delta}{\delta A^{\mu_3 a_3}(\x_3)}\Big)\Big(\frac{-1}{\sqrt{g(\x_2)}}\frac{\delta}{\delta A^{\mu_2 a_2}(\x_2)}\Big)\Big(\frac{-1}{\sqrt{g(\x_1)}}\frac{\delta}{\delta A^{\mu_1 a_1}(\x_1)}\Big)W\Big|_0\nn\\[1ex]
&\equiv\<J^{\mu_1 a_1}(\x_1)J^{\mu_2 a_2}(\x_2) J^{\mu_3 a_3}(\x_3)\>_{\mathrm{there}}, 
\\[3ex]
&\Big(\frac{-1}{\sqrt{g(\x_3)}}\frac{\delta}{\delta A^{\mu_3 a_3}(\x_3)}\Big)\Big(\frac{-1}{\sqrt{g(\x_2)}}\frac{\delta}{\delta A^{\mu_2 a_2}(\x_2)}\Big)\Big(\frac{-2}{\sqrt{g(\x_1)}}\frac{\delta}{\delta g^{\mu_1\nu_1}(\x_1)}\Big)W\Big|_0\nn\\[1ex]
&\equiv\<T_{\mu_1\nu_1}(\x_1)J^{\mu_2 a_2}(\x_2) J^{\mu_3 a_3}(\x_3)\>_{\mathrm{there}} \nn\\[1ex]&\quad
-\<\frac{\delta T_{\mu_1\nu_1}(\x_1)}{\delta A^{\mu_2 a_2}(\x_2)}J^{\mu_3 a_3}(\x_3)\>
-\<\frac{\delta T_{\mu_1\nu_1}(\x_1)}{\delta A^{\mu_3 a_3}(\x_3)}J^{\mu_2 a_2}(\x_2)\>,
\\[2ex]
&\Big(\frac{-2}{\sqrt{g(\x_3)}}\frac{\delta}{\delta g^{\mu_3\nu_3}(\x_3)}\Big)\Big(\frac{-2}{\sqrt{g(\x_2)}}\frac{\delta}{\delta g^{\mu_2\nu_2}(\x_2)}\Big)\Big(\frac{-2}{\sqrt{g(\x_1)}}\frac{\delta}{\delta g^{\mu_1\nu_1}(\x_1)}\Big)W\Big|_0\nn\\[1ex]
&\equiv\<T_{\mu_1\nu_1}(\x_1)T_{\mu_2\nu_2}(\x_2)T_{\mu_3\nu_3}(\x_3)\>_{\mathrm{there}} \nn\\[1ex]&\quad
-2\<\frac{\delta T_{\mu_1\nu_1}(\x_1)}{\delta g^{\mu_2\nu_2}(\x_2)}T_{\mu_3\nu_3}(\x_3)\>
-2\<\frac{\delta T_{\mu_1\nu_1}(\x_1)}{\delta g^{\mu_3\nu_3}(\x_3)}T_{\mu_2\nu_2}(\x_2)\>
-2\<\frac{\delta T_{\mu_2\nu_2}(\x_2)}{\delta g^{\mu_3\nu_3}(\x_3)}T_{\mu_1\nu_1}(\x_1)\>,
\end{align}
where the functional derivatives reflect the implicit dependence of the operators on the sources.  As 2-point functions of operators with different dimensions vanish in a CFT,  these terms can only appear when the corresponding dimensions match.

The correlators arising through these two definitions can then be related as follows. 
The current 3-point functions are identical,
\begin{align}
&\lla J^{\mu_1 a_1}(\bs{p}_1)J^{\mu_2 a_2}(\bs{p}_2) J^{\mu_3 a_3}(\bs{p}_3)\rra_{\mathrm{here}} = \lla J^{\mu_1 a_1}(\bs{p}_1)J^{\mu_2 a_2}(\bs{p}_2) J^{\mu_3 a_3}(\bs{p}_3)\rra_{\mathrm{there}},
\end{align}
while that of one stress tensor and two currents differs by 
\begin{align}
&\lla T_{\mu_1 \nu_1}(\bs{p}_1)J^{\mu_2 a_2}(\bs{p}_2) J^{\mu_3 a_3}(\bs{p}_3)\rra_{\mathrm{here}} 
=\lla T_{\mu_1 \nu_1}(\bs{p}_1)J^{\mu_2 a_2}(\bs{p}_2) J^{\mu_3 a_3}(\bs{p}_3)\rra_{\mathrm{there}} \nn\\[1ex]&
\qquad\quad
-\lla\frac{\delta T_{\mu_1\nu_1}(\bs{p}_1)}{\delta A^{\mu_2 a_2}(\bs{p}_2)}J^{\mu_3 a_3}(\bs{p}_3)\rra
-\lla\frac{\delta T_{\mu_1\nu_1}(\bs{p}_1)}{\delta A^{\mu_3 a_3}(\bs{p}_3)}J^{\mu_2 a_2}(\bs{p}_2)\rra.
\end{align}
Permuting the order of functional derivatives, this relation can equivalently be written
\begin{align}
&\lla T_{\mu_1 \nu_1}(\bs{p}_1)J^{\mu_2 a_2}(\bs{p}_2) J^{\mu_3 a_3}(\bs{p}_3)\rra_{\mathrm{here}} \nn\\[1ex]&\quad
=\lla T_{\mu_1 \nu_1}(\bs{p}_1)J^{\mu_2 a_2}(\bs{p}_2) J^{\mu_3 a_3}(\bs{p}_3)\rra_{\mathrm{there}} \nn\\[1ex]&
\quad\quad
-2\lla \frac{\delta J^{\mu_2 a_2}(\bs{p}_2)}{\delta g^{\mu_1 \nu_1}(\bs{p}_1)}J^{\mu_3 a_3}(\bs{p}_3)\rra
-2\lla\frac{\delta J^{\mu_3 a_3}(\bs{p}_3)}{\delta g^{\mu_1 \nu_1}(\bs{p}_1)}J^{\mu_2 a_2}(\bs{p}_2)\rra\nn\\[1ex]&\quad\quad
+\delta_{\mu_1\nu_1}\lla J^{\mu_2 a_2}(\bs{p}_2)J^{\mu_3 a_3}(-\bs{p}_2)\rra
+\delta_{\mu_1\nu_1}\lla J^{\mu_2 a_2}(\bs{p}_3)J^{\mu_3 a_3}(-\bs{p}_3)\rra.
\end{align}
The stress tensor 3-point functions are related by
\begin{align} 
&\lla T_{\mu_1\nu_1}(\bs{p}_1)T_{\mu_2\nu_2}(\bs{p}_2)T_{\mu_3\nu_3}(\bs{p}_3)\rra_{\mathrm{here}}\nn\\[1ex]
&\quad =\lla T_{\mu_1\nu_1}(\bs{p}_1)T_{\mu_2\nu_2}(\bs{p}_2)T_{\mu_3\nu_3}(\bs{p}_3)\rra_{\mathrm{there}} \nn\\[1ex]&\quad\quad
-2\lla \frac{\delta T_{\mu_1\nu_1}(\bs{p}_1)}{\delta g^{\mu_2\nu_2}(\bs{p}_2)}T_{\mu_3\nu_3}(\bs{p}_3)\rra
-2\lla \frac{\delta T_{\mu_1\nu_1}(\bs{p}_1)}{\delta g^{\mu_3\nu_3}(\bs{p}_3)}T_{\mu_2\nu_2}(\bs{p}_2)\rra
-2\lla\frac{\delta T_{\mu_2\nu_2}(\bs{p}_2)}{\delta g^{\mu_3\nu_3}(\bs{p}_3)}T_{\mu_1\nu_1}(\bs{p}_1)\rra
\nn\\[1ex]&\quad\quad
+\delta_{\mu_3\nu_3}\lla T_{\mu_1\nu_1}(\bs{p}_2)T_{\mu_2\nu_2}(-\bs{p}_2)\rra
+\delta_{\mu_3\nu_3}\lla T_{\mu_1\nu_1}(\bs{p}_1)T_{\mu_2\nu_2}(-\bs{p}_1)\rra\nn\\[1ex]&\quad\quad
+\delta_{\mu_2\nu_2}\lla T_{\mu_1\nu_1}(\bs{p}_3)T_{\mu_3\nu_3}(-\bs{p}_3)\rra.
\label{relatedefTTT}
\end{align}
Using these formulae, one can straightforwardly convert the results of \cite{Bzowski:2013sza} to our present definitions.  
The form of the transverse and trace Ward identities, along with the associated reconstruction formulae, change to those listed here.
The primary and secondary CWI take the same form with either definition of the 3-point function, although as the transverse Ward identities are different, the specific coefficients appearing on the right-hand sides of the secondary CWIs may take different values.  The form factors then differ by terms that are at most semilocal.  
All such terms come from expanding the functional derivatives in a local basis of operators, since the 2-point functions above do not contribute to the transverse-traceless form factors. 

\subsection{Evaluation of counterterm contributions}
\label{sec:evalctcontr}

In this appendix, we evaluate the counterterm contributions to the form factors for the stress tensor 3-point function in four dimensions.
As these form factors are associated with the transverse-traceless part of the correlator, 
it suffices to work in a gauge where the {\it inverse} metric perturbation is transverse traceless,\footnote{In this section, all raised indices should be understood as being raised with the full perturbed metric. Repeated lowered indices should be summed using the flat background metric. Commas denote partial derivatives and $\p^2 = \p_\mu\p_\mu$.  Our conventions follow those of \cite{Wald}.} 
\[
g^{\mu\nu}=\delta_{\mu\nu}+\g_{\mu\nu}, \qquad \g_{\mu\mu}=0, \qquad \g_{\mu\nu,\nu}=0,
\]
since all other metric components are projected out in the calculation of form factors.

Writing $g_{\mu\nu}=\delta_{\mu\nu}+h_{\mu\nu}$, we then have
\[\label{deltagexp}
h_{\mu\nu} = -\g_{\mu\nu}+\g_{\mu\alpha}\g_{\alpha\nu}+O(\g^3)
\]
where
\[\label{newgauge}
h=h_{\mu\mu} = \g_{\mu\nu}\g_{\mu\nu}+O(\g^3), \qquad
h_{\mu\nu,\nu} = \g_{\mu\alpha,\nu}\g_{\alpha\nu}+O(\g^3).
\]
The Ricci curvature 
\begin{align}\label{Rexpressions}
R_{\mu\nu} &= -\frac{1}{2}\p^2h_{\mu\nu} -\frac{1}{2}h_{,\mu\nu}+h_{\alpha(\mu,\nu)\alpha} +\frac{1}{4}(h_{\alpha\beta}h_{\alpha\beta})_{,\mu\nu}-h_{\alpha\beta} \hat{S}_{\beta\mu\nu,\alpha}-\hat{S}_{\alpha\beta\mu}\hat{S}_{\beta\alpha\nu}+O(\g^3)
\end{align}
where
\[
\hat{S}_{\mu\nu\alpha} = \Gamma^{(1)\mu}_{\quad \nu\alpha} = \frac{1}{2}(h_{\mu\nu,\alpha}+h_{\mu\alpha,\nu}-h_{\nu\alpha,\mu})
\]
and we have used the fact that to $O(\g^3)$ we can treat $h_{\mu\nu}$ as transverse traceless where it appears quadratically.
We thus have
\begin{align}
R^{(1)}_{\mu\nu} &= \frac{1}{2}\p^2 \g_{\mu\nu}, \\
R^{(2)}_{\mu\nu} &= -\frac{1}{2}\p^2(\g_{\mu\alpha}\g_{\alpha\nu})+  (\g_{\alpha\beta}\g_{\alpha(\mu})_{,\nu)\beta}
-\frac{1}{4}(\g_{\alpha\beta}\g_{\alpha\beta})_{,\mu\nu}-\g_{\alpha\beta} S_{\beta\mu\nu,\alpha}-S_{\alpha\beta\mu}S_{\beta\alpha\nu}
\end{align}
where
\[
S_{\mu\nu\alpha} =  \frac{1}{2}(\g_{\mu\nu,\alpha}+\g_{\mu\alpha,\nu}-\g_{\nu\alpha,\mu}).
\]
The scalar curvature
\begin{align}
R^{(1)} &= 0,\\
R^{(2)} & = -\g_{\mu\nu}\p^2 \g_{\mu\nu}-\frac{5}{4}\g_{\mu\nu,\alpha}\g_{\mu\nu,\alpha}+\frac{1}{2}\g_{\mu\nu,\alpha}\g_{\mu\alpha,\nu},
\end{align}
while the Riemann curvature
\begin{align}\label{Riemannexp}
R^{(1)}_{\mu\nu\alpha\beta} &= -2 S_{\mu\nu[\alpha,\beta]} \\
R^{(2)}_{\mu\nu\alpha\beta} &= -2S_{\lambda\nu[\alpha}\gamma_{\mu\lambda,\beta]}+2S_{\lambda \nu[\beta}S_{\mu\lambda \alpha]}.
\end{align}
We now find the Ricci-squared counterterm,
\begin{align}
&\int\D^{4+2u\ep}\x\,\sqrt{g}\mu^{2u\ep} R_{\mu\nu}R^{\mu\nu} \nn\\[1ex]&\qquad
=\int\D^{4+2u\ep}\x\,\mu^{2u\ep}\Big[\frac{1}{4}\p^2\g_{\mu\nu}\p^2 \g_{\mu\nu}
+ \p^2 \g_{\mu\nu}\Big({-}\frac{1}{4}\g_{\alpha\beta,\mu}\g_{\alpha\beta,\nu}
+\g_{\alpha\beta,\nu}\g_{\mu\beta,\alpha}
\nn\\[1ex]&\qquad\qquad\qquad\quad
-\frac{1}{2}\g_{\mu\alpha}\p^2 \g_{\alpha\nu} +\frac{1}{2}\g_{\alpha\beta}\g_{\mu\nu,\alpha\beta}-\frac{1}{2}\g_{\alpha\mu,\beta}\g_{\beta\nu,\alpha}-\frac{1}{2}\g_{\alpha\mu,\beta}\g_{\alpha\nu,\beta}\Big)\Big],
\end{align}
generates the form factor contributions 
\begin{align}
A^{\Ct}_3 &= 4\mu^{2u\ep}(p_1^2+p_2^2+p_3^2),\\   
A^{\Ct}_4 &= 8\mu^{2u\ep} (p_1^2+p_2^2+p_3^2), \\
A^{\Ct}_5 &= -4 \mu^{2u\ep}(p_1^4+p_2^4+p_3^4),
\end{align}
while the Riemann-squared counterterm, 
\begin{align}
&\int\D^{4+2u\ep}\x\,\sqrt{g}\mu^{2u\ep} R_{\mu\nu\alpha\beta}R^{\mu\nu\alpha\beta} = \int \D^{4+2u\ep}\x\,\mu^{2u\ep} \Big[\p^2 \g_{\mu\nu}\p^2 \g_{\mu\nu} 
-2 \g_{\mu\alpha}\g_{\alpha\nu}\p^4 \g_{\mu\nu}
\nn\\[1ex]&\qquad\qquad\quad\qquad
-2S_{\lambda\mu\nu}S_{\lambda \beta\alpha }\g_{\alpha\beta,\mu\nu}
+4S_{\beta\mu\alpha,\nu}S_{\lambda\mu\alpha}S_{\lambda\nu\beta}-16S_{\beta\mu\alpha,\nu}S_{\lambda\mu [\nu,\alpha]}\g_{\beta\lambda}\Big],
\end{align}
generates the form factor contributions
\begin{align}
A^{\Ct}_2&=-16\mu^{2u\ep},\\
A^{\Ct}_3 &= 16\mu^{2u\ep}(p_1^2+p_2^2)+8\mu^{2u\ep}p_3^2,\\
A^{\Ct}_4&=24\mu^{2u\ep}(p_1^2+p_2^2)+40\mu^{2u\ep}p_3^2,\\
A^{\Ct}_5& = -12\mu^{2u\ep}(p_1^4+p_2^4+p_3^4)-8\mu^{2u\ep}(p_1^2p_2^2+p_1^2 p_3^2+p_2^2p_3^2).
\end{align}
Since the counterterm ${\int}\D^{4+2u\ep}\x\sqrt{g}\mu^{2u\ep}\,R^2$ vanishes at cubic order in $\g_{\mu\nu}$, the counterterm action
\begin{align}
S_{\Ct}=\int\D^{4+2u\ep}\x\, \sqrt{g} \mu^{2u\ep}\,(\mathfrak{a} E_4 + \mathfrak{b}R^2+\mathfrak{c} W^2),
\end{align}
where $E_4$ and $W^2$ are defined in \eqref{Eulerdef} and \eqref{Weyldef},
therefore generates a form factor contribution
\begin{align}
A^{\Ct}_1&= 0,\\
A^{\Ct}_2 &= -16 (\mathfrak{c}+\mathfrak{a})\mu^{2u\ep},\\
A^{\Ct}_3 &= 8 \mathfrak{c}\mu^{2u\ep}(p_1^2+p_2^2)-8 \mathfrak{a} \mu^{2u\ep} p_3^2,\\
A^{\Ct}_4 &= 
8\mathfrak{c}\mu^{2u\ep}(p_1^2+p_2^2+3p_3^2)-8\mathfrak{a}\mu^{2u\ep}(p_1^2+p_2^2-p_3^2),\\
A^{\Ct}_5 &=-4\mathfrak{c}\mu^{2u\ep}(p_1^2+p_2^2+p_3^2)^2-4\mathfrak{a}\mu^{2u\ep}J^2.
\end{align}
As expected, the Euler counterterm generates no contribution at quadratic order in $\gamma_{\mu\nu}$.  Note its variation can also be computed as described in section \ref{sec:Eulerdeg}; projecting  \eqref{Eulerctcontr5form} into a transverse traceless basis leads to the same form factors as we find here.

\subsection{Degeneracies in three dimensions}
\label{app:3ddeg}

 In three dimensions, as described in appendix B of \cite{Bzowski:2013sza}, the existence of a vector cross-product reduces the number of independent form factors in the stress tensor 3-point function from five to two.
In this appendix, we show this result can also be understood as a dimension-dependent identity analogous to our four-dimensional discussion in section \ref{sec:Eulerdeg}.
 
In three dimensions, we have the identity
\[\label{3ddeg}
0=6\,\delta_{[\alpha_1}^{\beta_1}\delta_{\alpha_2}^{\beta_2}K_{\alpha_4\alpha_5]}{}^{\alpha_4\alpha_5} = K_{\alpha_1\alpha_2}{}^{\beta_1\beta_2}-4\delta_{[\alpha_1}^{[\beta_1}K_{\alpha_2]}^{\beta_2]}+K\delta_{[\alpha_1}^{\beta_1}\delta_{\alpha_2]}^{\beta_2},
\]
where $K_{\alpha_1\alpha_2}{}^{\beta_1\beta_2}$ is as defined in \eqref{ourK}.  As in section \ref{sec:Eulerdeg}, the left-hand side vanishes due to a necessary repetition of indices in the totally antisymmetrised product.  In fact, this identity is simply the trace of our four-dimensional identity \eqref{Edgar}.
Unlike in four dimensions, however, in three dimensions there are no divergences once the primary constants have been suitably redefined (as described in section \ref{sec:tttodd}).  
We can thus set $d=3$ exactly rather than considering the dimensionally regulated theory.

Taking the transverse-traceless projection of \eqref{3ddeg} yields the three-dimensional degeneracy
\begin{align}
0 & =4\,\Pi_{\mu_1\nu_1}{}^{\alpha_1}{}_{\beta_1}(\bs{p}_1)\Pi_{\mu_2\nu_2}{}^{\alpha_2}{}_{\beta_2}(\bs{p}_2)\Big[K_{\alpha_1\alpha_2}{}^{\beta_1\beta_2}-4\delta_{[\alpha_1}^{[\beta_1}K_{\alpha_2]}^{\beta_2]}+K\delta_{[\alpha_1}^{\beta_1}\delta_{\alpha_2]}^{\beta_2}\Big]
\nn\\[1ex]&
= \Pi_{\mu_1\nu_1\alpha_1\beta_1}(\bs{p}_1)\Pi_{\mu_2\nu_2\alpha_2\beta_2}(\bs{p}_2)\Big[p_2^{\alpha_1}p_2^{\beta_1}p_3^{\alpha_2}p_3^{\beta_2}\nn\\&\qquad\qquad\qquad
+(p_3^2-p_1^2-p_2^2)\delta^{\alpha_1\alpha_2}p_2^{\beta_1}p_3^{\beta_2} -\frac{J^2}{4}\delta^{\alpha_1\alpha_2}\delta^{\beta_1\beta_2}\Big].
\end{align}
Let us now multiply this identity by $f\Pi_{\mu_3\nu_3\alpha_3\beta_3}(\bs{p}_3) p_1^{\alpha_3}p_1^{\beta_3}$, where  $f\equiv f(p_1,p_2,p_3)=f(p_2,p_1,p_3)$ is an arbitrary function of dilatation weight minus two, as required on dimensional grounds.  After symmetrising under permutations, we then find the degenerate form factor combination
\begin{align}\label{fdeg}
A_1&=f+f(p_1\leftrightarrow p_3)+f(p_2\leftrightarrow p_3),\nn\\ A_2 &= (p_3^2-p_1^2-p_2^2)f, \nn\\ A_3 &= -\frac{1}{4}J^2 f,\nn \\ A_4&=0,\nn\\ A_5&=0.
\end{align} 
Two further degenerate form factor combinations can be found by considering instead 
\begin{align}
 0 & =-4\,\Pi_{\mu_1\nu_1}{}^{\alpha_1}{}_{\beta_1}(\bs{p}_1)\Pi_{\mu_2\nu_2}{}^{\alpha_2}{}_{\beta_2}(\bs{p}_2)\Pi_{\mu_3\nu_3}{}^{\alpha_3}{}_{\beta_3}(\bs{p}_3) \Big[K_{\alpha_1\alpha_3}{}^{\beta_2\beta_3}-4\delta_{[\alpha_1}^{[\beta_2}K_{\alpha_3]}^{\beta_3]}+K\delta_{[\alpha_1}^{\beta_2}\delta_{\alpha_3]}^{\beta_3}\Big]\nn\\[1ex]&
 =\Pi_{\mu_1\nu_1\alpha_1\beta_1}(\bs{p}_1)\Pi_{\mu_2\nu_2\alpha_2\beta_2}(\bs{p}_2)\Pi_{\mu_3\nu_3\alpha_3\beta_3}(\bs{p}_3) \Big[
 p_2^{\alpha_1}p_3^{\beta_2}p_1^{\alpha_3}p_1^{\beta_3}+p_3^2\delta^{\alpha_1\beta_2}p_1^{\alpha_3}p_1^{\beta_3}\nn\\[0.5ex]&\qquad +\frac{1}{2}(p_1^2-p_2^2+p_3^2)\delta^{\alpha_3\beta_2}p_2^{\alpha_1}p_1^{\beta_3}+\frac{1}{2}(-p_1^2+p_2^2+p_3^2)\delta^{\alpha_1\beta_3}p_3^{\beta_2}p_1^{\alpha_3}+\frac{J^2}{4}\delta^{\alpha_1\beta_3}\delta^{\alpha_3\beta_2}\Big].
\end{align}
We can now multiply by either $g p_2^{\beta_1}p_3^{\alpha_2}$ or else $h\delta^{\alpha_2\beta_1}$, where on dimensional grounds we have introduced the arbitrary functions $g\equiv g(p_1,p_2,p_3)=g(p_2,p_1,p_3)$ of weight minus two and $h\equiv h(p_1,p_2,p_3)=h(p_2,p_1,p_3)$ of weight zero.  Symmetrising under  permutations then yields  the additional degenerate form factor combinations
\begin{align}\label{gdeg}
A_1&=g+g(p_1\leftrightarrow p_3)+g(p_2\leftrightarrow p_3),\nn\\ A_2 &= p_3^2 g+\frac{1}{2}(p_1^2-p_2^2+p_3^2)g(p_1\leftrightarrow p_3)+\frac{1}{2}(-p_1^2+p_2^2+p_3^2)g(p_2\leftrightarrow p_3),\nn\\ A_3&=0,\nn \\ A_4 &= \frac{1}{4}J^2 g, \nn\\ A_5 &= 0,
\end{align}
and similarly,
\begin{align}\label{hdeg}
A_1&=0,\nn\\ 
A_2 &= h,\nn\\ 
A_3&=p_3^2 h,\nn\\ 
A_4&=\frac{1}{2}(p_1^2+p_2^2-p_3^2)\Big(h(p_1\leftrightarrow p_3)+h(p_2\leftrightarrow p_3)\Big), \nn\\ 
A_5&=\frac{1}{4}J^2\Big(h+h(p_1\leftrightarrow p_3)+h(p_2\leftrightarrow p_3)\Big).
\end{align}

As there are no further ways (modulo permutations) of contracting the four indices of $K_{\alpha_1\alpha_2}{}^{\beta_1\beta_2}$ with three transverse-traceless projectors, the three degeneracies \eqref{fdeg}, \eqref{gdeg} and \eqref{hdeg} exhaust the list of possibilities for the stress tensor 3-point function.
(An alternative derivation of these three degeneracies is also discussed below.)
Through appropriate choice of the arbitrary functions $f$, $g$ and $h$, we can, if we wish,  set any three of the five form factors to zero. The two remaining form factors are then non-degenerate.\footnote{For example, in appendix B of \cite{Bzowski:2013sza}, the choice  $A_3=A_4=A_5=0$ was made.    
From (B.6) and (B.7) of \cite{Bzowski:2013sza}, the two remaining form factors (labelled there $B_1$ and $B_2$) are then invariant under the degenerate combinations 
\eqref{fdeg}, \eqref{gdeg} and \eqref{hdeg}.  The primary constant $C_2$ drops out, since its contribution is purely degenerate as can be shown using a totally symmetric ansatz for $h$ in \eqref{hdeg}.}

As our three-dimensional identity \eqref{3ddeg} is simply the trace of its four-dimensional counterpart \eqref{Edgar}, we expect that our previous four-dimensional degeneracy \eqref{Eulerdeg} should also be valid in three dimensions.  This is indeed the case, as can be seen by setting 
\[
f = -\frac{4}{3J^2}p_3^2 H, \qquad g = \frac{4}{3J^2}(p_1^2+p_2^2-p_3^2)H, \qquad h=\frac{2}{3}H,
\]
where $H=H(p_1,p_2,p_3)$ is an arbitrary function of dilatation weight zero that is completely symmetric under permutation of all momenta.
Upon summing \eqref{fdeg}, \eqref{gdeg} and \eqref{hdeg}, we then recover
\begin{align}
A_1 = 0, \quad A_2 = 2H, \quad A_3= p_3^2 H, \quad A_4 = (p_1^2+p_2^2-p_3^2)H, \quad A_5 = \frac{1}{2}J^2 H.
\end{align}
This degenerate combination is equivalent to \eqref{Eulerdeg} multiplied by the arbitrary symmetric function $H$, which is the most general form of the degeneracy in four dimensions.
The four-dimensional degeneracy is thus a subset of those in three dimensions.

It is interesting to consider the three-dimensional degeneracy \eqref{3ddeg} from a  geometrical perspective.  Unlike in four-dimensions, where the degeneracy \eqref{Eulerdeg} derives from the existence of an evanescent  counterterm, in three dimensions there are no counterterms.  Instead, the identity \eqref{3ddeg} is analogous to the vanishing of the Weyl tensor: replacing $K_{\alpha_1\alpha_2}{}^{\beta_1\beta_2}$ in \eqref{3ddeg} by the Riemann curvature $R_{\alpha_1\alpha_2}{}^{\beta_1\beta_2}$, the right-hand side is equal to the Weyl tensor $W_{\alpha_1\alpha_2}{}^{\beta_1\beta_2}$, which vanishes in three dimensions by precisely this identity.

Finally, note that the degeneracies \eqref{fdeg}, \eqref{gdeg} and \eqref{hdeg} can also be derived by introducing appropriate antisymmetrisations in the tensor structures associated with the various form factors.
For example, we can eliminate the  form factor $A_3$ by writing
\begin{align}
&\Pi_{\mu_1\nu_1\alpha_1\beta_1}(\bs{p}_1)\Pi_{\mu_2\nu_2\alpha_2\beta_2}(\bs{p}_2)
\Pi_{\mu_3\nu_3\alpha_3\beta_3}(\bs{p}_3)\Big[
 \: A_3 \delta^{\alpha_1 \alpha_2} \delta^{\beta_1 \beta_2} p_1^{\alpha_3} p_1^{\beta_3} \nonumber \\[1ex]
& \qquad \qquad \qquad 
+ A_3(p_1 \leftrightarrow p_3) \delta^{\alpha_2 \alpha_3} \delta^{\beta_2 \beta_3} p_2^{\alpha_1} p_2^{\beta_1} 
 + \: A_3(p_2 \leftrightarrow p_3) \delta^{\alpha_1 \alpha_3} \delta^{\beta_1 \beta_3} p_3^{\alpha_2} p_3^{\beta_2} \Big] \nn\\[1ex]& = 
2\,\Pi_{\mu_1\nu_1\alpha_1\beta_1}(\bs{p}_1)\Pi_{\mu_2\nu_2\alpha_2\beta_2}(\bs{p}_2)
\Pi_{\mu_3\nu_3\alpha_3\beta_3}(\bs{p}_3)\Big[
 \: A_3 \delta^{\alpha_1 [\alpha_2} \delta^{\beta_1] \beta_2} p_1^{\alpha_3} p_1^{\beta_3}  \nonumber \\[1ex]
& \qquad \qquad \qquad 
+ A_3(p_1 \leftrightarrow p_3) \delta^{\alpha_2 [\alpha_3} \delta^{\beta_2] \beta_3} p_2^{\alpha_1} p_2^{\beta_1}
+ \: A_3(p_2 \leftrightarrow p_3) \delta^{\alpha_1 [\alpha_3} \delta^{\beta_1] \beta_3} p_3^{\alpha_2} p_3^{\beta_2} \Big].
\end{align}
Using the three-dimensional identity \eqref{3ddeg}, we can now reduce the right-hand side to terms involving at most one metric tensor (rather than the two we started with).  This allows us to rewrite $A_3$ in terms of the form factors $A_1$ and $A_2$, yielding the first degeneracy \eqref{fdeg}.  The form factor $A_4$ can be eliminated in a similar fashion, leading to  the second degeneracy \eqref{gdeg}.  The third degeneracy \eqref{hdeg} can be obtained by eliminating $A_2$.  To do this, we write  
\begin{align}
&\Pi_{\mu_1\nu_1\alpha_1\beta_1}(\bs{p}_1)\Pi_{\mu_2\nu_2\alpha_2\beta_2}(\bs{p}_2)
\Pi_{\mu_3\nu_3\alpha_3\beta_3}(\bs{p}_3)\Big[
 \: A_2 \delta^{\beta_1 \beta_2} p_2^{\alpha_1} p_3^{\alpha_2} p_1^{\alpha_3} p_1^{\beta_3}
\nonumber \\[1ex]
& \qquad \qquad \qquad   
  + A_2(p_1 \leftrightarrow p_3) \delta^{\beta_2 \beta_3} p_2^{\alpha_1} p_2^{\beta_1} p_3^{\alpha_2} p_1^{\alpha_3} +\: A_2(p_2 \leftrightarrow p_3) \delta^{\beta_1 \beta_3} p_2^{\alpha_1} p_3^{\alpha_2} p_3^{\beta_2} p_1^{\alpha_3} \Big]\nonumber \\[1ex]
&
=-4\,\Pi_{\mu_1\nu_1\alpha_1\beta_1}(\bs{p}_1)\Pi_{\mu_2\nu_2\alpha_2\beta_2}(\bs{p}_2)
\Pi_{\mu_3\nu_3\alpha_3\beta_3}(\bs{p}_3)\Big[
 \: A_2 \delta^{\beta_1 \beta_2} K^{\alpha_1\alpha_3\alpha_2\beta_3}
\nonumber \\[1ex]
& \qquad \qquad \qquad   
   A_2(p_1 \leftrightarrow p_3) \delta^{\beta_2 \beta_3} K^{\alpha_2\alpha_1\alpha_3\beta_1} +\:  A_2(p_2 \leftrightarrow p_3) \delta^{\beta_1 \beta_3} K^{\alpha_1\alpha_2\alpha_3\beta_2} \Big].
\end{align}
Using \eqref{3ddeg}, we can now reduce the right-hand side to terms involving two or three metric tensors.  In this fashion, the form factor $A_2$ can be rewritten in terms of the form factors $A_3$, $A_4$ and $A_5$. 
From this latter approach it is clear that similar degeneracies also exist for the form factors of other three-dimensional correlators besides that of three stress tensors.
In four dimensions, however, of the correlators we study here, only that of three stress tensors has degenerate form factors.

\bibliographystyle{JHEP}
\bibliography{cwis2017}

\end{document}